\documentclass{ws-rv9x6}
\usepackage{ws-rv-thm}   % comment this line when `amsthm / theorem / ntheorem` package is used
\usepackage[square]{ws-rv-van}   % numbered citation & references (default)

\usepackage{graphicx}
\usepackage{subcaption}
\usepackage{wrapfig}
\usepackage[colorlinks=true,linkcolor=blue,citecolor=blue,urlcolor=blue]{hyperref}

\makeindex

\begin{document}

\chapter[Image-Based Jet Analysis]{Image-Based Jet Analysis}\label{JI_ch}
\vspace{-1cm}
\author[M. Kagan]{Michael Kagan\footnote{makagan@slac.stanford.edu}}
\address{SLAC National Accelerator Laboratory}

\begin{abstract}
Image-based jet analysis is built upon the \textit{jet image} representation of jets that enables a direct connection between high energy physics and the fields of computer vision and deep learning. Through this connection, a wide array of new jet analysis techniques have emerged. In this text, we survey jet image based classification models, built primarily on the use of convolutional neural networks, examine the methods to understand what these models have learned and what is their sensitivity to uncertainties, and review the recent successes in moving these models from phenomenological studies to real world application on experiments at the LHC. Beyond jet classification, several other applications of jet image based techniques, including energy estimation, pileup noise reduction, data generation, and anomaly detection, are discussed.

\end{abstract}
%\markright{Customized Running Head for Odd Page} % default is Chapter Title.
\body

\tableofcontents
\clearpage

\section{Introduction}\label{JI_sec1}
The \textit{jet image}~\cite{Cogan2015JetimagesCV} approach to jet tagging is built upon the rapidly developing field of Computer Vision (CV) in Machine Learning (ML). Jets~\cite{LookingInsideJets,Shelton2013} are collimated streams of particles produced by the fragmentation and hadronizaton of high energy quarks and gluons. The particles are subsequently measured by particle detectors and clustered with jet clustering algorithms to define the jets. Jet images view the energy depositions of the stream of particles comprising a jet within a fixed geometric region of a detector as an image, thereby connecting particle detector measurements with an image representation and allowing the application of image analysis techniques from CV. In this way, models built upon advancements in deep convolutional neural networks (CNN) can be trained for jet classification, energy determination through regression, and the reduction of noise e.g. from simultaneous background  interactions at a high intensity hadron collider such as the Large Hadron Collider (LHC). Throughout this text, the focus will be on the use of jet image techniques studied within the context of hadron colliders like the LHC~\cite{Evans_2008}.

Jet images form a representation of jets highly connected with the detector; one can look at segmented detectors as imaging devices and interpret the measurements as an image. In contrast, other representations of jets exist that are built more closely from the physics of jet formation, such as viewing jets as sequences~\cite{Andreassen2019,ATL-PHYS-PUB-2020-014} or trees~\cite{Louppe2019} formed through a sequential emission process, or viewing jets as sets, graphs, or point clouds~\cite{Komiske2019,PhysRevD.101.056019} with the geometric relationship between constituents of the jet encoded in the adjacency matrix and node properties. There are overlaps in these approaches, for instance a graph can be defined over detector energy measurements, but these approaches will not be discussed in detail in this chapter. The utilization of an image-based approach comes with the major advantage that CV is a highly developed field of ML with some of the most advanced models available for application to jet analysis with jet images. From the experimental viewpoint, the detector measurements are fundamental to any subsequent analysis, and the detailed knowledge of the detector and its systematic uncertainties can be highly advantageous for analysis of LHC data.

Among the earliest use of jet images was for the classification of the parent particle inducing the jet~\cite{Cogan2015JetimagesCV}, and relied on utilizing linear discriminants trained on image representations of jets for this task. While the remainder of this text will focus on deep learning approaches to jet images, even this early work saw interesting discrimination power for this task. By utilizing the detector measurements directly, rather than relying on jet features developed using physics domain knowledge, additional discrimination power could be extracted. Deep learning approaches surpass such linear methods, but build on this notion of learning discriminating information from detector observables rather than engineered features.

\begin{figure}[h]
\begin{center}
\includegraphics[width=0.6\linewidth]{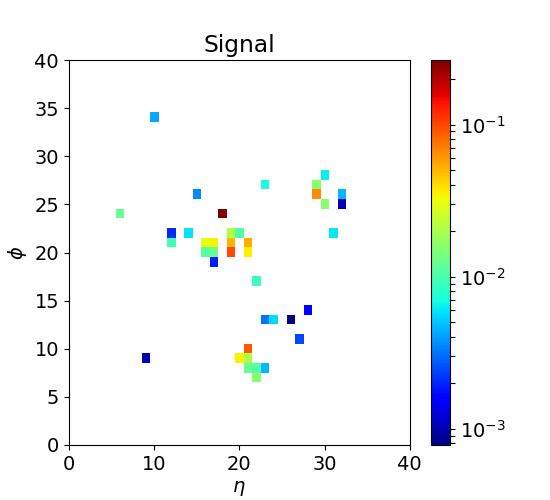}
\end{center}
\caption{An example jet image of a Lorentz boosted top quark jet after preprocessing has been applied~\cite{Kasieczka_2019}.}
\label{fig:singleimage}
\end{figure}%

While designed to take advantage of advances in computer vision, jet images have notable differences with respect to typical natural images in CV. Jet images are sparse, with most pixels in the image having zero content. This is markedly different from natural images that tend to have all pixels containing content. Moreover, jet images tend to have multiple localized regions of high density in addition to diffusely located pixels throughout the image, as opposed to the smooth structures typically found in natural images. An example top quark jet image illustrating these features can be seen in Figure~\ref{fig:singleimage}. These differences can lead to notable challenges, for instance the number of parameters used in jet image models (and consequently the training time) tend to be large to account for the size of the image, even though most pixels carry no information. Some techniques exist for sparse-image computer vision approaches~\cite{graham2017submanifold}, but have not been explored in depth within the jet image community.

This text will first discuss jets and typical jet physics in Section~\ref{sec:jets}. The formation of jet images and the jet image preprocessing steps before classification are discussed in Section~\ref{sec:jetimages}. A brief introduction to Computer Vision is found in Section~\ref{sec:CV}.   The application of jet images in various jet classification problems is then discussed in Section~\ref{sec:tagging}, followed by a discussion on the interpretation of information learned by jet image based classifiers in Section~\ref{sec:understand}. Some recent applications of jet images beyond classification are discussed in Section~\ref{sec:other}. A brief note on the notation used throughout the text follows below (Section~\ref{subsec:notation}).

It should be noted that the majority of the studies presented in this text relate to phenomenological work using simplified setting that often do not include a realistic modeling of a detector's impact on observables. These will frequently be denoted as \textit{phenomenological studies}, in contrast to the studies using realistic detector simulations or using real experiment data that are discussed mainly in Section~\ref{subsec:ji_exp}. 

\subsection{Notations and Definitions}\label{subsec:notation}
As we will focus on studies of jet images in the LHC setting, we will primarily utilize the hadron collider coordinate system notation. The beam-line defines the $z$-axis, $\phi$ indicates the azimuthal angle, $\eta = -\log\tan\frac{\theta}{2}$ is the pseudo-rapidity which is a transformation of the polar angle $\theta$. The rapidity is defined as $y=\frac{1}{2}\log\Big[ \frac{E+p_z}{E-p_z}\Big]$ is frequently used for the polar measurement of massive particles, such as jets, as differences in rapidity are invariant with respect to Lorentz boosts along the beam direction. The angular separation of particles is defined as $\Delta R(p_1 , p_2) = \sqrt{ (y_1 - y_2)^2 + (\phi_1 - \phi_2)^2}$.  The transverse momentum $p_T = \sqrt{p_x^2 + p_y^2} $, is frequently used as it is invariant with respect to Lorentz boosts along the beam direction. The transverse energy is defined as $E_T = E\sin(\theta)$.

\section{Jets and Jet Physics Challenges}\label{sec:jets}

Jets are collimated streams of particles produced by the fragmentation and hadronizaton of high energy quarks and gluons. Jet clustering algorithms are used to combine particles into clusters that define the jets (See references~\cite{LookingInsideJets,Shelton2013} for recent reviews). At the LHC, jet algorithms typically rely on sequential reclustering algorithms which, given a definition of distance, iteratively combine the closest two constituents (either particles or previously combined sets of particles denoted \textit{proto-jets}) until a stopping condition is met. Different distance metrics define different jet algorithms and perhaps the most commonly used algorithm at the LHC is the anti-$k_T$ algrithm~\cite{Cacciari_2008_akt} in which the distance between particle \textit{i} and particle \textit{j} is defined as $d_{ij}=\min\{k_{T, i}^{-2}\ , k_{T, j}^{-2}\}\Delta_{ij}^2 / R^2$. Here, $\Delta_{ij}^2 = (y_i - y_j)^2 + (\phi_i - \phi_j)^2$ and $y$, $\phi$, and $k_T$ are the particle rapidity, azimuth, and transverse momentum, respectively. The parameter $R$ of the jet algorithm has the effect of defining the spatial span, or approximate ``radius" (though the jet is not necessarily circular), of the jet. Jets and jet algorithms are required to be \textit{IRC safe}, i.e. insensitive to additional infrared radiation or collinear splittings of particles, in order for the jet properties to be calculable in fixed-order perturbation theory. This allows comparison between jets clustered on partons from the hard scattering process, referred to as parton jets, on final state particles after showering and hadronization simulation, referred to as particle jets, and on reconstructed particles in detectors, referred to as reconstructed jets. 

Most of the work presented in this text are phenomenological studies outside the context of any individual experiment. These studies primarily utilize particle level simulation after fragmentation and hadronization and thus study particle jets defined after clustering the final state particles. These studies typically do not use a simulation of a detector and its impact on particle kinematic measurements. Studies of jets and jet images after real detector simulation or in real detector data are discussed in Section~\ref{subsec:ji_exp}. In the detector setting, various inputs to jet algorithms can be used to define jets: (i) towers refer to a fixed spatial extent in $\eta$ and $\phi$ in which all energy within the longitudinal depth of the calorimeter is summed, (2) topological clusters~\cite{Aad2017} are used to cluster together  energy depositions in nearby calorimeter cells, (3) tracks, or charged particle trajectories, measured using tracking detectors. The Particle Flow (PF) algorithm~\cite{2017JInst..12P0003S} is used by the CMS collaboration to match charged particles with energy in the calorimeter in order to utilize both measurements to define PF candidates that can be used as inputs to jet algorithms.

The $R$-parameter of the jet is used to define the spatial extent to which particles are clustered into the jet. When studying quark and gluon jets, $R=0.4$ is frequently used. When studying the decay of Lorentz boosted heavy particles, in which multiple partons may be spatially collimated, \textit{large-$R$} jets are often used which have a larger $R=1.0$ or $R=1.2$. \textit{Subjets}, defined by running a jet clustering algorithm with smaller radius on the constituents of a jet, are frequently used to study the internal properties of a jet. More broadly, \textit{jet substructure} refers to the study of the internal structure of jets and the development of theoretically motivated jet features which are useful for discrimination and inference tasks (See references~\cite{Asquith:2018igt,Larkoski:2017jix} for recent reviews). 

One particularly important feature of a jet is the jet mass, computed as: $m^2 = \big( \sum_{i \in \textrm{jet}} p_i \big)^2$. The sum of four-vectors runs over all the constituents $i$ clustered into the jet. As different heavy resonances have different masses, this feature can be a strong discriminant between jet types. Note that any operation performed on a jet which alter the constituents, such as the pileup mitigation discussed in the next paragraph, may alter the jet mass. 

It is important to note that additional proton-proton interactions within a bunch crossing, or pileup, creates additional particles present within an event that can impact jet clustering and the estimation of jet properties. This is especially important for large-$R$ jets which cover large spatial extents. Dedicated pileup removal algorithms are used to mitigate the impact of pileup~\cite{Soyez_2019}. Jet trimming~\cite{Krohn_2010} is a jet grooming technique used to remove soft and wide angle radiation from jets, in which the constituents within the jet using a jet algorithm with a smaller radius to define subjets and remove subjets carrying a fraction of the jet energy below a threshold, and is frequently used on ATLAS to aid in pileup mitigation. The pileup per particle identification algorithm (PUPPI)~\cite{Bertolini_2014} is frequently used by CMS, in which for each particle a local shape parameter, which probes the collinear versus soft diffuse structure in the neighborhood of the particle, is calculated.  The distribution of this shape parameter per event is used to calculate per particle weights that describe the degree to which particles are pileup-like. Particle four-momenta are then weighted and thus the impact of (down-weighted) pileup particles on jet clustering is reduce~\cite{Bertolini_2014}. Pileup mitigation can greatly improve the estimation of the jet mass, energy and momentum by removing / downweighting the pileup particles clustered into a jet that only serve as noise in the jet properties estimation.

Jet identification, energy estimation, and pileup estimation / reduction are among the primary challenges for which the jet images approach has been employed: (i) Jet identification refers to the classification of the parent particle type that gave rise to the jet, and is needed to determine the particle content of a collision event. (ii) Jet energy estimation refers to the regression of the true jet total energy from the noisy detector observations, and is needed to determine the kinematic properties of an event. (iii) Jet pileup estimation and reduction refers to the determination of the stochastic contributions to detector observations arising from incident particles produced in proton-proton collisions that are not the primary hard scattering. This form of denoising is required to improve the energy and momentum resolutions of measurements of jets. 

Among the primary physics settings in which jet images have been used are in studies of jets produced by Lorentz boosted heavy particles, such as a $W$ or $Z$ boson, Higgs boson ($h$), top quark ($t$), or a hypothetical new beyond the Standard Model particle. When a heavy short-lived  particle is produced with a momentum on the order of twice its mass or more, the quark decay products of such a heavy particle have a high likelihood of a collimated emergence in which the subsequent hadronic showers produced by the quarks overlap. Jet clustering algorithms can capture the entirety of the heavy particle decay within one large-$R$ jet with an $R$-parameter typically between 0.8 and 1.0, though in some cases larger $R$ parameters have been used.  The internal structure of such a \textit{boosted jet} can be highly non-trivial and significantly different than a typical jet produced by a single quark or gluon. However, the production of quarks and gluons is ubiquitous at hadron colliders, and thus powerful discrimination methods, or \textit{taggers}, are needed to identify relatively clean samples of heavy-particle-induced boosted jets. Moreover, the mass scale of heavy hadronically decaying particles in the Standard Model is similar, from the $W$ boson mass of $\sim 80$ GeV~\cite{10.1093/ptep/ptaa104} up to the top quark mass of $\sim173$ GeV~\cite{10.1093/ptep/ptaa104}. Typical discrimination tasks thus include discriminating boosted $W$-, $Z$-, $h$-, or $t$-jets from quarks and gluons, but also in discriminating between boosted heavy particle jets.

Jet images have also been employed for studying jets from individual quarks and gluons. This includes discriminating between quark and gluon jets, and between jets produced by quarks of different flavour. In these cases, smaller jets typically with $R=0.4$ are used.

\section{Jet Images and Preprocessing} \label{sec:jetimages}

Jet images are built using a fixed grid, or pixelation, of the spatial distribution of energy within a jet. Early instances of such pixelation relied on energy depositions in calorimeter detectors, wherein the angular segmentation of the detector cells was used to define the ``pixels" of the jet image and the pixel ``intensity" was defined with the transverse energy in a cell. More recently, high resolution measurements of charged particles from tracking detectors have also been used to form images, wherein the transverse momentum of all particles found within the spatial extent of a jet image pixel are summed to define the pixel intensity. While calorimeter and tracking detectors typically span a large angular acceptance, a typical jet has limited angular span. The angular span of a jet is related to the $R$-parameter of the jet clustering algorithm. Jet images are thus designed to cover the catchment area of the jet~\cite{Cacciari_2008}. In many cases, the jet image is first defined to be slightly larger than the expected jet catchment area, to ensure that preprocessing steps (discussed in Section~\ref{subsec:preproc}) do not disrupt peripheral pixel estimates, and then after pre-processing are cropped. Nonetheless, only a slice of the angular space of the detector is used to define the jet image, with the image centered on the direction of the jet and the image size chosen to capture the extend of a jet with a given $R$ parameter. If depth segmentation is present in a calorimeter, the energy is often summed in depth. From this vantage point, a jet image can be viewed as a grey-scale image comprising the energy measurements encapsulated by the angular span of the jet. In some cases energy depositions from hadronic and electromagnetic calorimeters will be separated into different images, or separate images will be formed from both calorimeter cell measurements and the spatially pixelated charged particle measurement. In these cases, the set of jet images, each defining a view of the jet from a different set of measurements, can be seen as color channels of jet image.

It should be noted that jet pileup mitigation, such as the aforementioned trimming or PUPPI algorithms, is vital to reduce the impact of pileup on downstream jet image prediction tasks. While not explicitly discussed as a part of the jet image preprocessing, this step is almost always performed prior to jet image formation using the jet constituents, especially in the case of studying large-$R$ jets.

\subsection{Preprocessing}\label{subsec:preproc}
An important consideration in the training of a classifier is how to process data before feeding it to the classifier such that the classifier can learn most efficiently. For instance, a common preprocessing step in ML is to standardize inputs by scaling and mean shifting each input feature such that each feature has zero mean and unit variance. In this case,  standardization helps to ensure that features have similar range and magnitude so that no single feature dominates gradient updates.  In general, data preprocessing can help to stabilize the optimization process and can help remove redundancy in the data features to ease the learning of useful representations and improve the learning sample efficiency. However, data preprocessing may come at a cost if the preprocessing step requires approximations that lead to distortion of the information in the data. The primary jet preprocessing steps include:

\vspace{0.3cm} \noindent
\textbf{Translation:} An important consideration when preparing inputs to a classifier are the symmetries of data and transformations of inputs that should not affect the classifier prediction. In the case of jet images, these symmetries are related to the physical symmetries of the system. At a particle collider, there is no preferred direction transverse to the beam line, and the physics should be invariant to azimuthal rotations in the transverse plane. In terms of jet images, given a fixed parent particle, the distribution of jet images at a given azimuthal coordinate $\phi=\phi_a$ should not differ from the distribution at a different $\phi=\phi_b$. As such, an important preprocessing step is to translate all jet images to be ``centered" at $\phi=0$.  This is often performed by translating the highest $p_T$ subjet (formed by clustering the jet constituents with a small $R$-parameter jet algorithms), or the jet energy centroid, to be located at $\phi=0$. The same invariance is not generically true for changes in $\eta$, as translations in $\eta$ correspond to Lorentz boosts along the beam direction which could alter the jet properties if not handled carefully. When energy is used for jet image pixel intensities, a translation in $\eta$ while keeping  pixel intensities fixed will lead to a change in the jet mass. However, when the transverse momentum, which is invariant to boosts along the beam direction, is used to define pixel intensities, a translation in $\eta$ can be performed without altering the jet mass distribution. With this definition of pixel intensities, jet images are typically translated such that the leading subjet is located at $\eta=0$. By centering the jet on the leading $p_T$ subjet, the classifier can focus on learning the relative variations of a jet, which are key for classification.

\vspace{0.3cm} \noindent
\textbf{Rotation:} The radiation within a jet is also approximately symmetric about the jet axis in the $\eta - \phi$ plane. A common preprocessing step is thus to rotate jet images, after centering the image on the leading $p_T$ subjet, such that the second leading $p_T$ subjet or the first principle axis of spatial distribution of $p_T$ in the image is aligned along the $y$ axis of the image.  However, there are challenges with rotations. First, rotations in the $\eta-\phi$ plane can alter the jet mass, thus potentially impacting the classification performance\footnote{Alternative definitions of rotations have been proposed that preserve jet mass~\cite{pearkes2017jet} but may alter other key jet properties.}. Second, as jet images are discretized along the spatial dimensions, rotations by angles other than factors of $\pi/2$ can not be performed exactly. One approach is to perform a spline interpolation of the $p_T$ distribution within a jet image, apply a rotation to this spline function, and then impose an image grid to discretize the spline back to an image. The interpolation and the post-rotation discretization can spatially smear information in the jet and lead to aliasing. As such, there is varying use of rotation preprocessing in jet image research.

\vspace{0.3cm} \noindent
\textbf{Flipping:} A transformation $\phi \to -\phi$ should not affect the physics of the jet, and this transformation can be performed to ensure that positive $\phi$ contains the half of the jet with more energy, for instance due to radiation emission.

\vspace{0.3cm} \noindent
\textbf{Normalization:} A step often found in image preprocessing for computer vision tasks is image normalization, typically through taking an $L^2$ norm of the image such that $x_i \to x_i / \sum_j x_j^2$ where $x_i$ is a pixel intensity and the sum runs over all pixels in an image. However, in the case of jet images, such a normalization may be destructive, as it does not preserve the total mass of the jet (as computed from the pixels) and can deteriorate discrimination performance due to this loss of information~\cite{Oliveira2016JetimagesD}. As such, there is varying usage of image normalization in jet image research.

\begin{figure}[t]
\begin{center}
\includegraphics[width=0.5\linewidth]{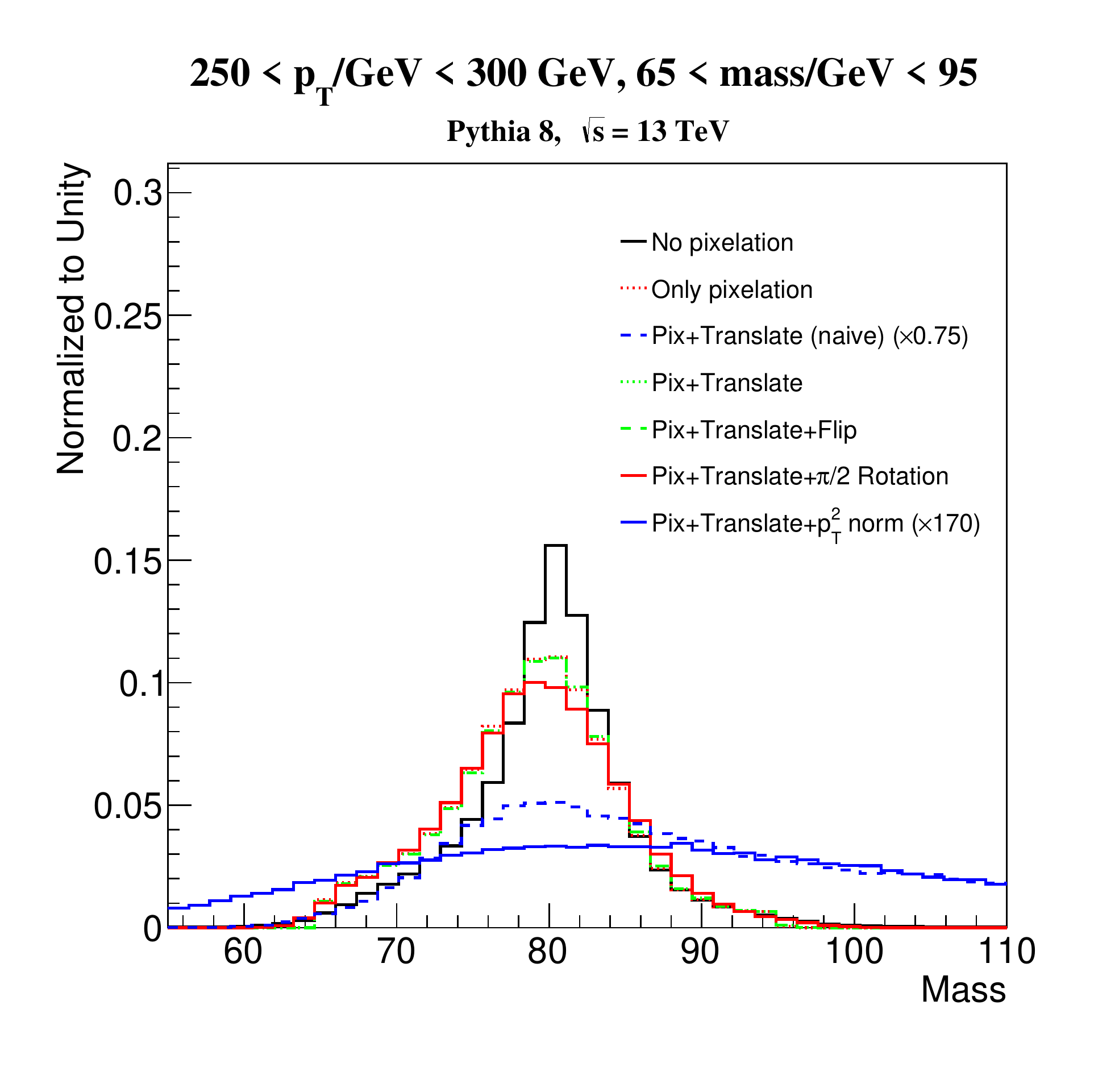}
\end{center}
\caption{The impact of various preprocessing steps on the distribution of estimated jet mass for boosted $W$ boson jet images~\cite{Oliveira2016JetimagesD}.}
\label{fig:preproc}
\end{figure}

\vspace{0.3cm} \noindent
The impact on the jet mass, as computed from the pixels of jet images, for $W$ boson jets within a fixed $p_T$ range and within a fixed pre-pixelation mass range can be found in Figure~\ref{fig:preproc}. The distortion on the jet mass from pixelation, rotations for images with energies as pixel intensities, and from $L^2$ normalization, can be seen clearly, whilst translation and flipping do not show distortions of the jet image mass. As expected, mild distortion of the mass can be seen when rotations are performed on jet images with transverse energy used for pixel intensities. These distortions may or may not be impactful on downstream tasks, depending on if the jet mass is a key learned feature for the downstream model.

\section{Computer Vision and Convolutional Neural Networks}\label{sec:CV}
Object classification in Computer Vision (CV) tasks served as a primary setting where Deep Learning had major early successes~\cite{AlexNet}, quickly surpassing then state of the art approaches and serving as one of the drivers for a deep learning revolution. While much of the work in CV has focused on understanding natural images, data collected by physics experiments come from heterogenous detectors, tend to be sparse, and do not have a clear similarity to natural images. Nonetheless, the success of DL in CV inspired a parallel effort in the collider physics community to explore applications of such techniques to HEP data. Below we present a brief introduction to convolutional neural networks (CNNs)~\cite{CRG-TR-89-4} and some of the state-of-the-art architecture variants in order to provide some background for the models used in jet tagging and other applications. For a more in depth pedagogical introduction to this material see for instance~\cite{Goodfellow-et-al-2016}.

Most of the models discussed in this text rely on the use of convolutional layers. However, it should be noted that some models make use of locally-connected layers~\cite{CRG-TR-89-4, PhysRevD.93.094034,Oliveira2017LearningPP}, in which a given neuron only has access to a small patch of an input but, unlike convolutional layers that rely on weight sharing (as discussed below), the neuron processing each image patch is associated with a different set of weights. 

\vspace{0.2cm}\noindent
\textbf{Convolutional neural networks} rely on neuron local spatial connectivity and weight sharing to produce translationally equivariant models that are well adapted to image analysis. A typical CNN is built by stacking one or more convolutional and non-linear activation layers often followed by a pooling layer. This structure is repeated several times. Fully connected layers, with full connections from all inputs to activations, are used to perform the final classification or regression prediction.  Images processed by CNNs are represented as 3D tensors with dimensions \textit{width} $\times$ \textit{height} $\times$ \textit{depth} and are often referred to as the image volume. The height and width dimensions correspond to the spatial extend of the image while the depth is typically the color channel.

\vspace{0.2cm}\noindent
\textbf{Convolutional layers} are composed of a set of \textit{filters}, where each filter applies an inner product between a set of weights and small patch of an input image. The filter is scanned, or \textit{convolved}, across the height and width of the image to produce a 2D map, often referred to as a response map or convolved image, that gives the response of applying the filter at each position of the image. The response at each position becomes large when the filter and the image patch match, i.e. when their inner product is large. The filters will thus learn to recognize visual features such as edges, textures, and shapes, and produce large responses when such visual features are present in a patch of an image. The spatial extent of the input patch is known as the receptive field or filter size, and the filters extend to the full depth of the image volume.  Several filters are learned simultaneously to respond to different visual features. The response map of the filters are then stacked in depth, producing an output convolved image volume. Finally, the response maps are passed through point-wise (i.e. per pixel) non-linear activations to produce an activation map.

By sharing weights between neurons, i.e. by scanning and applying the same filter at each image location, it is implicitly assumed that it is useful to apply the same set of weights to different image locations. This assumption is reasonable, as a visual feature may be present at any location in an image and the filter is thus testing for that feature across the image. This results in the convolutional layers being translationally equivariant, in that if a visual feature is shifted in an image, the response to that feature will be shifted in the activation map. In addition, parameter sharing results in dramatic reduction in the number of free parameters  in the network relative to a fully connected network of the same number of neurons.

\vspace{0.2cm}\noindent
\textbf{Pooling layers} reduce the spatial extent of the image volume while leaving the depth of the volume unchanged~\cite{pooling}. This further reduces the number of parameters needed by the network and helps control for overfitting. Pooling is often performed with a \textit{max} operation wherein only the largest activation in a region, typically $2\times2$, is kept for subsequent processing.

\vspace{0.2cm}\noindent
\textbf{Normalization layers} may be used to adjust activation outputs, typically to control the mean and variance of the activation distribution. This helps ensure the that a neuron does not produce extremely large or small activations relative to other neurons, which can aid in gradient-based optimization and in mitigating  exploding / vanishing gradients. Batch normalization~\cite{pmlr-v37-ioffe15} is a common normalization method in which, for each mini-batch, the mean and variance within the mini-batch of each activation dimension are used to normalize the activation to have approximately zero and unit variance. A linear transformation of the normalized activation, with learnable scale and offset parameters, is then applied.

\vspace{0.2cm}\noindent
\textbf{Fully connected layers} are applied at the end of the network after \textit{flattening} the image volume into a vector in order to perform classification or regression predictions. At this stage, auxiliary information, potentially processed by a separate set of neural network layers, may be merged with the information gleaned from the processing by convolutional layers in order to ensure that certain features are provided for classification. Within the jet tagging context, such information may correspond to information about the jet, such as its mass, or global event information such as the number of interactions in a given collision.

\vspace{0.2cm}\noindent
\textbf{Residual Connections:} While CNNs encode powerful structural information into the model, such as translation equivariance, it has been noted that scaling up such models by stacking large numbers of convolutional layers can lead to large challenges in training~\cite{pmlr-v9-glorot10a}. In order to train large models using the backpropagation algorithm, the chain rule is used to compute the gradient from the model output back to the relevant weight. In early layers, the multiplication of many gradients can lead to vanishingly small or exploding gradients, thus resulting in unhelpful gradient updates. To overcome this challenge, the residual block~\cite{he2015deep} was proposed, and has led to the development of \textit{residual networks}. While a typical neural network layer passes input $z$ through a nonlinear function $f(\cdot)$ to produce an output $z^\prime = f(z)$, a residual block also uses a ``skip connection" to pass the input to the output in the form  $z^\prime_{\textrm{res}} = W_{s} z+f(z)$ where the weights $W_s$ can be used to project the channels $z$ to have the same dimension as the function $f(z)$. In this way, the function $f(\cdot)$ is tasked with learning the relative change to the input. Moreover, the skip connection provides a path to efficiently pass gradients backwards to earlier layers of the network without distortion through the nonlinearities, making gradient descent much easier and thus enabling the training of significantly deeper models. Note that the function $f(\cdot)$ can contain several layers of convolutions, nonlinearities, and normalization before being recombined with the input.

\vspace{0.2cm}\noindent
\textbf{Training} in supervised learning tasks is performed by minimizing an appropriate loss function that compares the CNN prediction with a true label. The loss is typically the cross-entropy in the case of binary classification, and the mean squared error in the case of regression. Minimization is performed using stochastic gradient descent (SGD)~\cite{bottou2016optimization}, or one of its variants such as ADAM~\cite{kingma2014adam} designed to improve the convergence of the learning algorithm.

\vspace{0.2cm}\noindent
\textbf{Evaluation Metrics} Receiver operating characteristic (ROC) curves are frequently used to examine and compare the performance of binary classification algorithms. Given a model which produces a classification prediction $c(x)$, where $x$ is the input features and $c(\cdot) \in [0,1]$, the model is applied to a set of inputs thus yielding a distribution of predictions. A threshold $\tau$ on the prediction is scanned from 0 to 1, and the fraction of inputs for each the signal and background classes above this threshold, i.e. the signal efficiency ($\epsilon_S$) and background efficiency ($\epsilon_B$) for surviving this threshold, defines a point on the ROC curve for each $\tau$ value. ROC curves thus display the background efficiency (or background rejection defined as 1 divided by the background efficiency) versus the signal efficiency.  When the ROC curve is defined as the background efficiency versus the signal efficiency, a metric commonly used to evaluate the overall model performance is the ROC integral, also known as the area under the curve (AUC).

Significance improvement characteristic (SIC) curves~\cite{2011JHEP...04..069G} are closely related to ROC curves, but display $\epsilon_S / \sqrt{\epsilon_B} $ as a function of the signal efficiency $\epsilon_S$. This curve targets displaying the potential improvement in statistical significance when applying a given discriminant threshold relative to not applying such a threshold.

\section{Jet tagging} \label{sec:tagging}

Jet tagging refers to the classification of the parent particle which gave rise to a jet. Linear discriminant methods were first applied to jet images defined using a single channel, or ``grey-scale", with pixel intensities defined as the calorimeter cell $p_T$~\cite{Cogan2015JetimagesCV}.  Subsequently, CNN based classifiers trained on single channel images were developed for discriminating between $W/Z$ jets and quark / gluon jets~\cite{Oliveira2016JetimagesD, PhysRevD.93.094034}, between top jets and quark / gluon jets~\cite{Kasieczka2017DeeplearningTT,Kasieczka_2019,diefenbacher2019capsnets,10.21468/SciPostPhys.8.1.006}, and for discriminating between quarks and gluons~\cite{Komiske2017DeepLI}. Quark / gluon discrimination with single-channel jet images has also been explored for use in heavy ion collisions~\cite{Chien:2018dfn}.  The extension to utilizing jet images ``in color" with multiple channels, defined for instance using charged particle information, has shown promising performance improvements over single channel approaches in many of these tasks~\cite{Komiske2017DeepLI,Fraser2018JetCA,Macaluso2018PullingOA,Kasieczka_2019,2019arXiv190808256C,2018JHEP...10..101L,2019JHEP...09..047K}, and has been explored in realistic experimental settings by the ATLAS and CMS collaborations~\cite{ATL-PHYS-PUB-2017-017,Sirunyan:2020lcu}.

\subsection{Jet Tagging on Single Channel Jet Images}\label{subsec:JIsingle}

\begin{figure}[t]
\begin{center}
\begin{subfigure}[h]{0.45\linewidth}
\includegraphics[width=\linewidth]{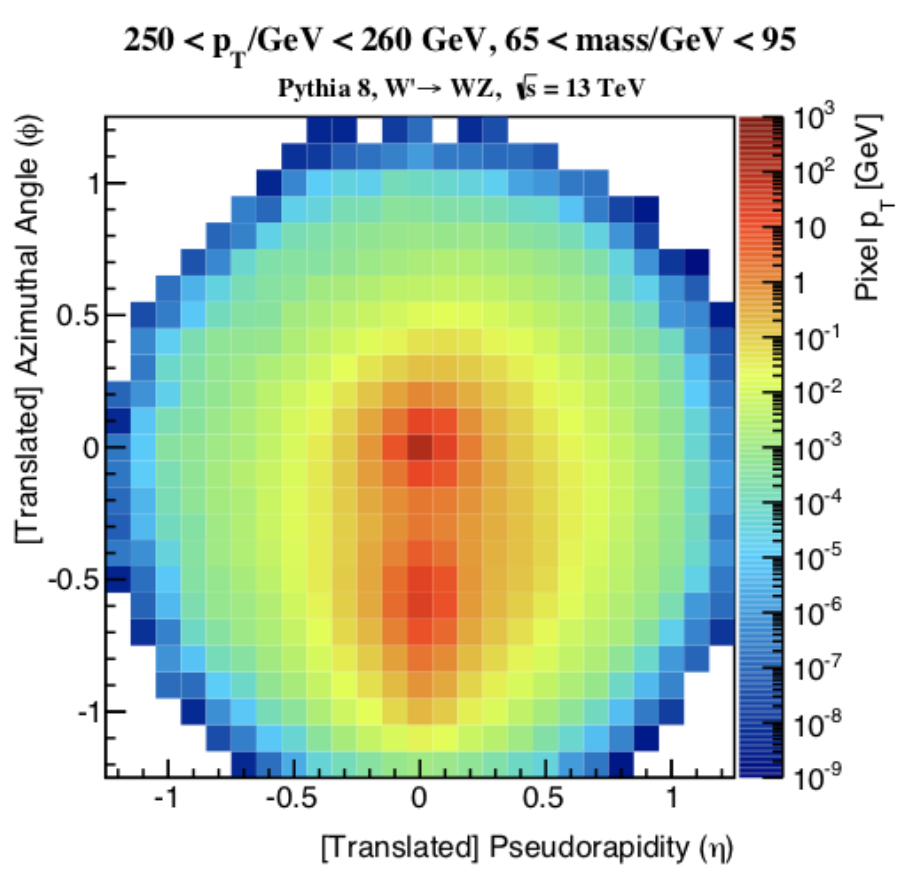}
%\caption{}
\label{fig:wjet}
\end{subfigure}\quad
\begin{subfigure}[h]{0.45\linewidth}
\includegraphics[width=\linewidth]{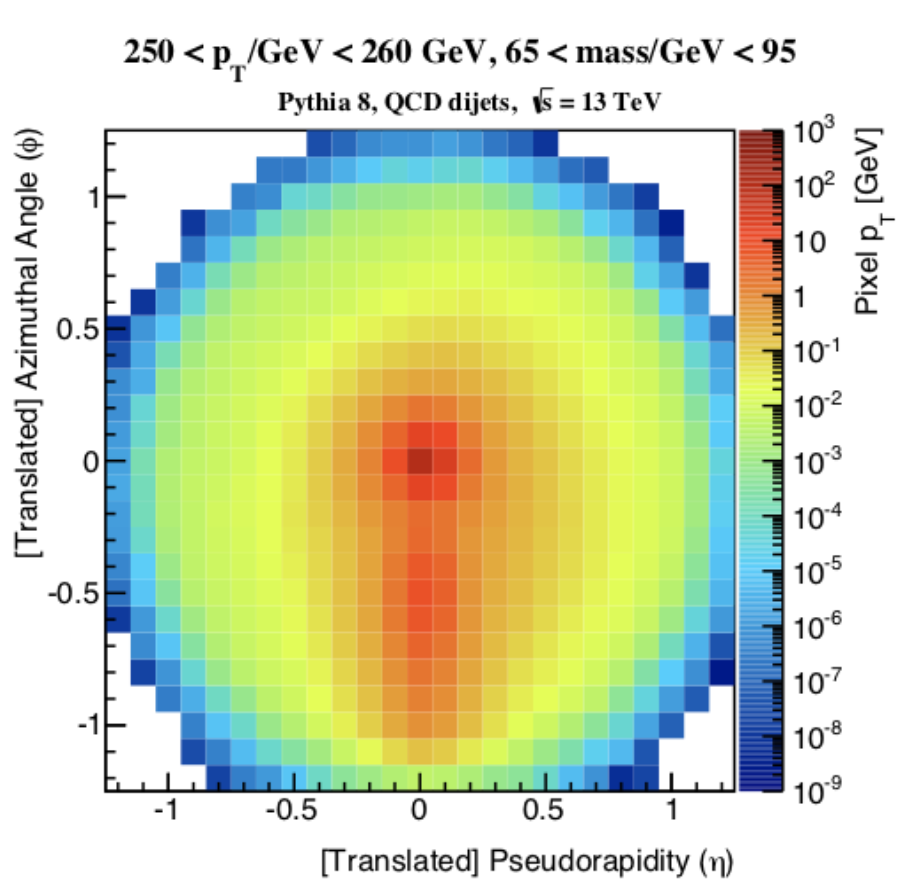}
%\caption{}
\label{fig:qcdjet}
\end{subfigure}
\end{center}
\caption{Average $W$ boson jet image (left) and average quark / gluon jet image (right) after preprocessing~\cite{Oliveira2016JetimagesD}.}
\label{fig:jetimages}
\end{figure}

\paragraph{$W/Z$ Tagging:} The discrimination of boosted $W$ and $Z$ vector boson initiated jets from quark / gluon jets has served as a benchmark task in boosted jet tagging. The color singlet nature of electroweak bosons decaying to quark pairs leads to an internal structure of boosted $W/Z$ jets in which there are typically two high energy clusters, or subjets, and additional (dipole) radiation tends to appear in the region between such subjets. The Higgs boson, also a color singlet with decays to quark pairs, has a similar substructure, although the decays of heavy flavour bottom and charm quark pairs can lead to some structural differences owing to the long lifetime of such quarks and their harder fragmentation than lighter quarks. In contrast, single quarks and gluons tend to produce jets with a high energy core, lower energy secondary subjets created through radiative processes, as well as diffuse wide angle radiation further from the core of the jet. These features can be seen clearly in Figure~\ref{fig:jetimages}, which shows the average $W$ boson jet image and average quark / gluon jet image after preprocessing. 

Building ML models applied to jet images for this discrimination task avoids the explicit design of physics-inspired features, and rather focuses on the learning task of identifying differences in the jet image spatial energy distributions.  In phenomenological studies, both fully convolutional~\cite{Oliveira2016JetimagesD} and models with locally connected layers~\cite{PhysRevD.93.094034} have been examined for discriminating jet images of boosted $W$ and $Z$ vector boson initiated jets from quark / gluon jets. The CNN models were examined in simulated samples of jets without pileup. The locally connected models were examined in events both with and without pileup, thus enabling the examination of the impact of pileup noise on jet image based tagging.

Within both the studies on convolutional~\cite{Oliveira2016JetimagesD} and locally connected~\cite{PhysRevD.93.094034} models, hyperparameter scans were performed to find model parameters that maximized performance\footnote{In the case of CNNs the AUC was maximized whilst the Spearmint Bayesian Optimization package~\cite{snoek2012} was used to optimize the model with locally connected layers}. The hyperparameters that were considered in the scans included the number of convolutional / locally connected layers, the number of hidden units per layer, and the number of fully connected layers. The resulting optimized models were similar, containing 3 to 4 convolutional or locally connected layers, as well as 2 to 4 fully connected layers with approximately 300 to 400 hidden units at the end of the network.  In the CNN, 32 filters were used in each convolutional layer, as well as $(2\times 2)$ or $(3\times 3)$ downsampling after each convolutional layer. One notable additional optimization performed for the CNN models was the size of the convolution filters in the first layer.  While filter sizes are typically $(3\times 3)$ or $(4\times 4)$ in standard CV applications, in the case of application to jet images it would found that a larger $(11\times11)$ filter in the first convolutional layer (with later layers using standard $(3\times 3)$ filter sizes) resulted in the best performance. It was hypothesized that the such large filters were beneficial when applied to sparse images~\cite{Oliveira2016JetimagesD}, in order to ensure that some non-zero pixels are likely to be found within the image patch supporting the filter application.

\begin{figure}[htbp]
\begin{center}
\begin{subfigure}[h]{0.45\linewidth}
\includegraphics[width=\linewidth]{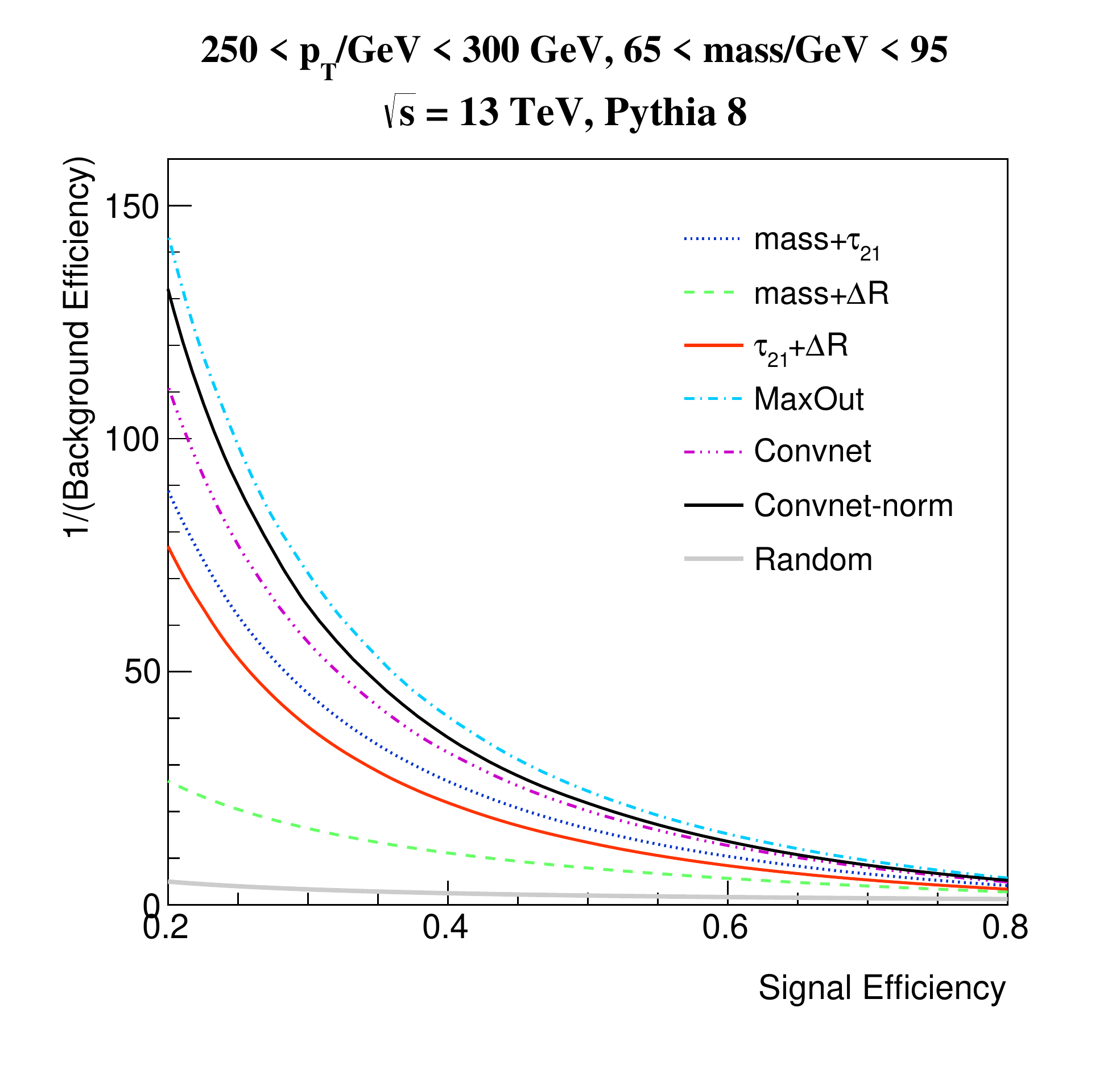}
\caption{}
\label{fig:Wroc_cnn}
\end{subfigure}\quad
\begin{subfigure}[h]{0.45\linewidth}
\includegraphics[width=\linewidth]{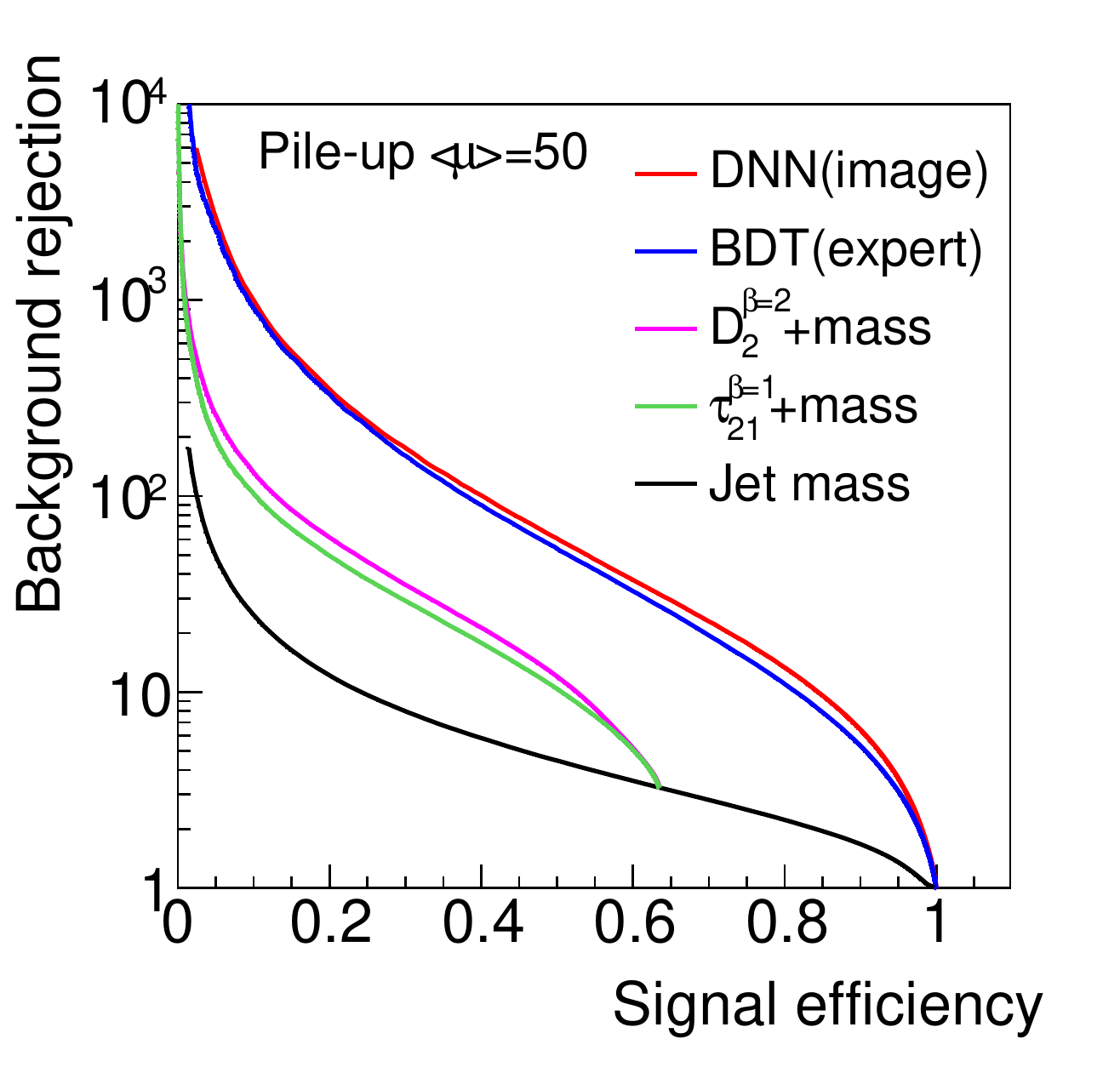}
\caption{}
\label{fig:Wroc_lc}
\end{subfigure}
\end{center}
\caption{ROC curves for quark/gluon background rejection versus boosted $W$ boson tagging efficiency for (a) events without pileup~\cite{Oliveira2016JetimagesD}, and (b) events with pileup~\cite{PhysRevD.93.094034}. The jet image based CNN taggers are seen to outperform combinations of jet substructure features, and to be stable with respect to the addition of pileup.}
\label{fig:Wroc}
\end{figure}

The ROC curves indicating the performance of the CNN model and locally connected model (applied to jets with pileup included) are shown in Figure~\ref{fig:Wroc}. It should be noted that the jets in these figures correspond to different $p_T$ ranges, with jets of $p_T \in (250, 300)$ GeV for the CNN model and of $p_T \in [300, 400]$ GeV for the locally connected model, and thus are not directly comparable. Also shown are combinations of common physics expert engineered jet substructure features, such as the jet mass, the distance between the two leading $p_T$ subjets, the $\tau_{21}$ n-subjetiness~\cite{Thaler_2011}, and the energy correlation function $D_2^{\beta=2}$~\cite{Larkoski_2014_D2}. Two variable combinations were computed using 2D binned likelihood ratios. Both the CNN and locally connected model significantly outperform the 2D jet substructure feature combinations. It can also be seen that the jet image approach is not overly sensitive to the effects of pileup as the large performance gain over jet substructure features persists both with and without the presence of pileup, owing to the use of jet trimming to reduce the impact of pileup noise in the jets.  In addition, a Boosted Decision Tree (BDT) classifier~\cite{ESL} combining six substructure features was compared with the locally connected model and found to have similar performance. While these early jet image based models did not significantly outperform  combinations of several jet substructure features, this may be due to their relatively small model structure. As will be seen, more complex architectures and the use of multi-channel jet images can lead to large gains over combinations of jet substructure features. 

One can also see the effect of $L^2$ image normalization on CNN models, which appears to improve performance over unnormalized images. This effect was found to occur because the CNN model output was observed to have only small correlation with the jet mass, and thus was not learning to be heavily reliant on the jet mass information that is distorted by normalization. As a result, the regulation of the image variations due to normalization was found to be beneficial enough to overcome the induced distortion of the jet mass. With more powerful models that learn representations more correlated to the jet mass, this balance may not occur. 

\paragraph{Top Tagging:} The discrimination between boosted top quark jets and  quark / gluon jets using CNNs applied to jet images has also been examined both in phenomenological studies~\cite{Kasieczka2017DeeplearningTT} and in realistic simulations by the CMS experiment~\cite{Sirunyan:2020lcu}. Top quark jet images are structurally more complex than $W / Z / h$ jet images as hadronic decays of top quarks contain three quarks. This can have implications on both the preprocessing and the tagging performance. That is, some of the pre-processing steps previously defined will lead to uniformity among jet images for two quark systems, such as the rotation step which aligns the leading two subjets, but may not lead to the same level of uniformity for three quark systems. 

The DeepTop~\cite{Kasieczka2017DeeplearningTT} model is a CNN applied to single channel jet images after the preprocessing described above, including image normalization. Hyperparameter optimization yielded a model with four convolutional layers, each with 8 filters of size $(4\times4)$, MaxPooling for image downsampling after the second convolutional layer, and three dense layers of 64 hidden units each for classification. For these phenomenological studies, the model was trained with approximately 150k jets using the mean squared error (MSE) loss. While structurally similar to the single channel CNN used for $W/Z$ tagging in reference~\cite{Oliveira2016JetimagesD} there are some notable differences such as the use of fewer numbers of filters (8 rather than 32) and the smaller filter size in the first layer of convolution. The reason for these difference may be due to (a) the presence of three quarks in the top quark decay leads to more pixel-populated images and thus allowed for the use of smaller initial filter sizes, or (b) the global nature of the hyperparameter scan wherein the number of filters and the size of the filters was fixed to be the same across all convolutional layers. 

\begin{figure}[htbp]
\begin{center}
\begin{subfigure}[h]{0.48\linewidth}
\begin{center}
\includegraphics[width=\linewidth]{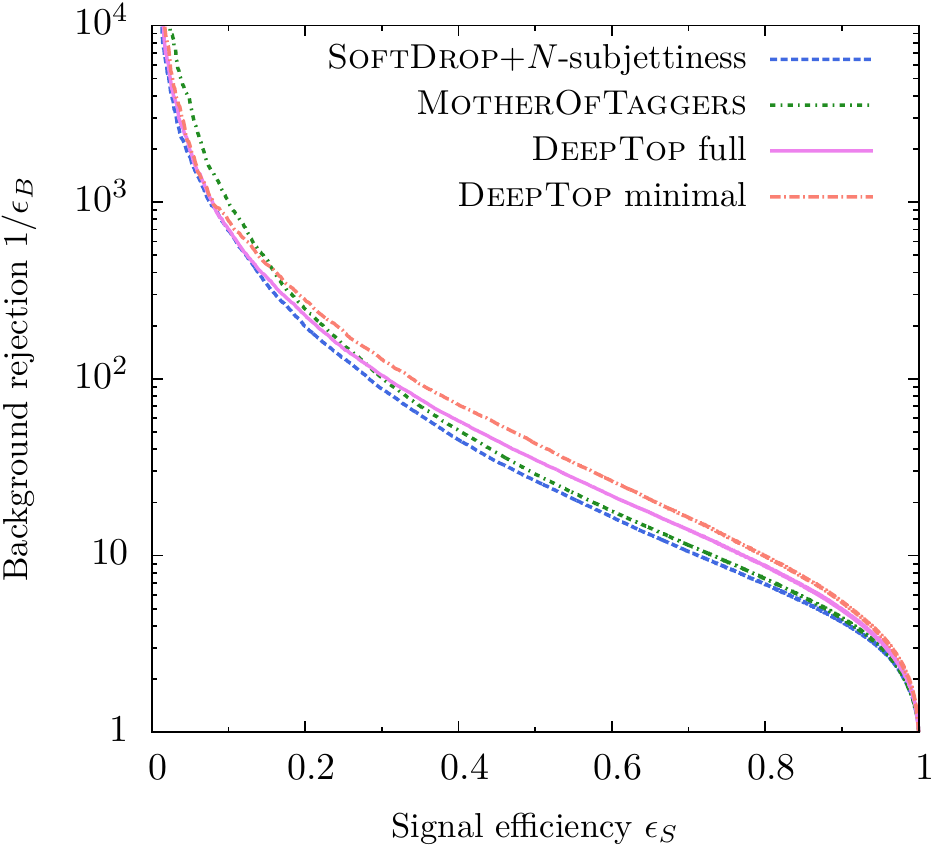}
\caption{}
\label{subfig:deeptop1}
\end{center}
\end{subfigure}\quad
\begin{subfigure}[h]{0.48\linewidth}
\begin{center}
\includegraphics[trim={0 0 0 0.6cm},clip,width=0.9\linewidth]{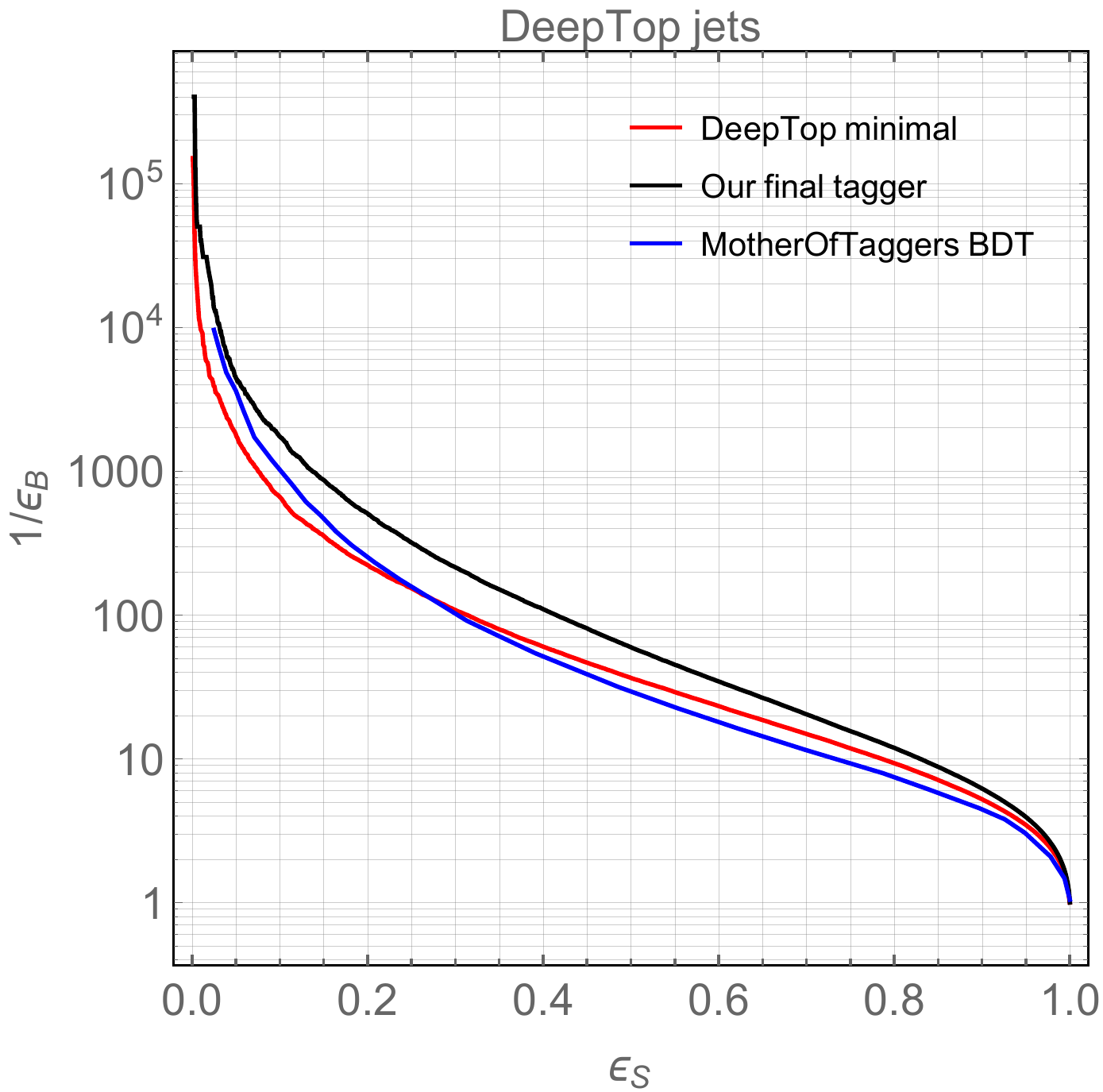}
\caption{}
\label{subfig:deeptop2}
\end{center}
\end{subfigure}
\end{center}
\caption{ROC curves for quark / gluon jet rejection versus boosted top efficiency for (a) the DeepTop model ~\cite{Kasieczka2017DeeplearningTT}, and (b) the updated DeepTop model from reference~\cite{Macaluso2018PullingOA}. In both cases, the CNN based DeepTop models outperform individual and BDT combinations of substructure features, while the updated model in (b) is also seen to significantly improve the DeepTop performance.}
\label{fig:deeptop}
\end{figure}%

The performance of the DeepTop model can be found in Figure~\ref{subfig:deeptop1} in terms of the ROC curve comparing the quark / gluon rejection versus the boosted top jet tagging efficiency for jets with $p_T \in [350, 450]$ GeV. In this momentum range, the decay products of the top quark may not be contained in a single jet, and such a containment was not required for the jets under study. DeepTop was compared with a combination of mass and $n$-subjettiness, as well and a BDT, denoted  MotherOfTaggers, combining several jet substructure features.  The jet image based DeepTop algorithm showed clear performance gains over substructure approaches across most of the signal efficiency range. As previously mentioned,  pre-processing steps have the potential to be beneficial for the learning process by producing more uniform images, but may also lead to performance degradation.  This was studied within the scope of the DeepTop algorithm, by examining the tagging performance using full preprocessing  and a minimal preprocessing that only performed centering but not the rotation or the flipping. This can be seen in Figure~\ref{subfig:deeptop1}, where a clear performance benefit was observed when utilizing only minimal pre-processing. While the full pre-processing may be beneficial for small sample sizes, with sufficient sample sizes and model complexity the CNN models appear able to learn well all the variations in jet images. In this case, the approximations introduced by pre-processing steps appear to be more detrimental than the benefits from uniformization of the jet image distributions. 

Building upon the DeepTop design, developments in architecture design, jet image preprocessing, and optimization were introduced in the phenomenological study of reference~\cite{Macaluso2018PullingOA}. These developments include: (i) the cross entropy loss function, rather than the mean squared error loss,  was used as it is more suitable to binary classification problems, (ii) a learning rate adaptive optimizer, AdaDelta~\cite{2012arXiv1212.5701Z}, and small mini-batch sizes of 128 was used rather than vanilla stochastic gradient descent and large mini-batches of 1000, (iii) larger numbers of filters per convolutional layer, between 64 and 128 rather than 8, and 256 neurons in the dense layers instead of 64, (iv) preprocessing is performed before pixelation under the assumption that one would have access to high resolution particle momentum measurements, for instance using Particle Flow~\cite{2017JInst..12P0003S} approaches to jet reconstruction, and (v) the training set size was increased by nearly a factor of 10. While the individual effects of these developments will be examined further in Section~\ref{subsec:tagincolor} when discussing top tagging on multi-channel jet images, the combination of these developments can be seen to provide large performance improvements over DeepTop of nearly a factor of two in background rejection at fixed signal efficiencies in Figure~\ref{subfig:deeptop2}. 

In terms of more complex architectures, the ResNeXt-50 architecture~\cite{Xie2017AggregatedRT} was adapted to boosted top jet tagging task using single channel jet images in the phenomenological studies in reference~\cite{Kasieczka_2019}. ResNeXt-50 utilizes blocks containing parallel convolutional layers that are aggregated and merged also with a residual connection at the end of the block. As the jet images typically have fewer pixels than natural images, the architecture was adapted to the top tagging dataset by reducing the number of filters by a factor of four in all but the first convolutional layer, and dropout was added before the fully connected layer. In addition, smaller pixel sizes in the jet images were utilized in this model, with a granularity of 0.025 radians in $\eta-\phi$ space (whereas the jet image granularities typically used in other models is 0.1 radians in $\eta-\phi$ space). 

\begin{figure}[htbp]
\begin{center}
\includegraphics[width=0.5\linewidth]{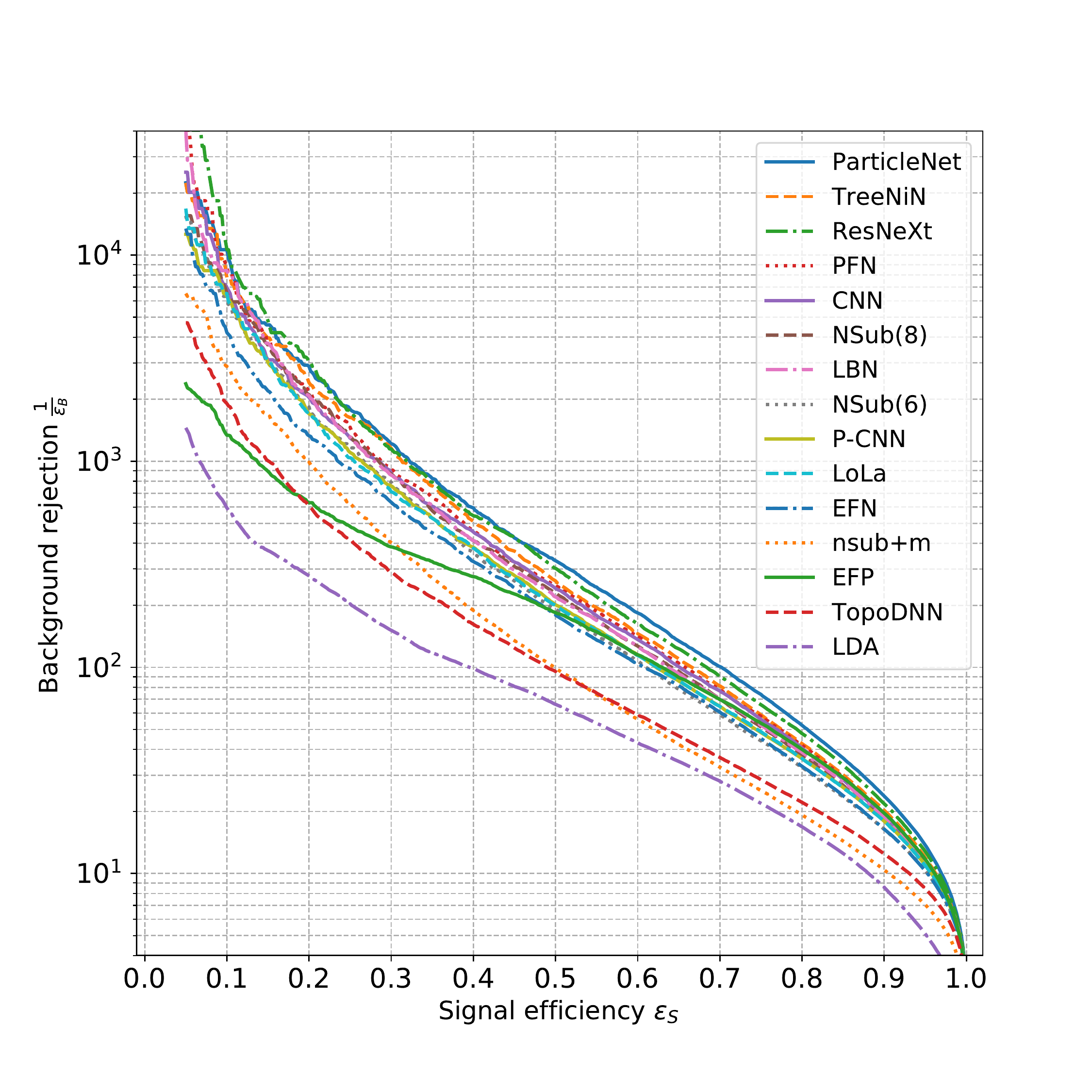}
\end{center}
\caption{ROC curve comparisons of various boosted top tagging models is shown~\cite{Kasieczka_2019}. Both ResNeXt and CNN curves are jet image based taggers using CNN based architectures.}
\label{fig:toplandscape}
\end{figure}%

The ROC curve comparing the ResNeXt-50 model to a CNN based on references~\cite{Kasieczka2017DeeplearningTT,Macaluso2018PullingOA}, and comparing to  several other neural network models with varying architectures can be found in Figure~\ref{fig:toplandscape}.  The ResNeXt-50 model provides approximately 20\% improvement in background rejection for fixed signal efficiency over the CNN model, and is among the most performant algorithms explored. This is notable as many of the other neural network models utilize particle 4-vectors as inputs, rather than aggregated particle information with a pixel cell, and make use of particle charge information, while the ResNeXt model only utilizes the distribution of energy within the jet. However, the ResNeXt-50 model contains nearly 1.5 million parameters, which is far more than other models such as the CNN which contains $\approx$610k parameters and the tree structured neural network (TreeNiN)  which contains $\approx$34k parameters.   Thus powerful information for discrimination can be extracted with jet image based models even from single channel images, but it may come with the price of models with large parameter counts.

This model comparison study has been performed in a phenomenological setting on particle level simulations, and the ultimate question remains as to the suitability for using these models in real experiment settings. In experimental settings, realistic detector noise, detection efficiency, detector heterogeneity, and data taking conditions such as pileup, underlying event, and beam evolution will impact the model performance.  Powerful models, including the large ResNext and CNN models, will likely have sufficient flexibility to learn powerful discriminators even in these more challenging settings. However, in general it remains to be seen if these models can be accurate whilst maintaining a low calibration error (where calibration in this context refers to the criteria that the predicted class probabilities correspond to the true probabilities of a given data input having a given label)~\cite{guo17a}, or if additional care is needed to ensure calibration. Moreover, applications in real experimental settings must consider systematic uncertainties associated with training ML models in (high fidelity) simulation but applying them in real data with potentially different feature distributions. The relationship between model complexity and sensitivity to systematic uncertainties in real experiment settings still remains to be thoroughly explored. The potential benefits in terms of sensitivity to systematic uncertainties when using neural networks with different structural assumptions, such as convolutional versus graph models, also requires further study and will likely depend on the details of how a given systematic uncertainty effects the feature distributions. Some exploration of these challenges can be found in Section~\ref{subsec:ji_theory_unc} examining model sensitivity to theoretical uncertainties and in Section~\ref{subsec:ji_exp} examining applications of these models in HEP experiments. Nonetheless, these remain important and exciting avenues of future work.

\paragraph{Decorrelated tagging with Jet Images:} A common strategy in HEP  to search for a particle is the so-called \textit{bump hunt} in which the  particle would give rise to a localized excess on top of a smoothly falling background in the distribution of the mass of reconstructed particle candidates. For instance, one may aim to identify the $W$-boson mass peak over the quark and gluon background from the distribution of jet mass. In addition to the particle mass being localization, a key to this strategy is that the smoothly falling background mass distribution can typically be characterized with simple parametric functions, thus facilitating a fit of the data to identify the excess above this background. Jet classification methods can cause challenges in the aforementioned strategy, as the classifier may preferentially select for jets with a specific mass, thereby sculpting the selected jet mass distribution of the background and rendering the search strategy unusable. As a result, one line of work has focused on de-correlating classifiers from a sensitive feature (e.g. mass) such that the sensitive feature is not sculpted by the application of the tagger. Such methods tend to rely on data augmentation or regularization, and overviews of these methods can be found for instance in references~\cite{ATL-PHYS-PUB-2018-014,Bradshaw:2019ipy}. Two recent regularization techniques that have seen strong de-correlation capability include (i) adversarial techniques~\cite{Louppe2017LearningTP,Shimmin:2017mfk}, wherein a second neural network is trained simultaneously with the jet classifier to penalize the jet classifier when the value of the sensitive feature can be predicted from the classifier's output or its hidden representations, and (ii) distance correlation regularizers~\cite{kasieczka2020disco}, wherein the jet classifier loss is augmented with an additional regularization which explicitly computes the correlation between the classifier predictions and the sensitive feature.  In both cases, the amount of penalization from the regularization can be varied through a hyperparameter scaling the relative size of the regulation term to the classification loss.

\begin{figure}[htbp]
\begin{center}
\begin{subfigure}[h]{0.52\linewidth}
\vspace{1.2cm}
\includegraphics[width=\linewidth]{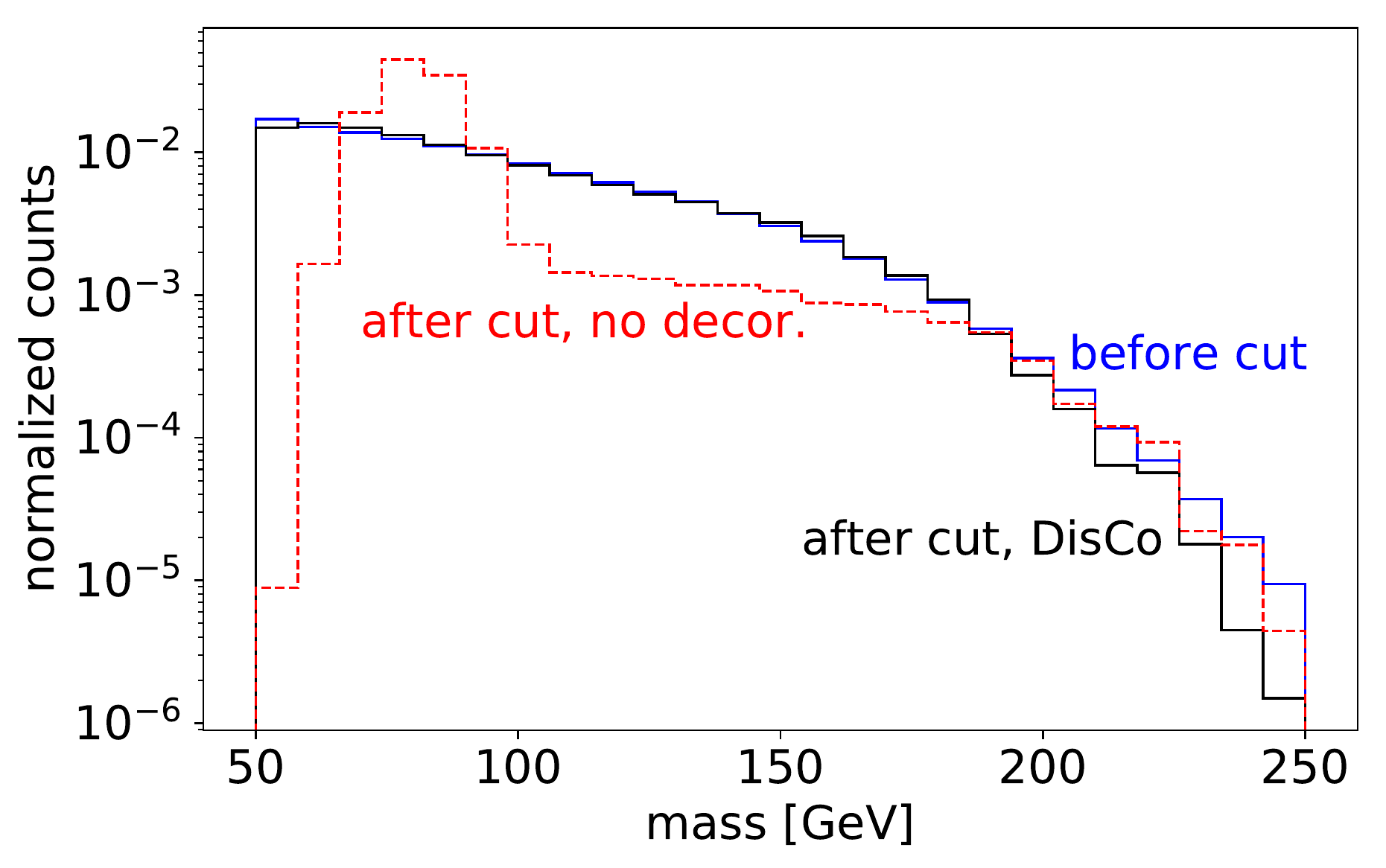}
\caption{}
\label{subfig:disco_mass}
\end{subfigure}\quad
\begin{subfigure}[h]{0.44\linewidth}
\includegraphics[width=\linewidth]{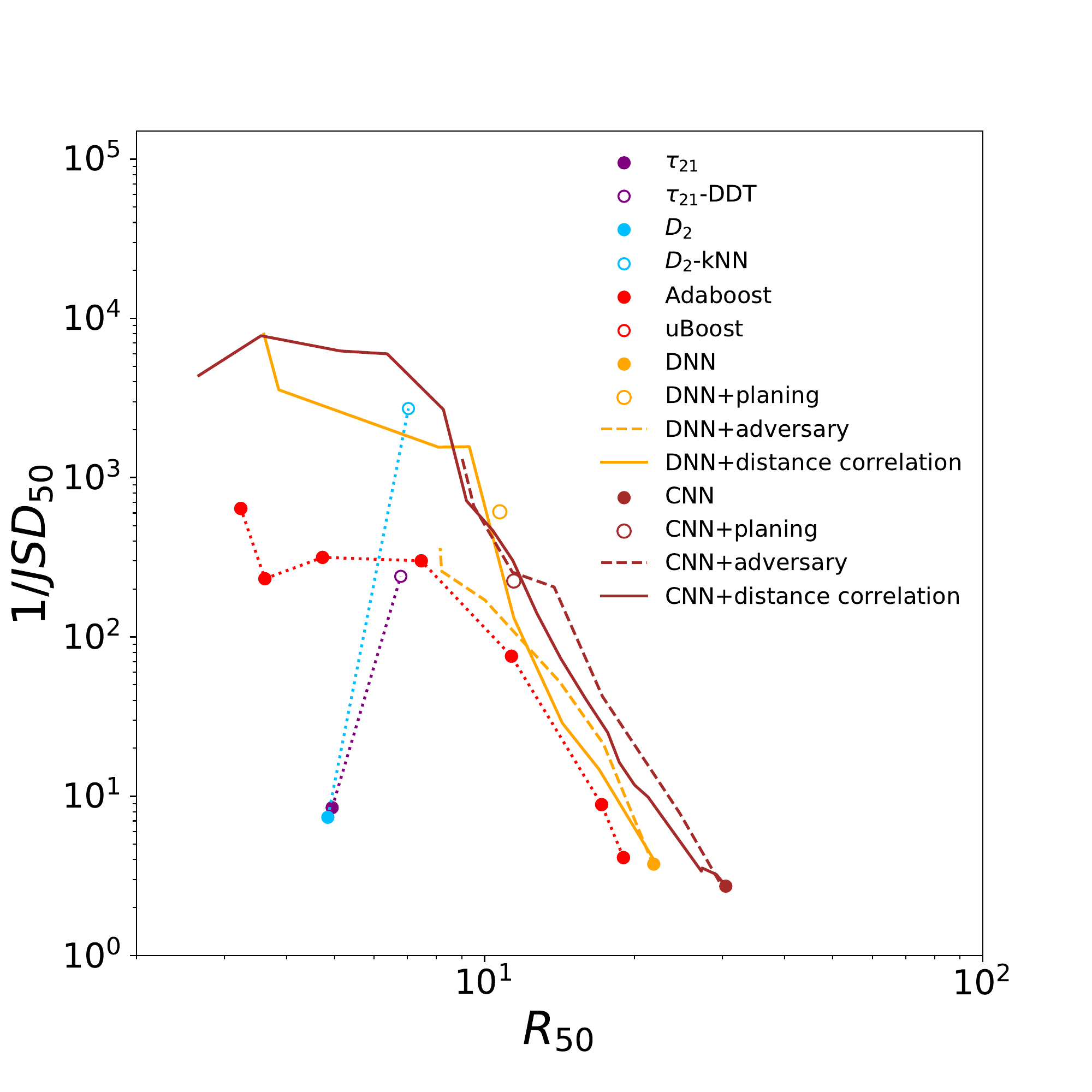}
\caption{}
\label{subfig:disco_perf}
\end{subfigure}
\end{center}
\caption{(a) For boosted $W$ boson tagging, the jet mass is shown before applying a threshold on a trained CNN tagger and after applying a threshold on a standard and mass decorrelated tagger~\cite{kasieczka2020disco}. A clear reduction in mass sculpting is observed. (b) The rejection at 50\% signal efficiency versus one over the Jensen-Shannon divergence, computed on the binned jet mass distribution before and after tagging, is shown for various taggers~\cite{kasieczka2020disco}. Jet image based CNN taggers are seen to outperform other methods, either using adversarial or distance-correlation based mass decorrelation.}
\label{fig:disco}
\end{figure}

De-correlation for $W$-boson jet tagging with jet images using CNNs was examined in phenomenological studies in reference~\cite{kasieczka2020disco}, using a CNN architecture similar to the model described in~\cite{Macaluso2018PullingOA}. The quark and gluon background jet mass distribution before and after applying a threshold on the output of a CNN can be seen in Figure~\ref{subfig:disco_mass}, showing a clear sculpting of the mass distribution. However, when the distance correlation regularization, or \textit{Disco}, is used during training, the mass distribution remains largely unsculpted after applying a classification threshold. The level of de-correlation can be estimated by examining the agreement between the mass distribution before and after applying a classifier threshold, for instance using the Jensen-Shannon divergence (JSD) computed between the binned mass distributions. For classifier thresholds fixed to 50\% signal efficiency, Figure~\ref{subfig:disco_perf} shows the JSD as a function of the background rejection where the curves are produced through training with varying sizes of the regularization hyperparameter. The CNN models are compared with neural networks trained on substructure features and other classifiers with de-correlation methods applied. The CNN models, both the adversarial and distance correlation regularization, are seen to typically provide the highest background rejection for a given level of de-correlation compared to other models.

\subsection{Multi-Channel Jet Tagging with CNNs}\label{subsec:tagincolor}

Recent work on jet image based tagging has shown performance gains through the use of multi-channel images. While single channel jet images have provided gains in classification performance over individual, or pairings of, engineered substructure features, the performance benefits were typically smaller when compared to ML models trained on larger groups of substructure features (except when very large models were used, as in~\cite{Kasieczka_2019}). Multi-channel jet images use calorimeter images as only a single input image channel, with additional channels computed from charged particle features such as the momentum, multiplicity, or charge. There is a significant amount of freedom in choosing the definition of the additional image channels, allowing for a flexibility in the choice of inductive bias to deliver relevant information to the CNN. 

One challenge in combining charged particle trajectory information and calorimeter images is the mismatch in resolution; charged particle trajectories tend to have a significantly finer spatial resolution than calorimeters, thus leading to the questions of how to combine such information. As charged particles are not measured on a regular grid, often the same spatial grid for the calorimeter component is used for the charged particle image and the energy of the constituents is summed within each pixel. Alternatively, separate CNN blocks (or upsampling procedures) can be used to process charged and calorimeter images separately into a latent representation of equal size such that they can be merged for further processing. Note that when Particle Flow objects are used, and thus both neutral and charged particle measurements do no necessarily fall on a grid, a fine grid can be used to exploit the better charged particle momentum resolution.  It should also be noted that while phenomenological studies at particle-level often use fixed grids to emulate the discretization of real detectors, different inputs (i.e. charge vs neutral) in real detector settings have different resolutions which may be difficult to account for in simple discretization approaches. 

Multi-Channel jet image based tagging was introduced in phenomenological studies of discriminating between quark initiated and gluon initiated jets~\cite{Komiske2017DeepLI,Fraser2018JetCA} and has since been explored within the \textit{quark vs. gluon} context on the ATLAS experiment~\cite{ATL-PHYS-PUB-2017-017}, in CMS Open Data~\cite{OpenDataPortal,Andrews_qvg_2020}, and for tagging in heavy ion collision environments~\cite{Chien:2018dfn}. More broadly, multi-channel jet image tagging has lead to improved performance in phenomenological studies of boosted top quark jet tagging~\cite{Macaluso2018PullingOA,diefenbacher2019capsnets,Kasieczka_2019}, as well as in boosted $W/Z$ jet tagging~\cite{2019arXiv190808256C} and in boosted Higgs boson tagging~\cite{2018JHEP...10..101L,2019JHEP...09..047K}. Notably, multi-channel jet image based boosted top tagging has been explored on the CMS experiment~\cite{Sirunyan:2020lcu} including the comparison and calibration of this discriminant with respect to CMS collision data, thus adding additional insights into the usability of such models within LHC data analysis. 

The use of multi-channel jet images built from charged particle momentum and multiplicity information within the context of discriminating between quarks and gluons is natural, as the number of charged particles within such a jet is known to be a powerful discriminant for this challenging task~\cite{Gallicchio2013QuarkAG}. As such, in the phenomenological studies of reference~\cite{Komiske2017DeepLI} three jet image channels were defined: (1) the transverse momentum of charged particles within each pixel, (2) the transverse momentum of neutral particles within each pixel, and (3) the charged particle multiplicity within each pixel. The same pixel size was used in each image, thus facilitating the direct application of multi-channel CNNs. This approach thus relies on the ability to separate the charged and neutral components of a jet; while the charged component is measured using tracking detectors, the unique identification of the neutral component of a jet is significantly more challenging task. However, advancements in Particle Flow~\cite{2017JInst..12P0003S} aid in such a separation, albeit not perfectly and with differing resolutions between charged and neutral measurements.

\begin{figure}[htbp]
\begin{center}
\begin{subfigure}[h]{0.48\linewidth}
\includegraphics[width=\linewidth]{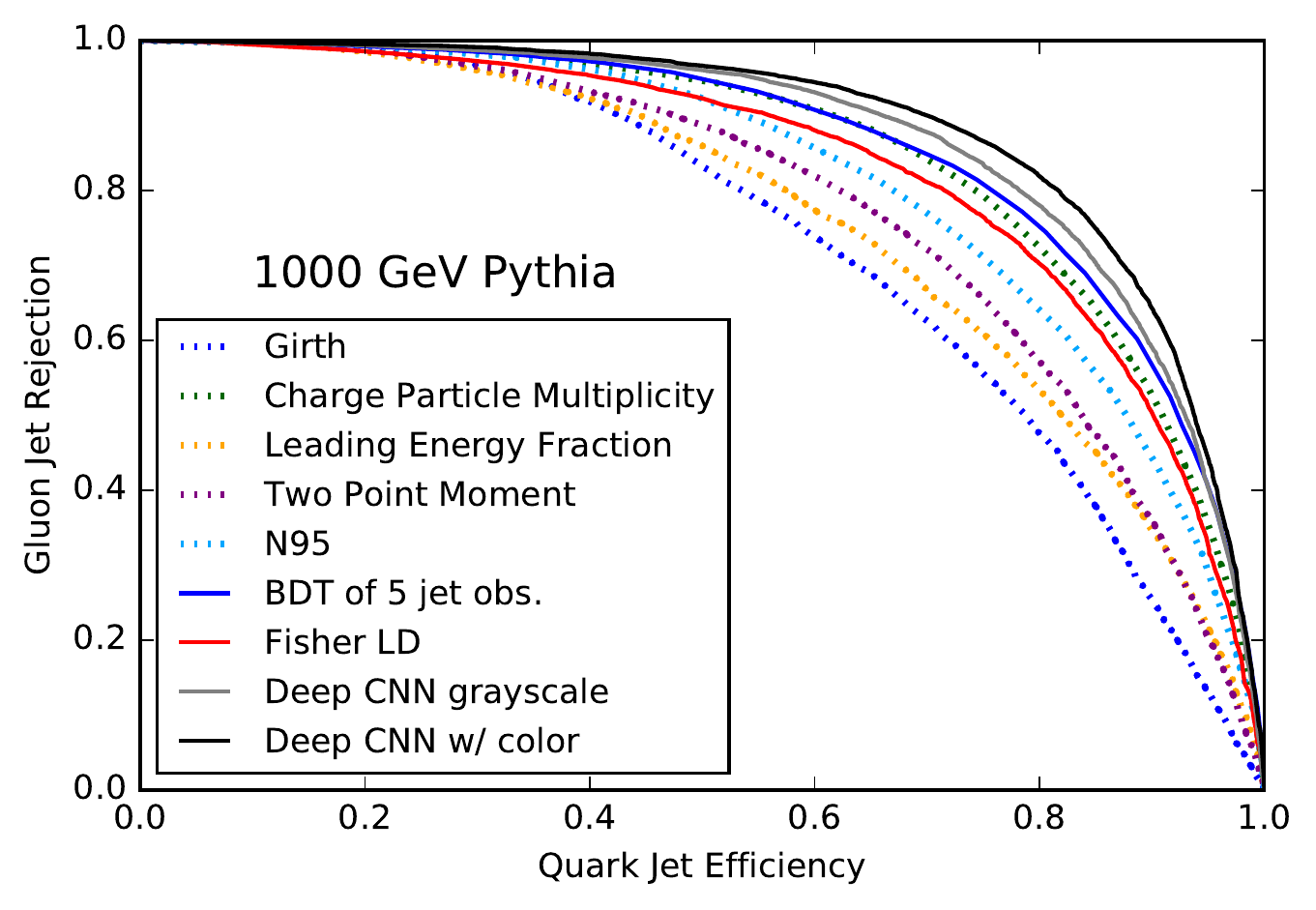}
\caption{}
\label{subfig:qvg_roc}
\end{subfigure}\quad
\begin{subfigure}[h]{0.48\linewidth}
\includegraphics[width=\linewidth]{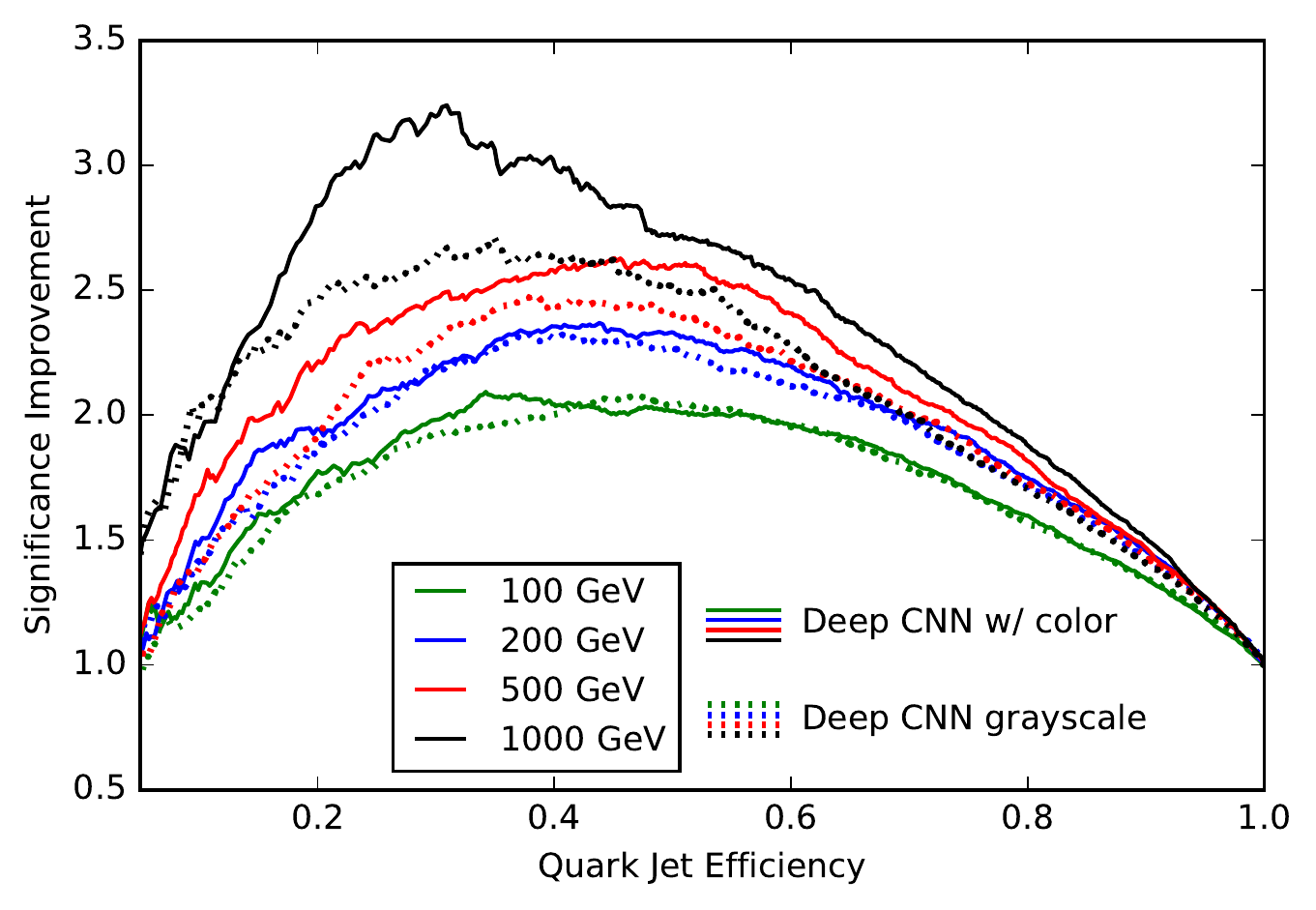}
\caption{}
\label{subfig:qvg_sic}
\end{subfigure}
\end{center}
\caption{ROC curve (a) and SIC curve (b) for quark versus gluon tagging using multi-channel jet images~\cite{Komiske2017DeepLI}. Comparisons with jet substructure  based discriminants is shown in (a), while comparison between single channel and multi-channel jet image based tagging with CNNs is shown in (b).}
\label{fig:qvg}
\end{figure}

The benefit of the multi-channel approach for quark vs. gluon discrimination can be seen in the ROC and SIC curves in Figure~\ref{fig:qvg}. Both the calorimeter only approach, denoted Deep CNN grayscale, as well as the multi-channel approach, denoted Deep CNN w/ color, outperform single features engineered for this task, BDTs trained using five of such features, and a linear discriminant trained on the greyscale jet images. In addition, the multi-channel model is seen to dominate over the single channel model in both the ROC curve for jets with a momentum of $p_T \approx 1000$ GeV and in the SIC curve across a range of jet momentum.  The multi-channel approach is found to be especially beneficial at higher momentum where the jets have a large charged particle multiplicity.

\begin{figure}[htbp]
\begin{center}
\includegraphics[width=0.5\linewidth]{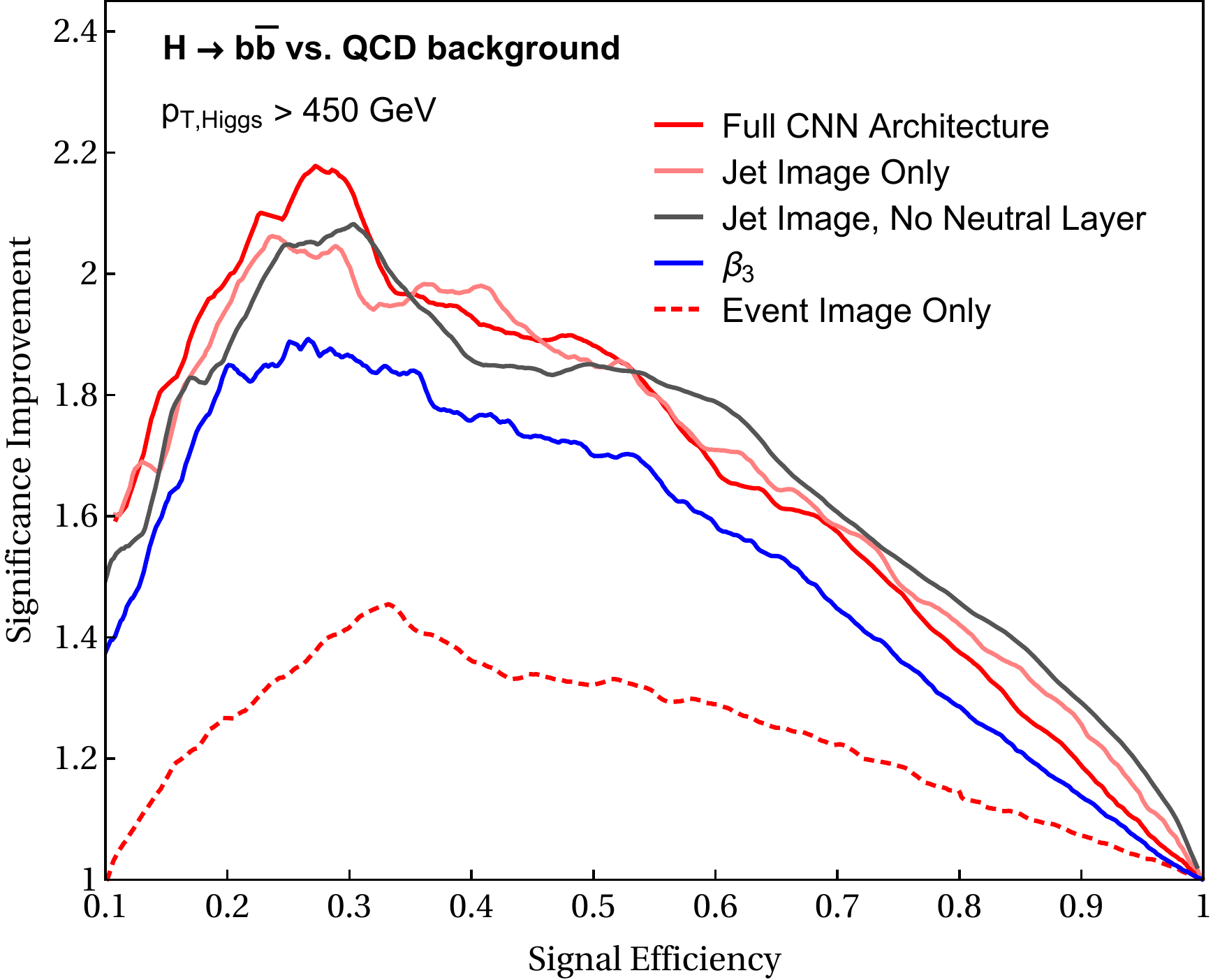}
\end{center}
\caption{SIC curve for boosted Higgs to $b\bar{b}$ tagging using multi-channel jet images~\cite{2018JHEP...10..101L}. Models using only jet images, and models using both jet images and ``event images" are shown. }
\label{fig:hbb}
\end{figure}

This multi-channel approach using charged, neutral, and multiplicity channels was also found to be powerful in phenomenological studies of discriminating between boosted Higgs boson jets and a background of gluon splitting to $b\bar{b}$ jets in multi-jet events~\cite{2018JHEP...10..101L}. In addition to a CNN focused on discrimination based on jet images, this work also explored simultaneously processing an \textit{event image}, defined using the aforementioned three channels over the entire calorimeter, through a separate set of convolutional layers and combining with the output of the convolutional processing of jet image before discrimination. By including such an event image, one may explore the potential benefits of event topology information outside of the jet image for discrimination. The SIC curve for this discrimination task can be seen in Figure~\ref{fig:hbb}, where the CNN approaches were seen to significantly outperform single engineered features. CNNs using only the jet image, event image, or both (denoted ``Full CNN Architecture" in Figure~\ref{fig:hbb}) were compared, showing that much of the discrimination power rests in the jet image whilst the event image may provide some modest improvements. In addition, the jet image discrimination without the neutral particle channel was also found to be comparable to one using the neutral channel, indicating that much of the discrimination power lies in the charged particle information within the jet.

\begin{figure}[htbp]
\begin{center}
\begin{subfigure}[h]{0.52\linewidth}
\includegraphics[width=\linewidth]{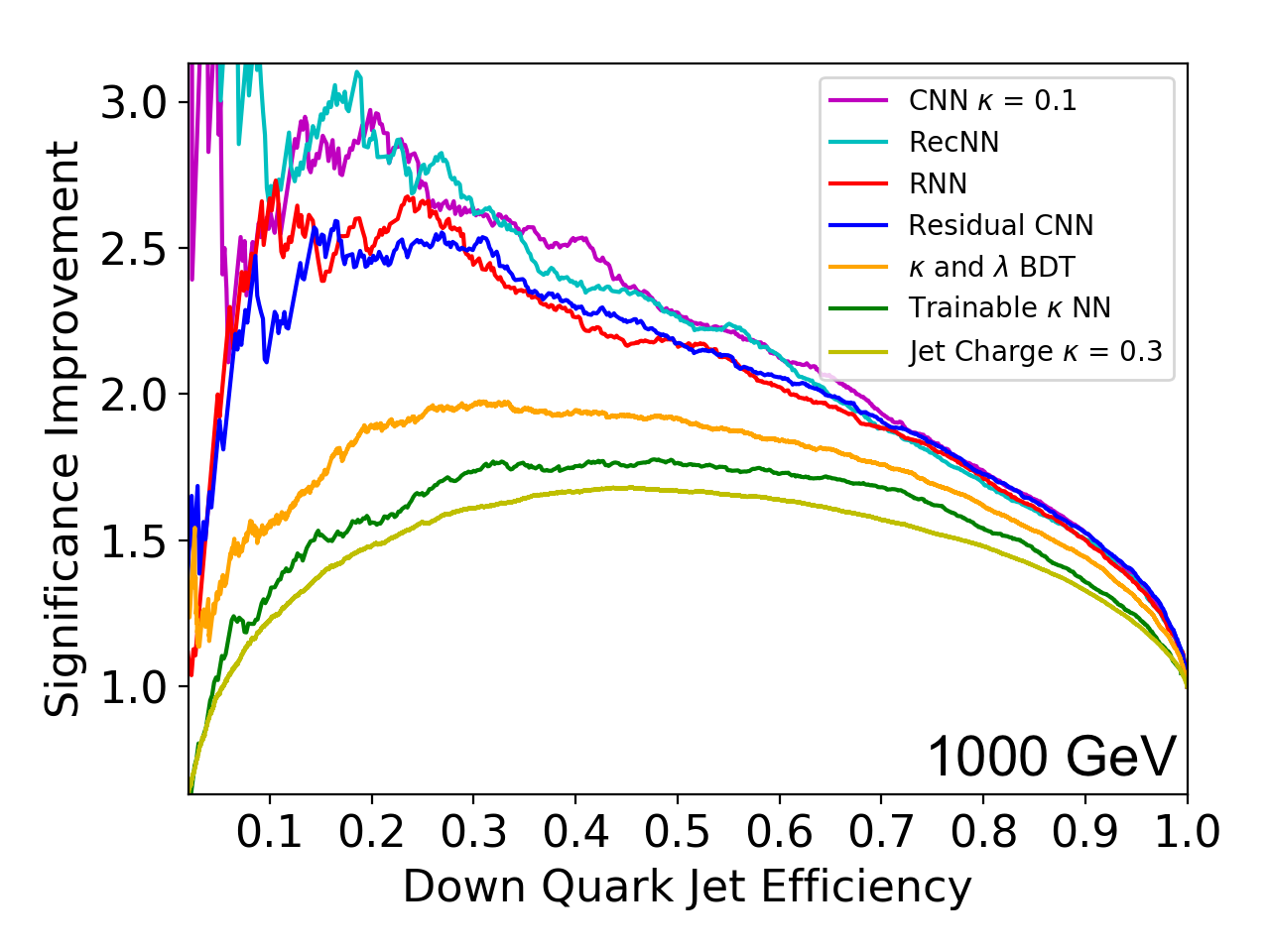}
\caption{}
\label{subfig:uvd_sic}
\end{subfigure}\quad
\begin{subfigure}[h]{0.44\linewidth}
\includegraphics[trim={0 0 0 1.2cm},clip,width=\linewidth]{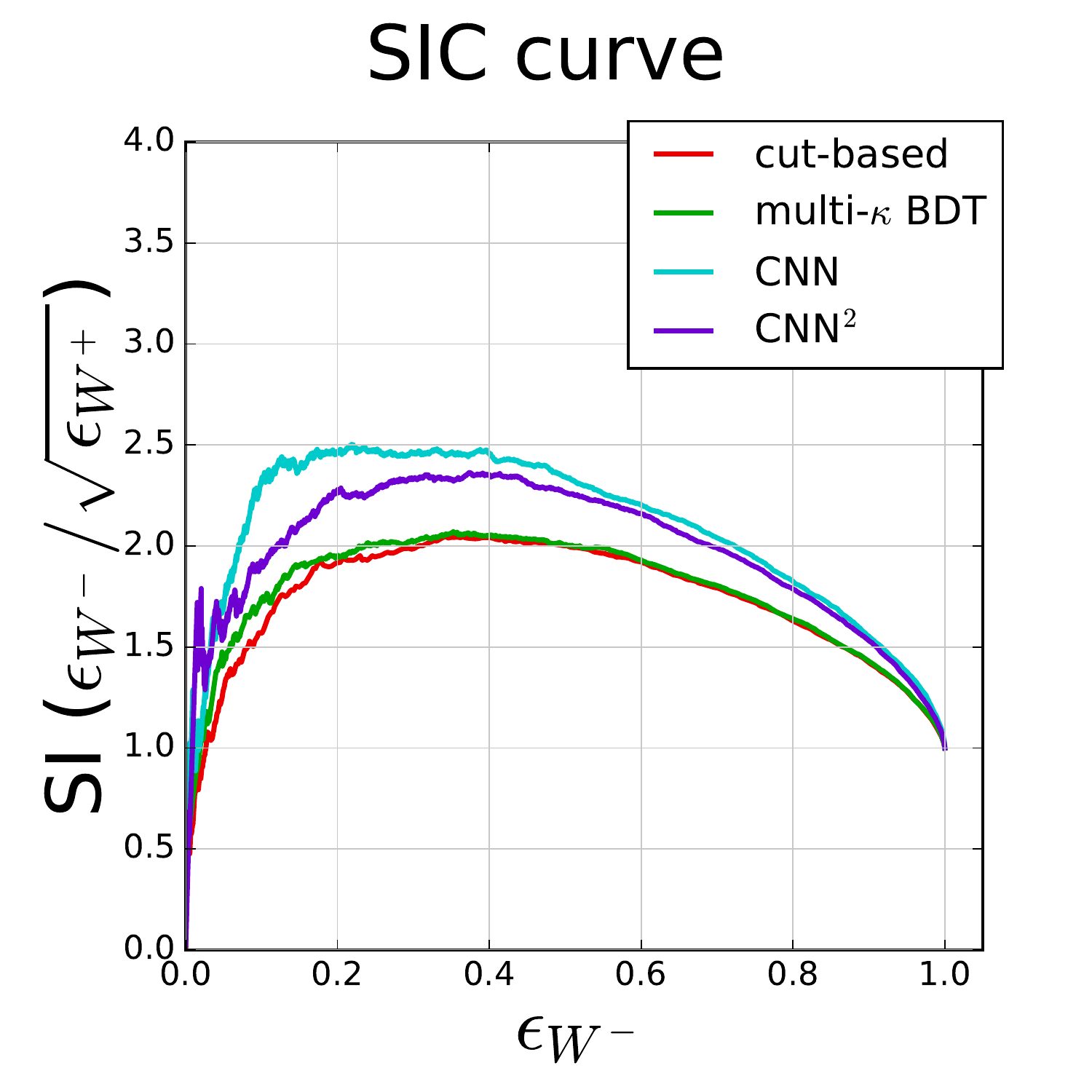}
\caption{}
\label{subfig:ww_sic}
\end{subfigure}
\end{center}
\caption{SIC curves for (a) discriminating down quarks from up quarks~\cite{Fraser2018JetCA}, and (b) discriminating between $W^+$ and $W^-$ bosons~\cite{2019arXiv190808256C}. The CNN $\kappa=0.1$ model in the left figure and the CNN models of the right Figure utilize the per pixel $p_T$ weighted charge image. }
\label{fig:charge}
\end{figure}

While a clear approach to extending jet images to contain multiple channels is to sum the momentum of the charged particles or compute multiplicities in each pixel to form an image channel, the high resolution of the charged particle information allows for the introduction of additional inductive bias. More specifically, given the set of charged particles contained in the region of a pixel, one may compute pixel-level features that may be more amenable to a given discrimination task. This approach was followed for building CNNs to discriminate between (a) up and down type quarks, and (b) quarks and gluons~\cite{Fraser2018JetCA}. In these phenomenological studies, knowledge of the utility of the jet charge feature~\cite{FIELD19781,PhysRevLett.110.212001,PhysRevD.86.094030} for discriminating jets of different parent particle charge inspired the development of the jet image channel computed per pixel as the $p_T$ weighted charge $Q_\kappa = \frac{1}{(\sum_j p_T^{(j)})^\kappa} \sum_j Q^{(j)} (p_T^{(j)})^\kappa$.  The SIC curve showing the performance of the CNN trained on the two channel jet images, one channel for $p_T$ and one for $Q_\kappa$ per pixel, is shown in Figure~\ref{subfig:uvd_sic}. The two channel CNN significantly outperformed the total jet charge and classifiers trained on engineered features, and is comparable to other deep architectures trained for this task.

\begin{figure}[htbp]
\begin{center}
\includegraphics[width=0.32\linewidth]{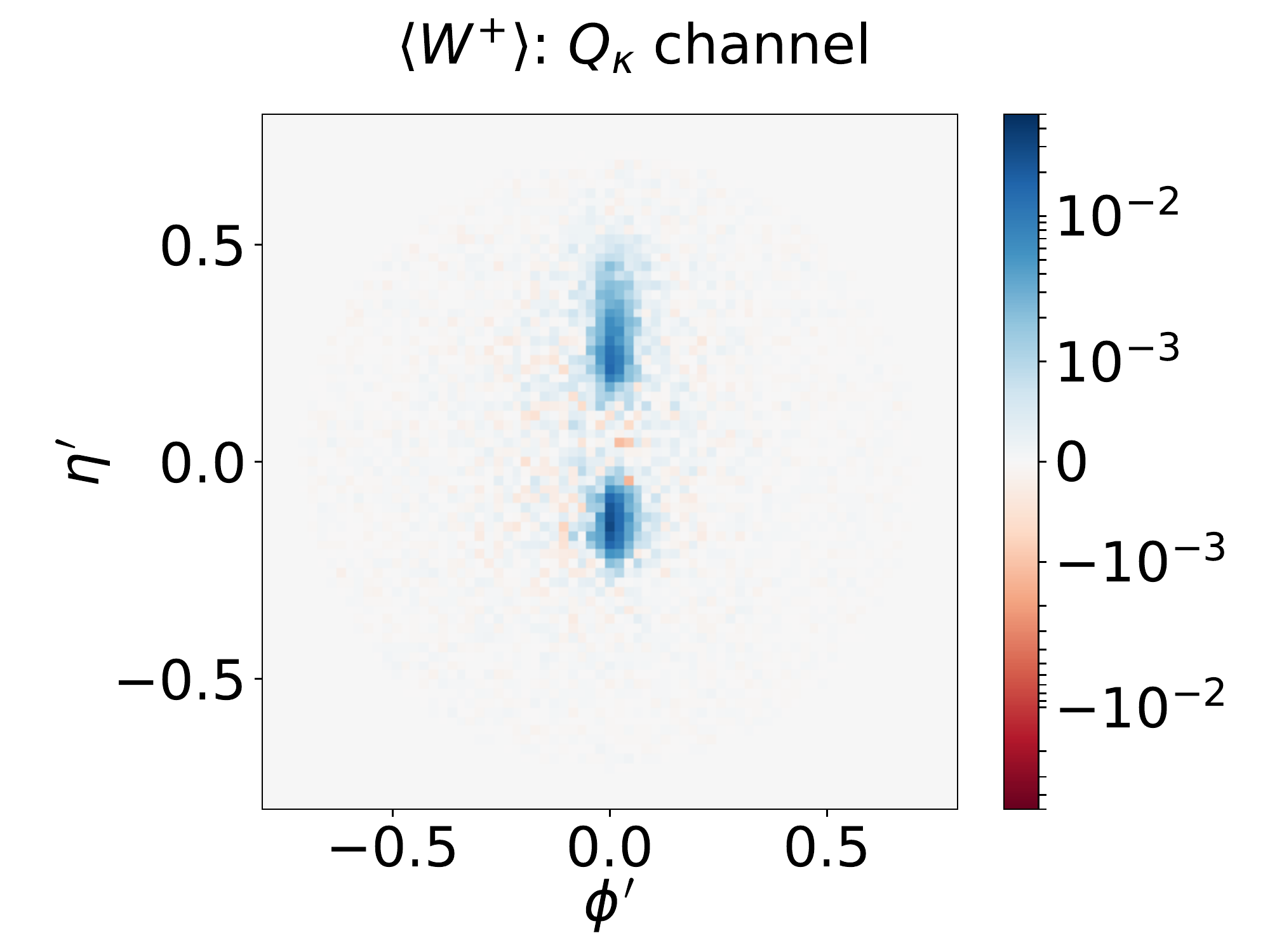}
\includegraphics[width=0.32\linewidth]{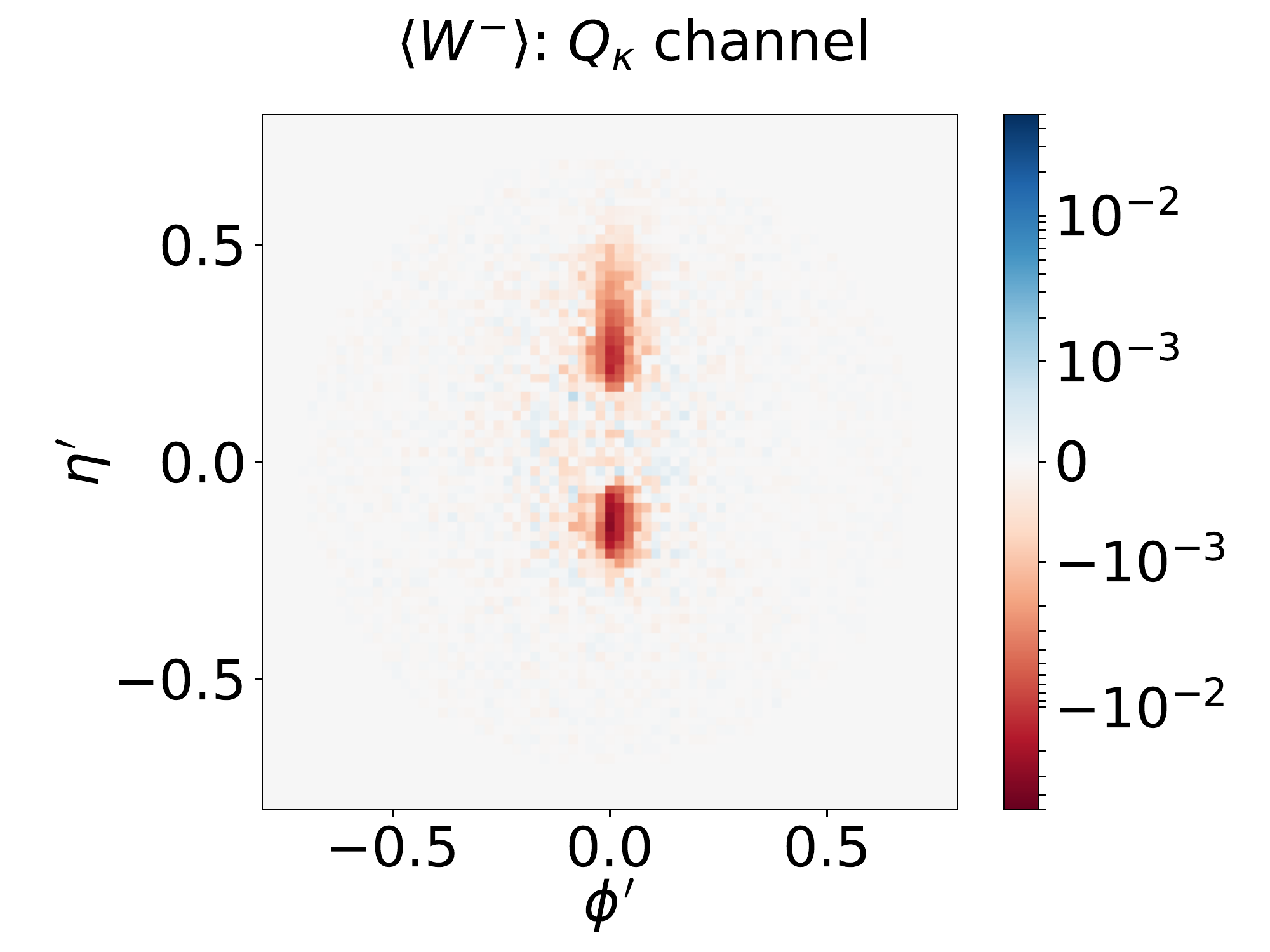}
\includegraphics[width=0.32\linewidth]{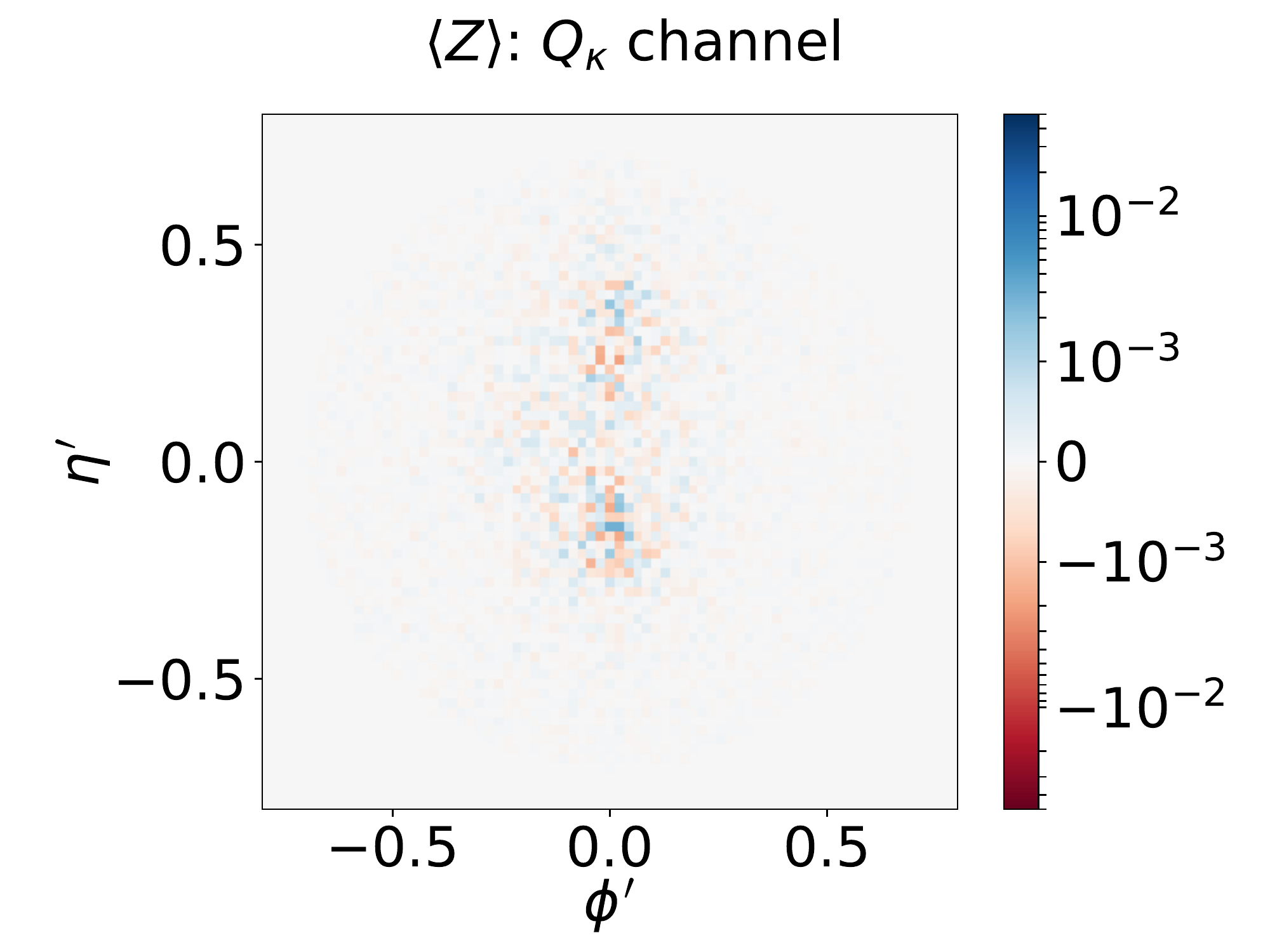}
\end{center}
\caption{Average image of the per pixel $p_T$ weighted charge $Q_\kappa$ is shown for $W^+$ (left), $W^-$ (middle), and $Z$ bosons (right)~\cite{2019arXiv190808256C}.}
\label{fig:Wmulti_q_image}
\end{figure}

A similar jet charge based multi-channel CNN was explored for discriminating between boosted $W^{+} / W^{-} / Z$ boson jets in the phenomenological studies of reference~\cite{2019arXiv190808256C}.  The per pixel charge image averaged over the test set for $W^{+}$, $W^{-}$, and $Z$ jet images is shown in Figure~\ref{fig:Wmulti_q_image}. The geometry of all three images is similar, but the average per pixel charge differs significantly as expected, with the typical $W^{+}$ image carrying a positive pixel value, the typical $W^{-}$ image carrying a negative pixel value, and the typical $Z$ image having charge close to zero.  The SIC curve for discriminating between $W^{+}$ and $W^{-}$ jets can be seen in Figure~\ref{subfig:ww_sic}. Two CNNs were explored in this work, one denoted \textit{CNN} in which both a $p_T$ and $Q_\kappa$ image were processed together (i.e. as a single multi-channel image processed by convolutional layers) and one denoted \textit{CNN$^2$} in which each channel is processed by a separate stack of convolutional layers and then combined before the classification layers. Both CNNs significantly outperform methods based on engineered features.

\begin{figure}[htbp]
\begin{center}
\includegraphics[width=0.5\linewidth]{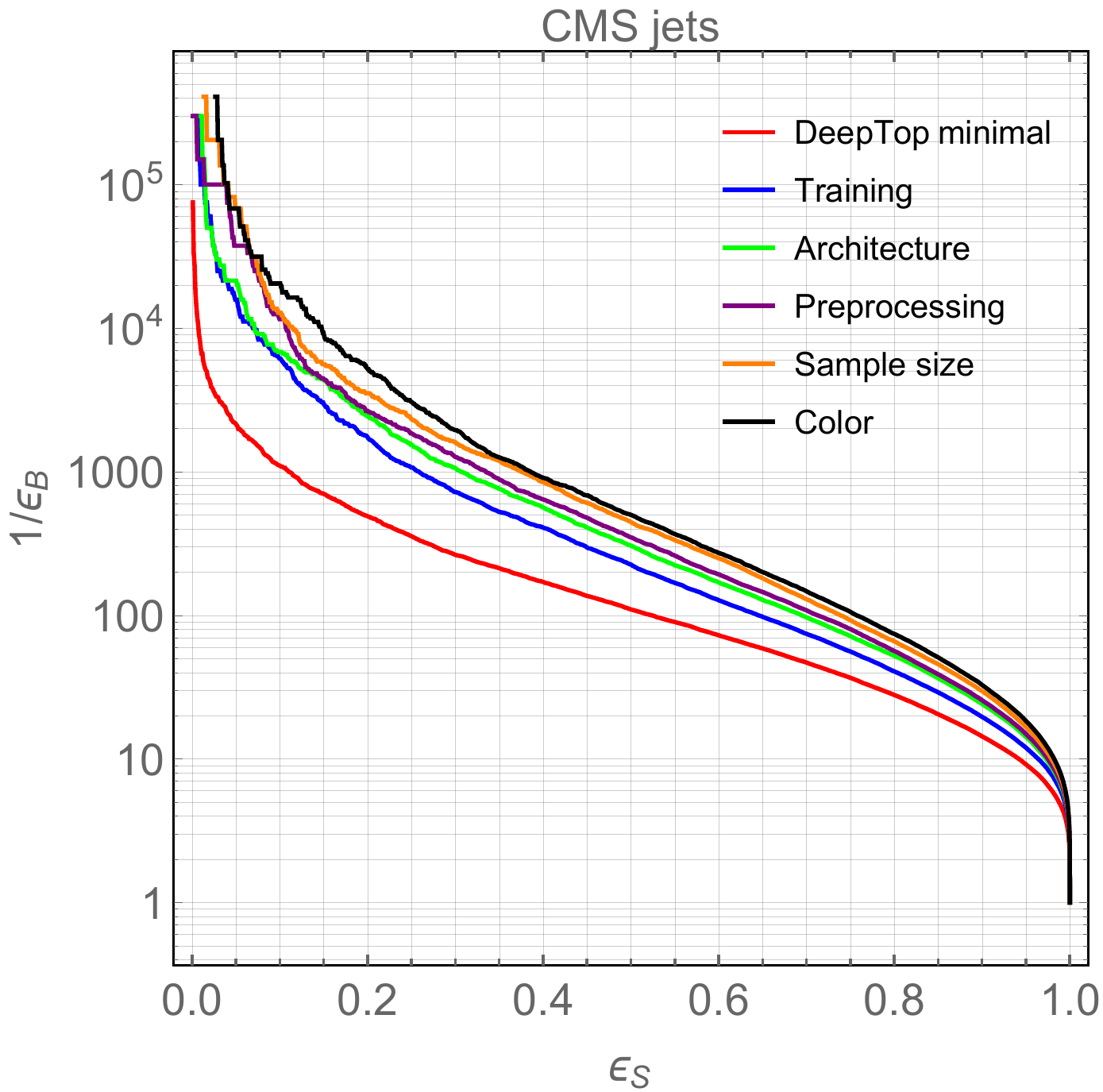}
\end{center}
\caption{ROC curve of boosted top jet tagging efficiency versus background quark and gluon rejection for the minimal DeepTop model~\cite{Kasieczka2017DeeplearningTT} compared with models sequentially including the changes proposed in~\cite{Macaluso2018PullingOA}.}
\label{fig:deeptopcolor}
\end{figure}

Multi channel jet images were explored for top tagging in the phenomenological studies of reference~\cite{Macaluso2018PullingOA}, using four channel jet images defined with the neutral jet component as measured by the calorimeter, the charged particle sum $p_T$ per pixel, the charged particle multiplicity per pixel, and the muon multiplicity per pixel. The architecture is discussed in Section~\ref{subsec:JIsingle} within the context of single channel jet images. The inclusion of the muon image channel targets the identification of $b$ quark initiated subjets within the top jet as muons can be produced in $b$-hadron decays. As noted in Section~\ref{subsec:JIsingle}, several changes to the model architecture, preprocessing, and training procedure relative to the first proposed DeepTop model~\cite{Kasieczka2017DeeplearningTT} were included in this work. The impact of these individual changes can be see in Figure~\ref{fig:deeptopcolor} wherein developments on top of the first proposed DeepTop model~\cite{Kasieczka2017DeeplearningTT} are sequentially added to the model and the resulting ROC curve is shown. The inclusion of multiple ``color" channels was only seen to provide modest performance gains over single channel jet images. Notable among changes that led to the largest improvements were changing the optimization objective to be more suitable for classification tasks and changing the optimizer to ADAM (denoted training in the figure),  increasing the model size (denoted architecture in the figure), and increasing the sample size. In agreement with these results, recent CNN models built for processing jet images have also tended to focus on larger models with large samples for training.

\subsection{Sensitivity to Theory Uncertainties}\label{subsec:ji_theory_unc}

While matrix element and parton shower Monte Carlo generators often provide high fidelity predictions of the data generation process, they provide only approximations to the scattering and showering processes and empirical models of the hadronization process. As such, uncertainties in the theoretical predictions of these generators must be propagated to downstream analyses. One mechanism for doing this is to compare an observable computed with samples from different Monte Carlo generators. While not a precise estimation of theoretical uncertainty, this comparison can provide a test of whether an observable is potentially sensitive to the different approximations of the different generators. 

\begin{figure}[htbp]
\begin{center}
\begin{subfigure}[h]{0.45\linewidth}
\includegraphics[width=0.95\linewidth]{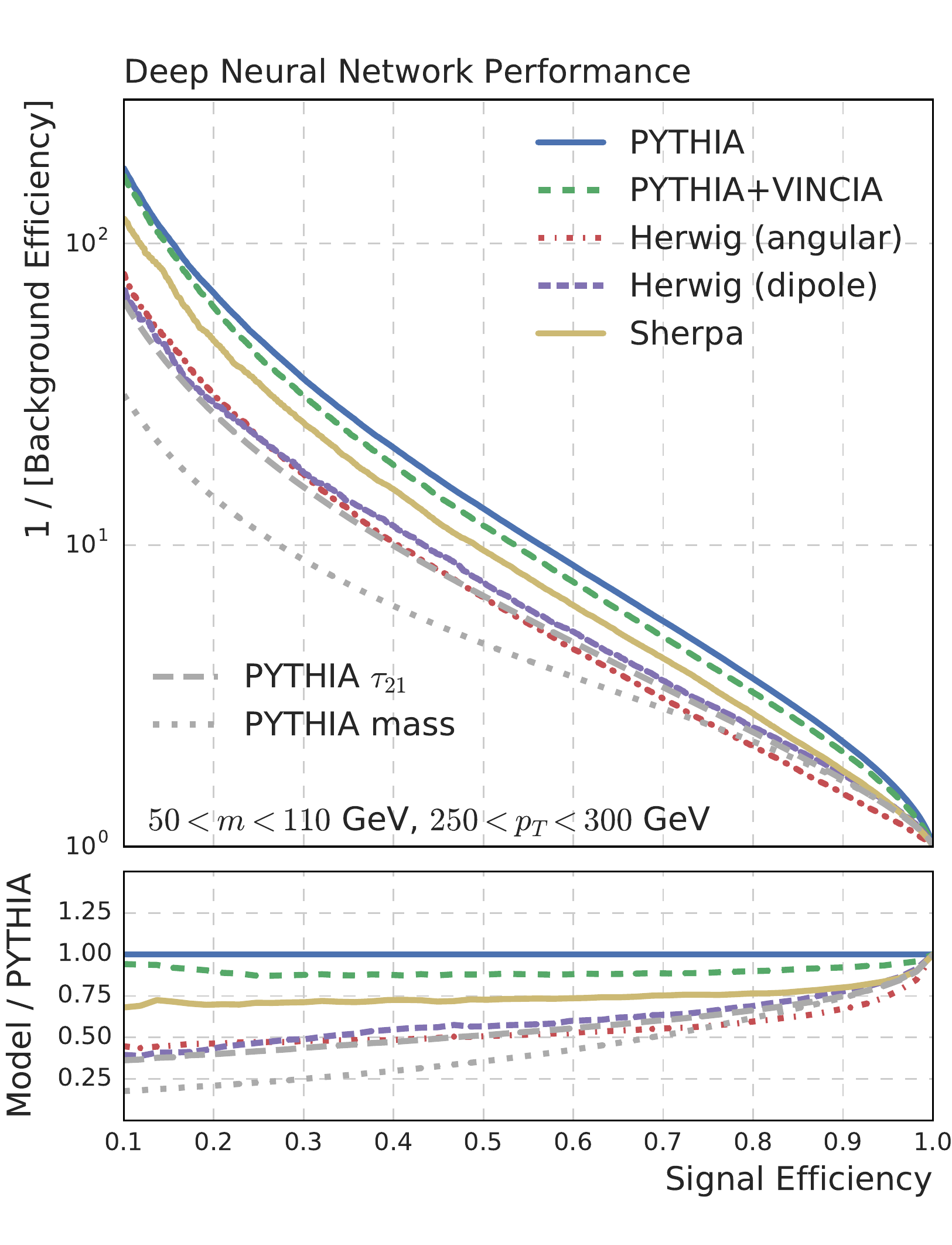}
\caption{}
\label{subfig:gen_var_W}
\end{subfigure}\quad
\begin{subfigure}[h]{0.51\linewidth}
\begin{center}
\vspace{2cm}
\includegraphics[width=\linewidth]{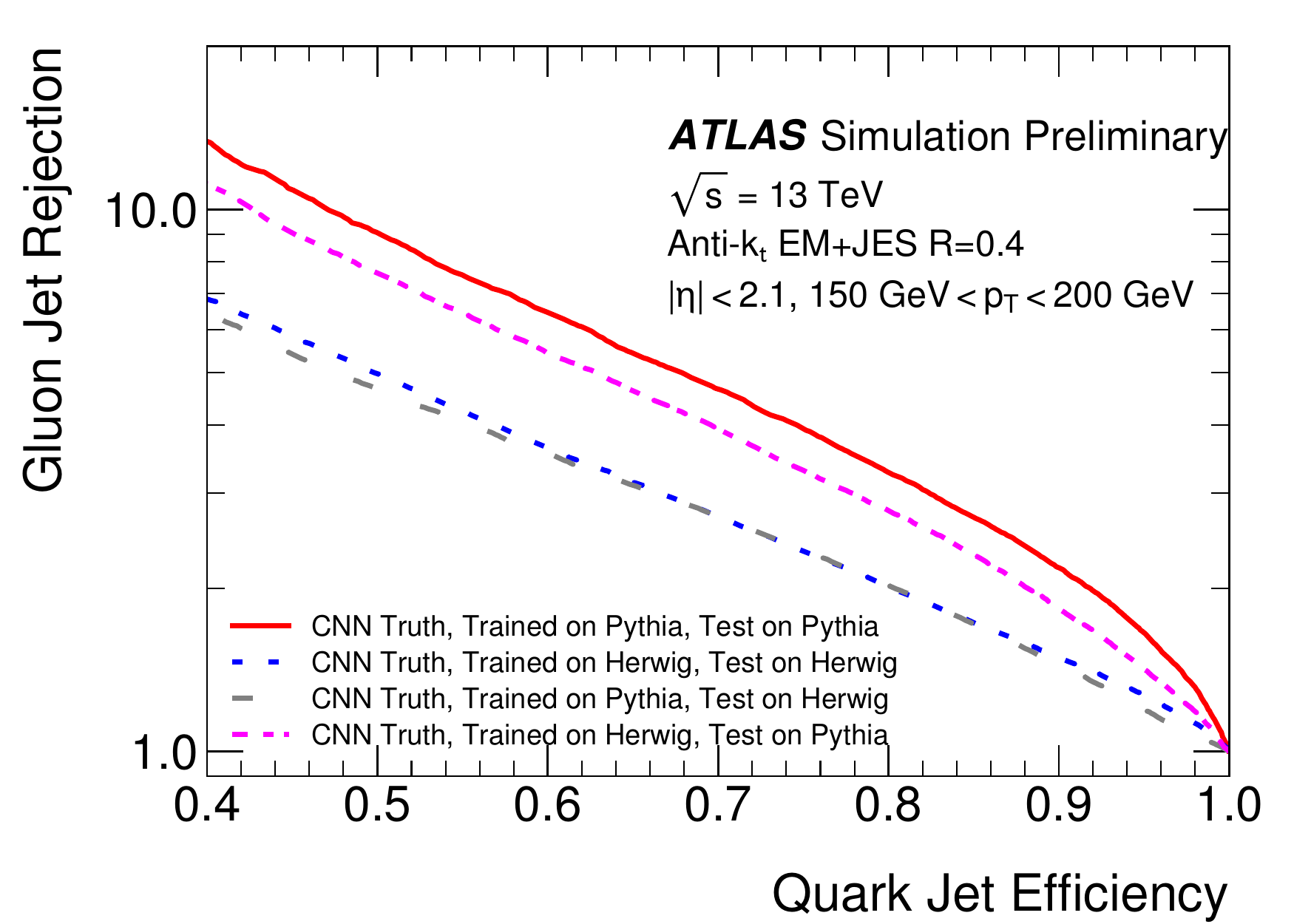}
\caption{}
\label{subfig:gen_var_atlas_qvg}
\end{center}
\end{subfigure}
\end{center}
\caption{(a) ROC curves for boosted $W$ tagging for a jet image based CNN tagger trained on Pythia generated jet images and applied to jet images from various generators are shown~\cite{PhysRevD.95.014018}. (b) In events with the ATLAS full detector simulation, ROC curve for quark versus gluon tagging for jet image based CNN tagger trained and applied on Pythia and Herwig based jet images are shown~\cite{ATL-PHYS-PUB-2017-017}.}
\label{fig:gen_var}
\end{figure}

This sensitivity has been examined for CNN-based taggers operating on jet images in several works, and we focus here on $W$-tagging in a phenomenological study~\cite{PhysRevD.95.014018} and on quark / gluon tagging using ATLAS simulation~\cite{ATL-PHYS-PUB-2017-017}. As the CNN + jet image approaches utilize the distribution of energy throughout a jet image to  discriminate, one concern is that the differences in modeling of the jet formation process by different generators may lead to large performance variations. To study this, reference~\cite{ATL-PHYS-PUB-2017-017} trained a CNN model on boosted $W$ boson jet images generated by Pythia~\cite{SJOSTRAND2008852,Sj_strand_2006} and applied this trained model on samples of  boosted $W$ boson jet images generated by different Monte Carlo generators. The ROC curves of the performance can be see in Figure~\ref{subfig:gen_var_W}, wherein, at the same signal efficiency, reductions of background rejection of up to 50\% can be seen when this tagger is applied to different generators. While such a variation is not ideal, it should be noted that similar variations were seen when a tagger of only substructure features, a binned two dimensional signal over background likelihood ratio of the distribution of jet mass and $\tau_{21}$, is applied for the same tagging task. Similar levels of performance variation are also seen in the ROC curves built for quark vs gluon tagging in ATLAS simulation with a CNN trained on Pythia jet images applied to Herwig~\cite{bellm2013herwig} generated jet images, as seen in Figure~\ref{subfig:gen_var_atlas_qvg}. Interestingly, when the test is reversed and the CNN is trained on jet images from Herwig and applied to jet images from Pythia, the tagging performance is similar to the CNN trained and applied to Pythia jet images. This suggests that the CNNs in both cases are learning similar representations of information useful for quark vs gluon tagging, but the amount this information is expressed in the jet images varies between generators~\cite{ATL-PHYS-PUB-2017-017}. Thus while these studies show that CNNs applied to different samples may vary in performance, there may be an underlying robustness to the information learned by CNNs for jet tagging.

\begin{figure}[htbp]
\begin{center}
\begin{subfigure}[h]{0.48\linewidth}
\includegraphics[width=\linewidth]{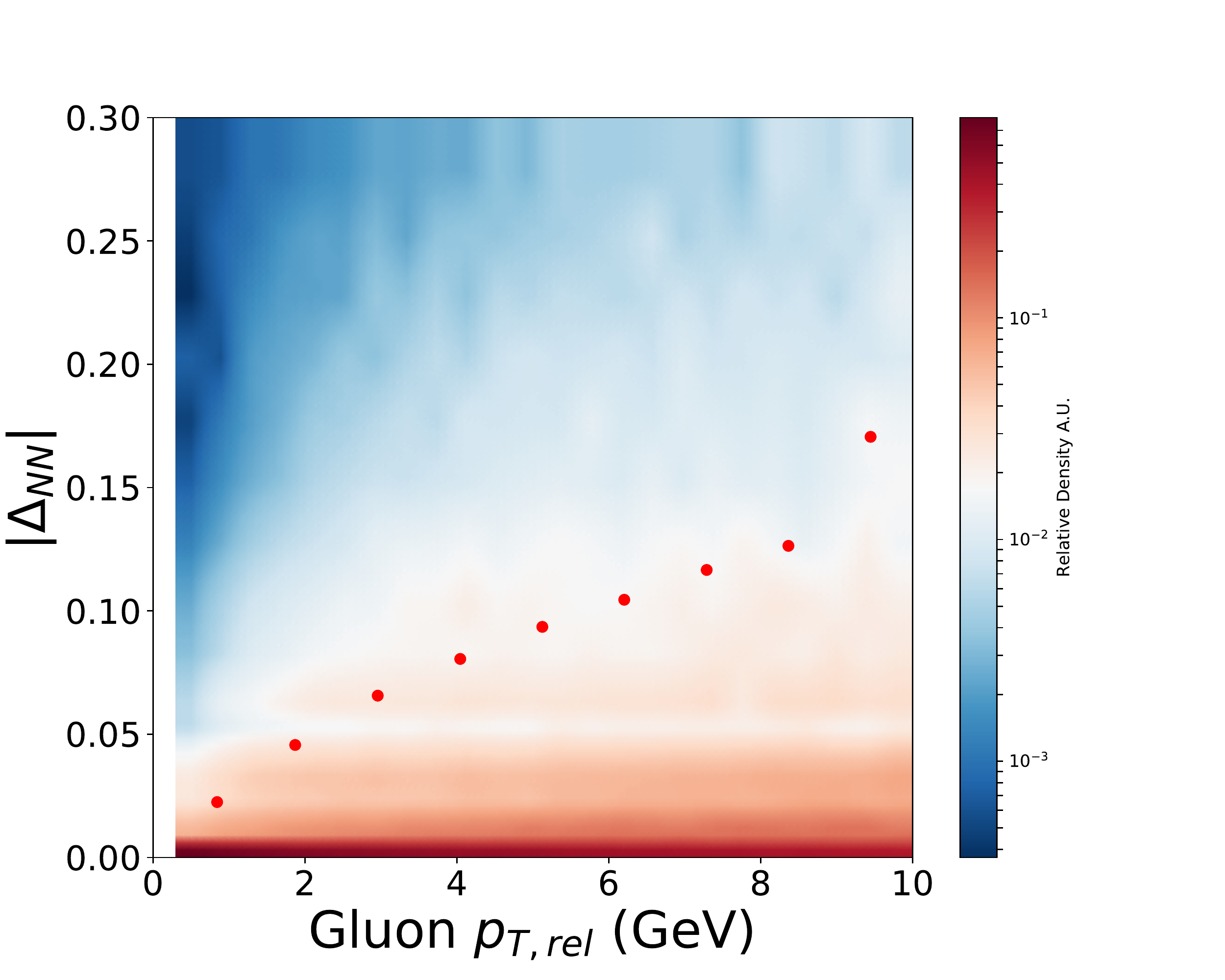}
\caption{}
\label{subfig:irc_ptrel}
\end{subfigure}\quad
\begin{subfigure}[h]{0.48\linewidth}
\begin{center}
\includegraphics[width=\linewidth]{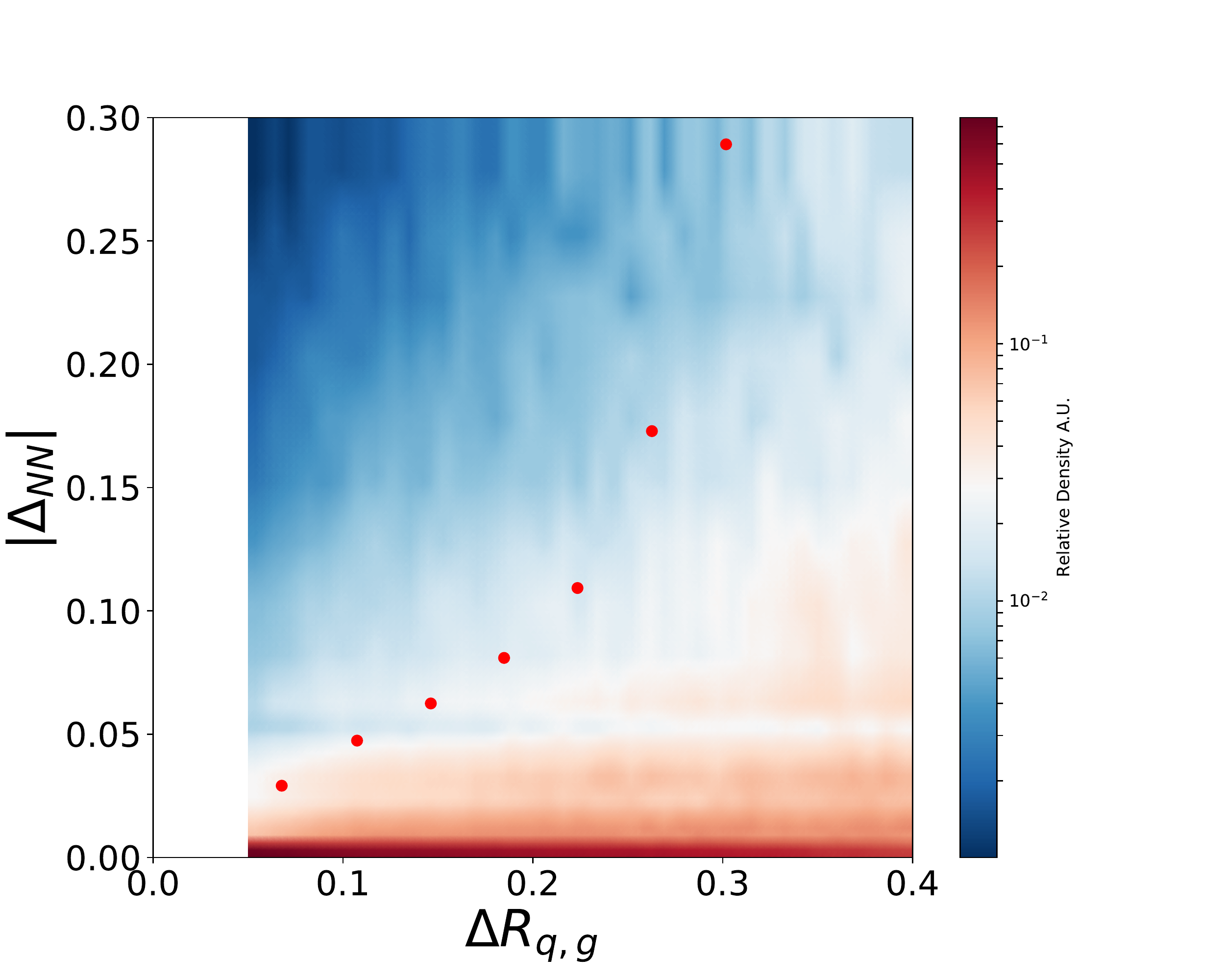}
\caption{}
\label{subfig:irc_dr}
\end{center}
\end{subfigure}
\end{center}
\caption{$\Delta_{NN}$ versus (a) gluon relative transverse momentum and (b) $\Delta R$ between the gluon on the nearest top decay parton. $\Delta_{NN}$ is the difference between particle jet trained CNN output applied on parton level jet images with and without merging the closest gluon with a top decay parton~\cite{Choi2018InfraredSO}. Red points denote the point at which 90\% of events within a vertical slice of the distribution are contained.}
\label{fig:irc}
\end{figure}

Beyond the potential tagging variations due to generator uncertainties, a key question when developing a jet observable of any kind is whether such an observable is theoretically sound and calculable. This is often expressed as whether the observable is infrared and collinear (IRC) safe. IRC safety for jet image based tagging of boosted top jets with CNNs has been examined empirically in the phenomenological studies of reference~\cite{Choi2018InfraredSO}. In this work, within the context of boosted top jet tagging using a jet image based CNN, a feature denoted $\Delta_{NN}$ is studied which explores the impact of merging soft/collinear radiation with nearby partons. $\Delta_{NN}$ is constructed as follows: (a) a CNN was trained on particle level jet images for boosted top tagging, (b) parton level jet images are generated for boosted top decays without (unmerged) and with (merged)  adding the closest gluon to a top quark parton together before forming the image, (c) the difference in CNN output between unmerged and merged jet images is defined as $\Delta_{NN}$. By examining the distribution of $\Delta_{NN}$ and its variations with features that explore soft or collinear effects, the sensitivity of the CNN tagger to IRC effects can be studied empirically.  This can be seen in Figure~\ref{fig:irc}, where the 2D distribution of $\Delta_{NN}$ and the gluon relative transverse momentum, and the $\Delta R$ to the parton, are shown. As either the gluon relative momentum or the $\Delta R$ tend to zero, the $\Delta_{NN}$ distribution tends towards a sharp peak at 0, which would be indicative of the CNN being insensitive to IRC perturbations.

\subsection{Jet images in LHC Experiments}\label{subsec:ji_exp}

The ultimate tests for the efficacy of jet image based tagging approaches are that the performance observed in phenomenological studies is also observed in realistic high fidelity simulations and that their performance generalizes to real data without large systematic uncertainties. With that in mind, jet image base tagging approaches have been examined for quark vs gluon tagging in ATLAS simulations~\cite{ATL-PHYS-PUB-2017-017} and in CMS Open Data~\cite{OpenDataPortal} simulated samples~\cite{Andrews_qvg_2020}, and for boosted top quark jet tagging in CMS simulation and real data~\cite{Sirunyan:2020lcu}.

\begin{figure}[htbp]
\begin{center}
\begin{subfigure}[h]{0.48\linewidth}
\begin{center}
\includegraphics[width=0.98\linewidth]{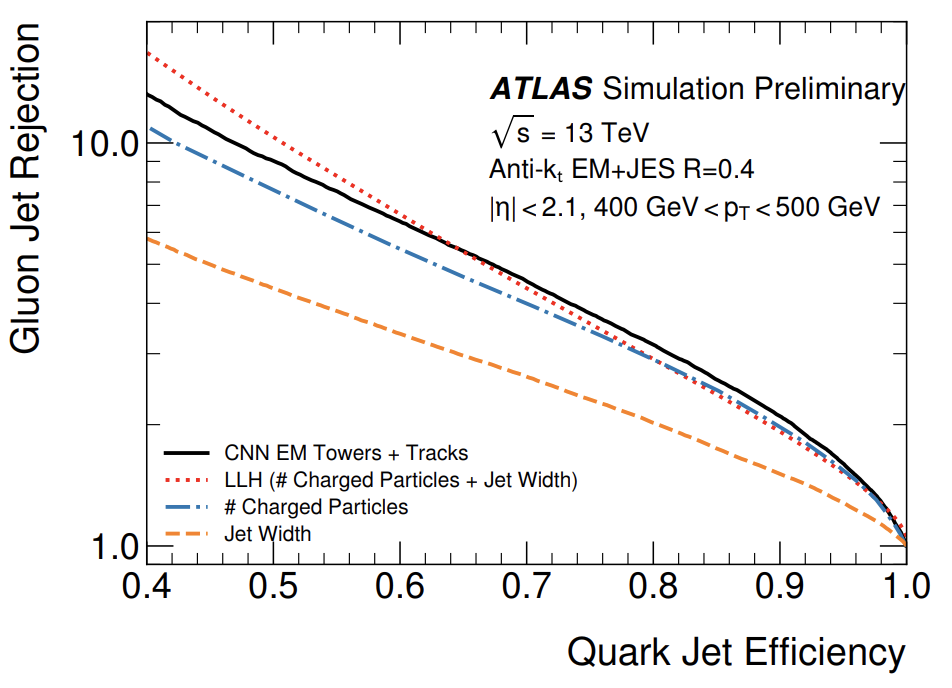}
\caption{}
\label{subfig:atlas_qvg_roc}
\end{center}
\end{subfigure}
\begin{subfigure}[h]{0.48\linewidth}
\includegraphics[width=\linewidth]{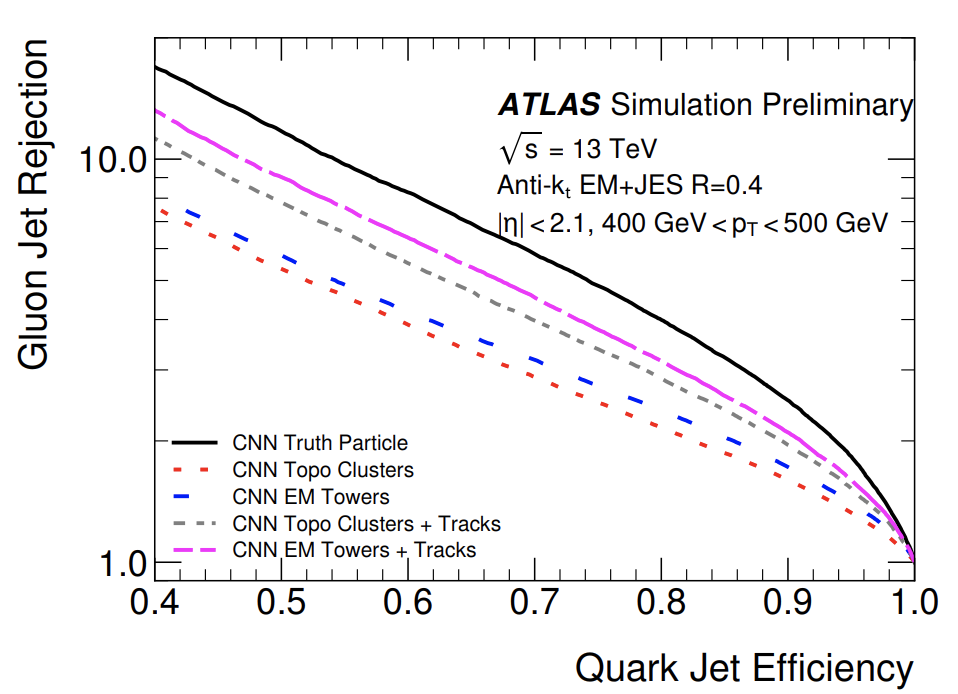}
\caption{}
\label{subfig:atlas_qvg_inputs}
\end{subfigure}\quad
\end{center}
\caption{ROC curves for quark jet efficiency versus gluon jet rejection in ATLAS fully simulated datasets showing comparisons of (a) jet image based CNN taggers against jet width and number-of-tracks discriminants, and (b) of jet image basd CNN taggers trained with different input images~\cite{ATL-PHYS-PUB-2017-017}.}
\label{fig:atlas_qvg}
\end{figure}

The ATLAS quark vs gluon jet image based CNN tagger~\cite{ATL-PHYS-PUB-2017-017} was trained using fully simulated ATLAS events~\cite{AGOSTINELLI2003250,Aad2010}. Multi-channel jet images were used, with one channel containing an image of the sum of measured charged particle track $p_T$ per pixel. A second image for calorimeter measurements was examined in two forms, a jet image containing either the transverse energy measured in calorimeter towers of size $\Delta \eta \times \Delta \phi = 0.1\times 0.1$ or a jet image containing a projection onto a fixed grid of topologically clustered calorimeter cells (topo-clusters)~\cite{Aad2017}. Translation, rotation, and normalization pre-processing was performed. A three layer CNN with filter sizes of $5\times5$, $5\times5$, and $3\times3$, respectively, and max pooling after each convolutional layer was used. As can be seen in the ROC curve in Figure~\ref{subfig:atlas_qvg_roc}, the CNN processing the track + tower jet images outperforms other standard taggers for quark vs gluon tagging. Interestingly, the standard tagger based on the combination of two jet substructure features (number of charged particles and the jet width) outperforms the CNN approach at low quark efficiency. This is likely due to the track image discretization that may result in multiple tracks falling in the same pixel. As track multiplicity is not stored in the images, this useful discriminating information is lost for the CNN.  In Figure~\ref{subfig:atlas_qvg_inputs}, the impact on performance of utilizing different jet image channels was examined, wherein utilizing only calorimeter based jet images provides significantly less performance than tagging using track and calorimeter images. In addition, topo-cluster based images, which are formed by projecting the continuous topo-cluster direction estimates into a discrete grid, are seen to have lower performance than tower based images. This is likely due to the projection  onto a fixed grid for use in a CNN, as this may cause a loss of information about the spatial distribution of energy within a topo-cluster and may result in the overlap of several clusters in the same pixel. Moreover, it can be seen that the track + calorimeter image approach does not reach the performance found when a CNN is trained on a jet image formed from truth particles (i.e. without the impact of detector smearing). It was noted in~\cite{ATL-PHYS-PUB-2017-017} that when comparing the performance of a CNN trained on only track images to  a CNN trained on only charged truth particles, the observed performances were extremely similar. This similarity is driven by the excellent charged particle track resolution, and further indicates the difference between the track + calorimeter jet image based CNN tagger and the truth particle based CNN tagger is driven by the low resolution, and thus loss of information, of the calorimeter.  

The CMS boosted top jet image based CNN tagger~\cite{Sirunyan:2020lcu}, denoted ImageTop, was trained on fully simulated CMS events~\cite{AGOSTINELLI2003250}. Multi-channel jet images with six channels were built using particle flow (PF) objects found within an $R=0.8$ jet.  Before pixelation, particle flow objects within the jet are pre-processed using translation, rotation, flipping, and normalization. The six channels were defined as the sum of PF candidate $p_T$ per pixel with one channel containing all PF candidates,  and one channel each for PF candidate flavor, i.e. charged, neutral, photon, electron, and muon candidates. ImageTop was based on the multi-Channel DeepTop algorithm~\cite{Macaluso2018PullingOA}, and comprises four convolutional layers each using $4\times4$ filter sizes and max pooling after two consecutive convolutional layers, followed by four dense layers before classification. To aid the classification of top quark decays containing $b$-quarks, a $b$-tagging identification score~\cite{CMS-DP-2018-033} evaluated on subjets of the large jet was also fed as input to the dense layers of the tagger before classification. In addition to a baseline ImageTop, a mass decorrelated  version denoted ImageTop-MD was also trained, wherein the mass decorrelation was performed by down-sampling the background quark and gluon jet samples to have the same mass distribution as the sample of boosted top jets used for training. In this way, the discriminating information from the jet mass is removed to first order~\footnote{As the authors note, though this method is not guaranteed to remove tagger mass dependence, it was found to work sufficiently well in this case as the baseline tagger inputs were not observed to have a strong correlation to mass.}.

\begin{figure}[htbp]
\begin{center}
\begin{subfigure}[h]{0.48\linewidth}
\includegraphics[width=\linewidth]{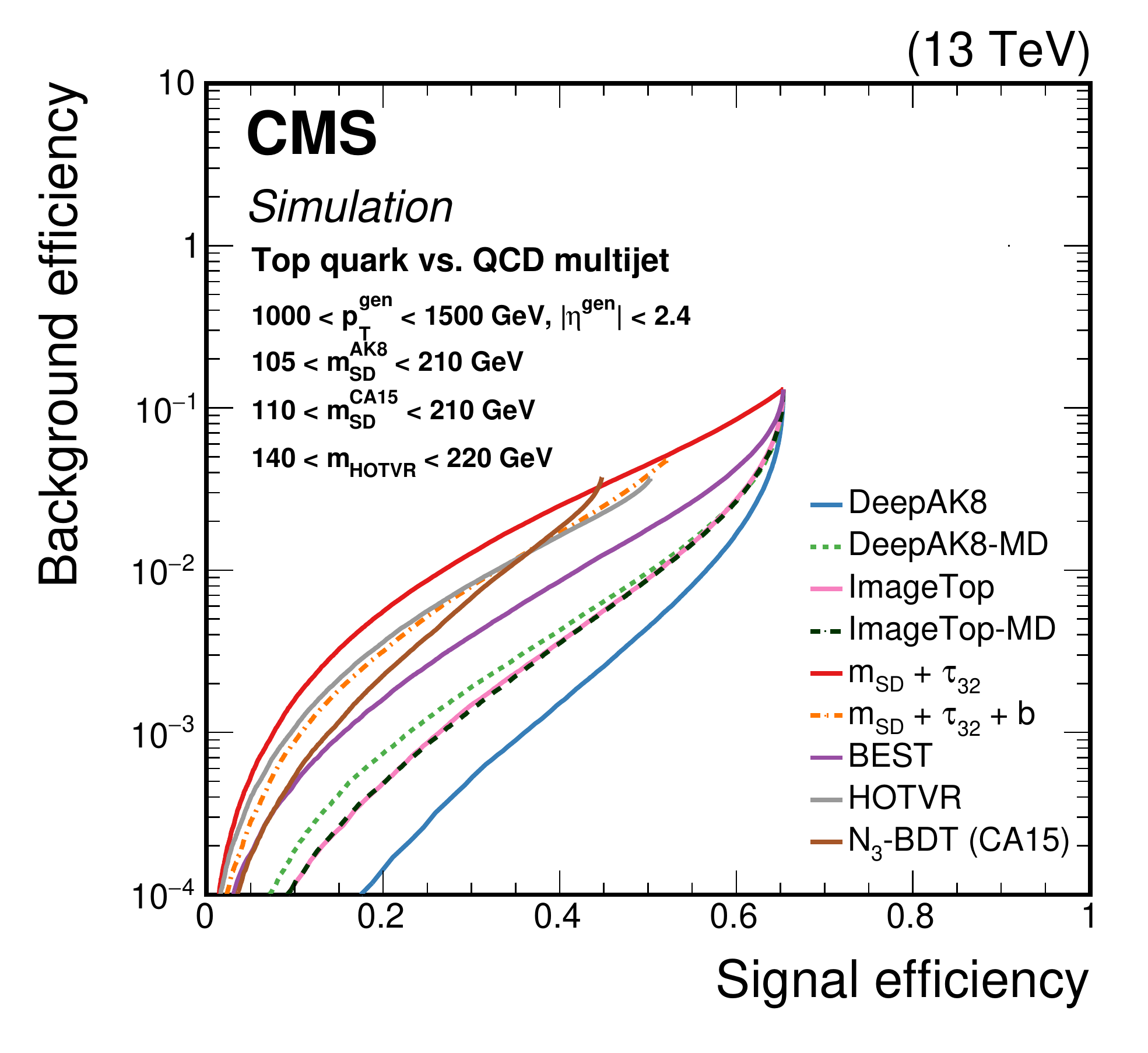}
\caption{}
\label{subfig:cms_top_roc}
\end{subfigure}\quad
\begin{subfigure}[h]{0.48\linewidth}
\begin{center}
\includegraphics[width=0.9\linewidth]{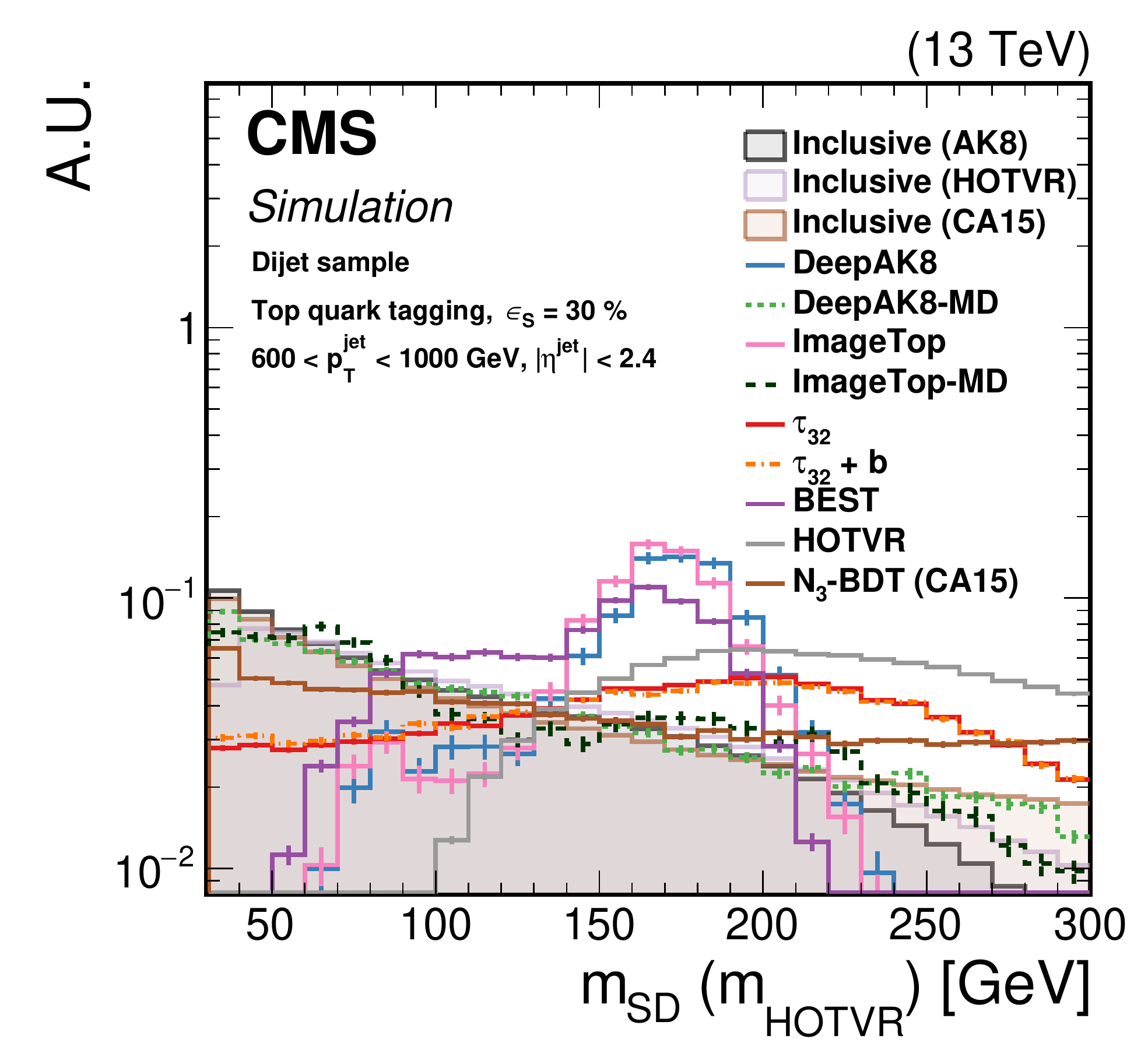}
\caption{}
\label{subfig:cms_top_md}
\end{center}
\end{subfigure}
\end{center}
\caption{Examination of the CMS ImageTop tagger~\cite{Sirunyan:2020lcu} trained on fully simulated CMS events in (a) ROC curves of the quark / gluon jet efficiency versus boosted top jet tagging efficiency  comparing several taggers and showing the dominant performance of the deep neural net based taggers, and (b) the impact of applying a threshold on tagger outputs to the background jet mass distribution wherein the mass decorrelated taggers show significantly less sculpting.}
\label{fig:cms_top}
\end{figure}

The ROC curve showing the performance of the ImageTop model is seen in Figure~\ref{subfig:cms_top_roc}. Several algorithms were compared to ImageTop, including several jet substructure feature based taggers and a deep neural network, denoted DeepAK8, based on processing PF candidates. ImageTop is seen to outperform all other algorithms except DeepAK8, and generally the deep network based taggers are found to significantly outperform other algorithms. Moreover, once mass decorrelation is included, the ImageTop-MD is found to be the highest performing mass-deccorrelated model. The smaller change in performance due to mass decorrelation of ImageTop relative to other algorithms such as DeepAK8 may be due the the image preprocessing; images are both normalized and ``zoomed" using a Lorentz boost determined by the jet $p_T$ to increase uniformity of jet images over the $p_T$ range. These steps can result in a reduction of mass information in the images and thus a reduction of the learned dependence of ImageTop on the mass. The mass spectrum for background quark and gluon jets before (in grey) and after applying a 30\% signal efficiency tagging threshold for ImageTop and ImageTop-MD (in green) can be seen in Figure~\ref{subfig:cms_top_md}. The decorrelation method greatly helped to preserve the mass distribution and was not seen to significantly degrade performance, as seen in the ROC curves of Figure~\ref{subfig:cms_top_roc}.

\begin{figure}[htbp]
\begin{center}
\begin{subfigure}[h]{0.48\linewidth}
\begin{center}
\includegraphics[width=\linewidth]{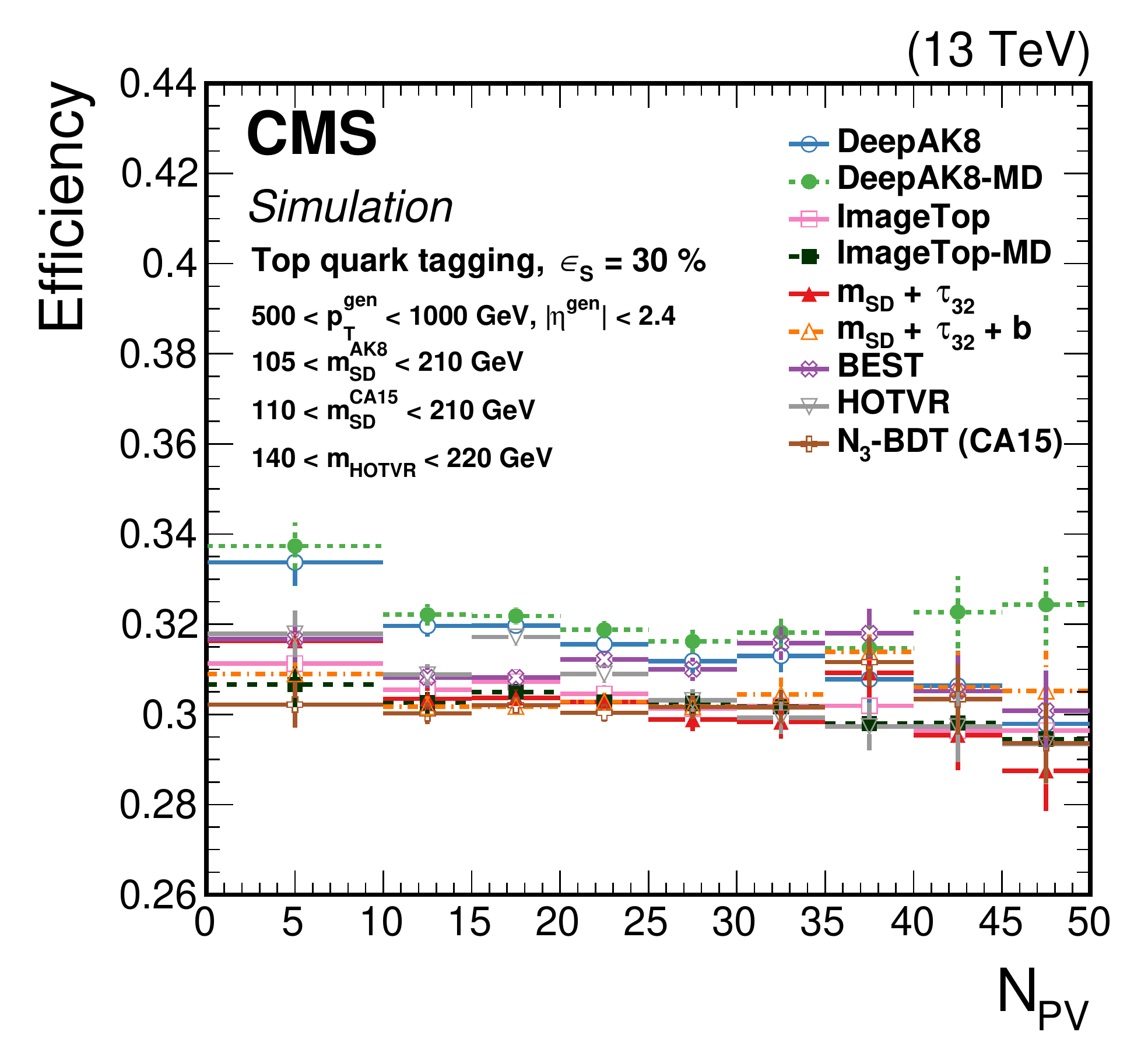}
\end{center}
\end{subfigure}
\end{center}
\caption{Variations as a function of the number of reconstructed vertices in an event of boosted top jet tagging efficiency after applying a fixed tagger output threshold on the CMS ImageTop tagger~\cite{Sirunyan:2020lcu}, as well as several other taggers, trained on fully simulated CMS events.}
\label{fig:cms_top_eff_npv}
\end{figure}

As noted early, one concern with jet image based approaches to jet tagging is their potential dependence on pileup conditions. For a fixed ImageTop tagging threshold giving an inclusive 30\% top jet tagging efficiency, the variations of the top jet tagging efficiency as a function of the number of primary vertices in the event can be seen in Figure~\ref{fig:cms_top_eff_npv}. Efficiency variations for both ImageTop and ImageTop-MD were found to be small, at the level of less than 1\%, across the values of number of primary vertices. A similar level of stability was observed for the background mis-identification rate. This stability draws largely from the pileup mitigation applied to the jet before creating the jet images, and this stability is not disturbed by the CNN discriminant.

While the simulation based training of classifiers can lead to powerful discriminants, differences in feature distributions between data and simulation could cause the tagger to have differing performance between data and simulation. As such, the discriminant is typically calibrated before application in data. Calibration entails defining control samples of jets in data where the tagging efficiency and mis-identification rate can be measured in data and simulation. The efficiency of the tagger as a function of jet $p_T$ is evaluated in data and simulation, and a $p_T$ dependent ratio of efficiencies known as a Scale Factor (SF) is derived. This SF can then be used to weight events such that the simulation trained tagger efficiency matches the data. The SFs for the ImageTop signal efficiency were estimated in a sample of single muon events selected to have a high purity of top-pair events in the 1-lepton decay channel, while quark and gluon background mis-identification rates were estimated in dijet samples and samples of photons recoiling off of jets. Systematic uncertainties were evaluated on the data based estimation of the tagging efficiency and propagated to SF uncertainties. These systematic uncertainties included theory uncertainties in the parton showering model, renormalization and factorization scales, parton distribution functions, as well as experimental uncertainties on the jet energy scale and resolution, $p_T^{\textrm{miss}}$ unclustered energy, trigger and lepton identification, pileup modeling, and integrated luminosity, as well as statistical uncertainties of simulated samples.

\begin{figure}[htbp]
\begin{center}
\begin{subfigure}[h]{0.48\linewidth}
\includegraphics[width=\linewidth]{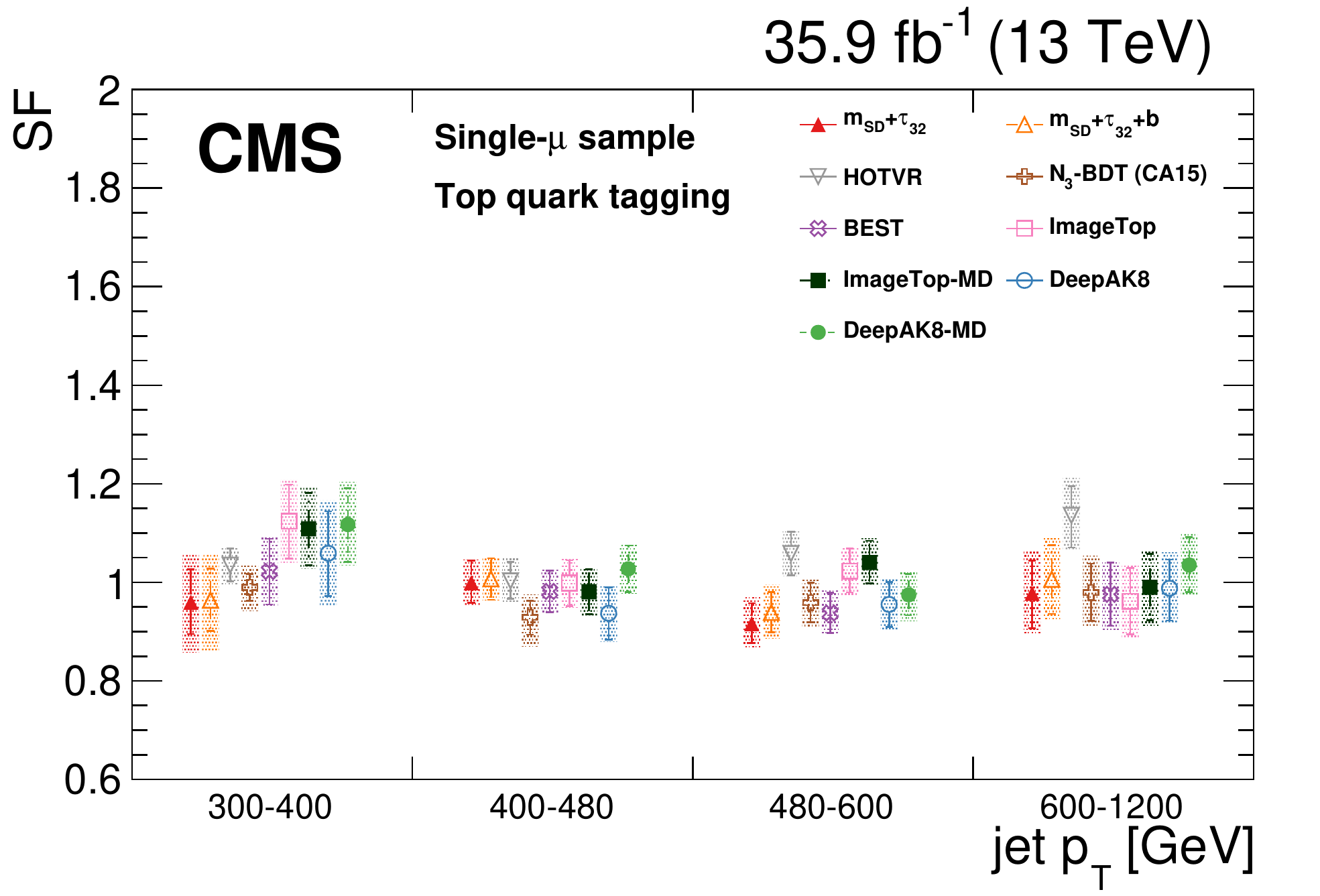}
\caption{}
\label{subfig:cms_top_sf_sig}
\end{subfigure}\quad
\begin{subfigure}[h]{0.48\linewidth}
\begin{center}
\includegraphics[width=\linewidth]{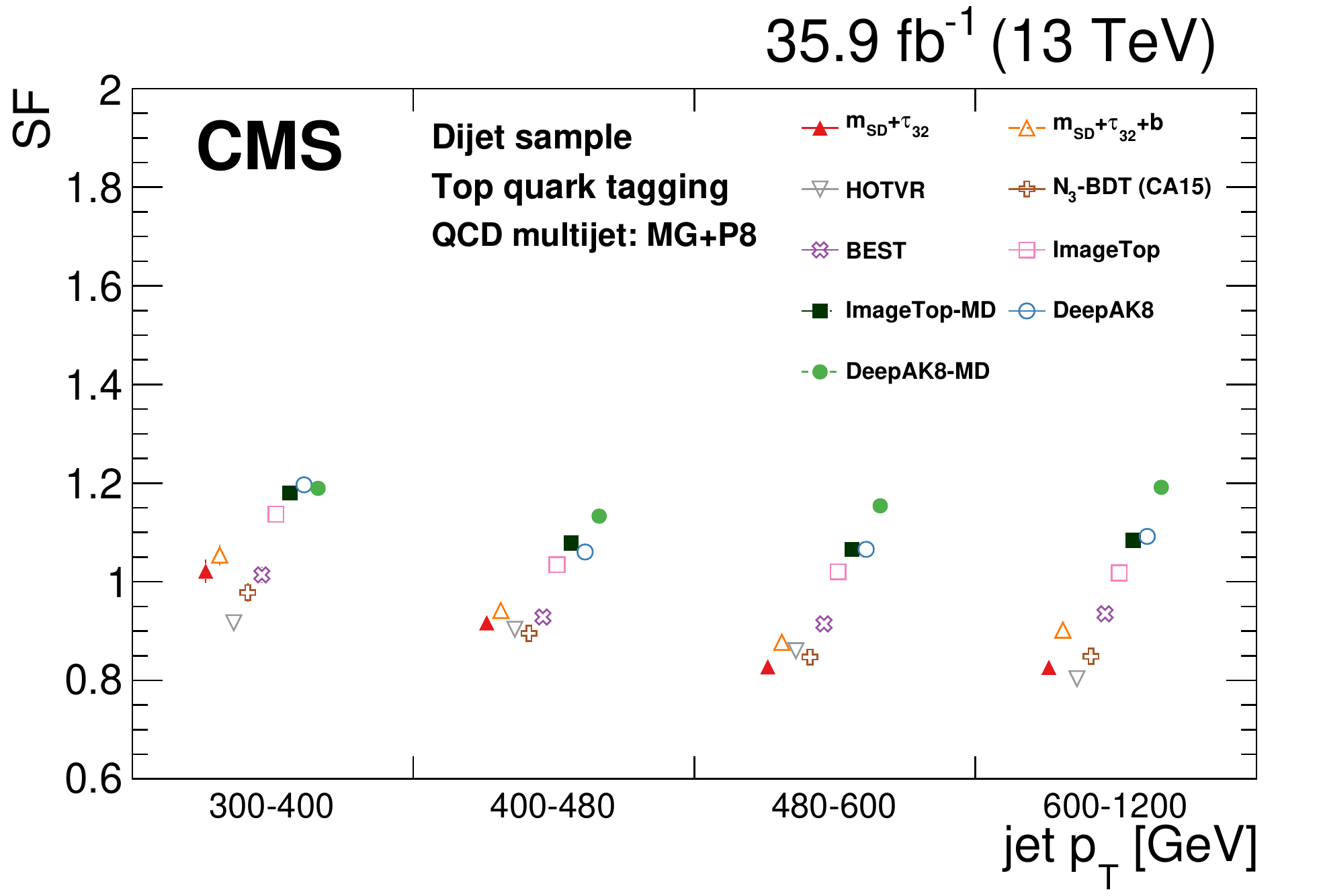}
\caption{}
\label{subfig:cms_top_sf_bkg}
\end{center}
\end{subfigure}
\end{center}
\caption{Calibration scale factors as a function of jet $p_T$ for (a) the top jet tagging efficiency in single muon events, and (b) the quark / gluon jet mistag efficiency in dijet events, for the CMS ImageTop tagger~\cite{Sirunyan:2020lcu} trained on fully simulated CMS events and calibrated to data.}
\label{fig:cms_top_sf}
\end{figure}%

The scale factors for ImageTop and ImageTop-MD for both the top tagging efficiency and the background mis-identification rate can be found in Figure~\ref{fig:cms_top_sf}. The signal efficiency scale factors were largest at low momentum, showing a departure from unity of around 10\%, but were significantly closer to unity in essentially all other $p_T$ ranges. The systematic uncertainties ranged from approximately 5-10\%, with the largest uncertainties at low $p_T$. The scale factors for the mis-identification rate tended to be larger, up to a 20\% departure from unity in dijet samples but with smaller scale factors in the photon+jet samples. These calibrations indicate that while some departures from unity of the scale factors are observed, they are largely consistent with observations from other taggers. The situation is similar in terms of the scale factor uncertainties. As such, the jet image and CNN based tagging approach can be seen to work well in data, without extremely large calibrations and uncertainties, thus indicating its viability for use in analysis.

\section{Understanding jet Image based tagging}\label{sec:understand}

Interpretability and explainability are vital when applying ML methods to physics analysis in order to ensure that (i) reasonable and physical information is being used for discrimination rather than spurious features of the data, and (ii) when training models on simulation, models are not highly reliant on information that may be mismodeled with respect to real data. Interpretability and explainability of deep neural networks is highly challenging and is an active area of research within the ML community~\cite{Bhatt2020ExplainableML}. While a large number of techniques exist for examining CNNs, a subset of the techniques from the ML community have been applied within the study of jet images. A benefit of the computer vision approach to jet analysis is that while the data input to ML models may be high dimensional, in this case with a large number of pixels, they can be visualized on the image grid for inspection and interpretation. Thus the tools for interpreting CNN models applied to jet images tend to center on this aspect with tools such as pixel-discriminant correlation maps, filter examination, and finding images that maximally activate neurons.

Given a jet image $x$ with pixel values $\{x_{ij}\}$ and discriminant $c(x)$, one can examine how changes to the input may effect the discriminant prediction. Correlation maps examine Pearson correlation coefficients between each pixel and the discriminant prediction, thus probing how each input feature is correlated with increases and decreases in prediction over a sample of inputs. For a sample of $N$ inputs, the correlation map is computed as $\rho_{ij} = \frac{1}{\sigma_{x_{ij}} \sigma_{c}} \sum_{k=1}^N (x_{ij}^{(k)} - \bar{x}_{ij}) (c(x^{(k)}) - \bar{c})$, where $\bar{x}_{ij} = \frac{1}{N} \sum_{k=1}^N x_{ij}^{(k)}$ and $\bar{c} = \frac{1}{N} \sum_{k=1}^N c(x^{(k)})$ are the mean feature and prediction values, while $\sigma_{x_{ij}} = \frac{1}{N} \sum_{k=1}^N (x_{ij}^{(k)} - \bar{x}_{ij})^2$ and $\sigma_{c}  = \frac{1}{N} \sum_{k=1}^N (c(x^{(k)}) - \bar{c})^2$ are the variances of the feature and prediction values.

The filters of a CNN perform local feature matching and are applied directly to the pixels of the image (or convolved image), and thus one may plot each filter as an image and examine what features each filter is targeting. As there can be a large number of filters at each CNN layer as well as a large number of channels in layers deep within a CNN, this approach tends to be easiest at the first layers of the CNN. In addition, rather than examining the filters themselves, after processing an image by a CNN model, one may examine the output of any given filter. This will produce a convolved image in which the local feature matching has been applied at each position of the image and will highlight the location of the image in which a given filter has become active. In order to highlight difference in convolved images between classes, the difference between average convolved image between two classes can highlight relative differences in the spatial location of information relevant for discrimination. 

Maximally activating images or image patches correspond to applying a CNN model on a large set of images and finding the images, or image patches, that cause a given neuron to output a large activation. In the case of neurons in the fully connected layers at the end of the network, this corresponds to full images, whilst for neurons in convolutional layers this corresponds to image patches in which the neuron is most active.

\subsection{Probing CNNs} 

\begin{figure}[t]
\begin{center}
\begin{subfigure}[h]{0.49\linewidth}
\begin{center}
\vspace{0.46\linewidth}
\includegraphics[width=0.23\linewidth]{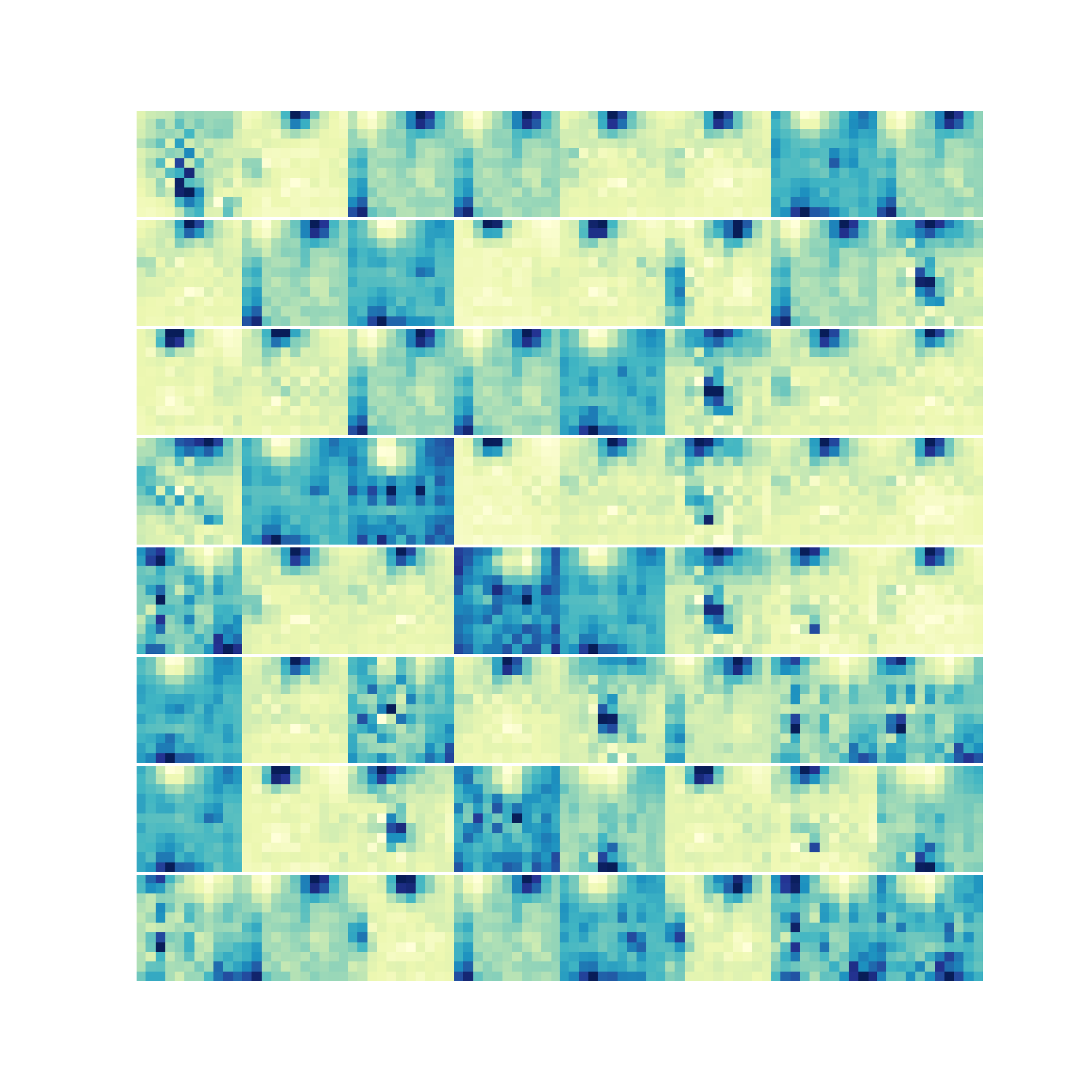}
\includegraphics[width=0.23\linewidth]{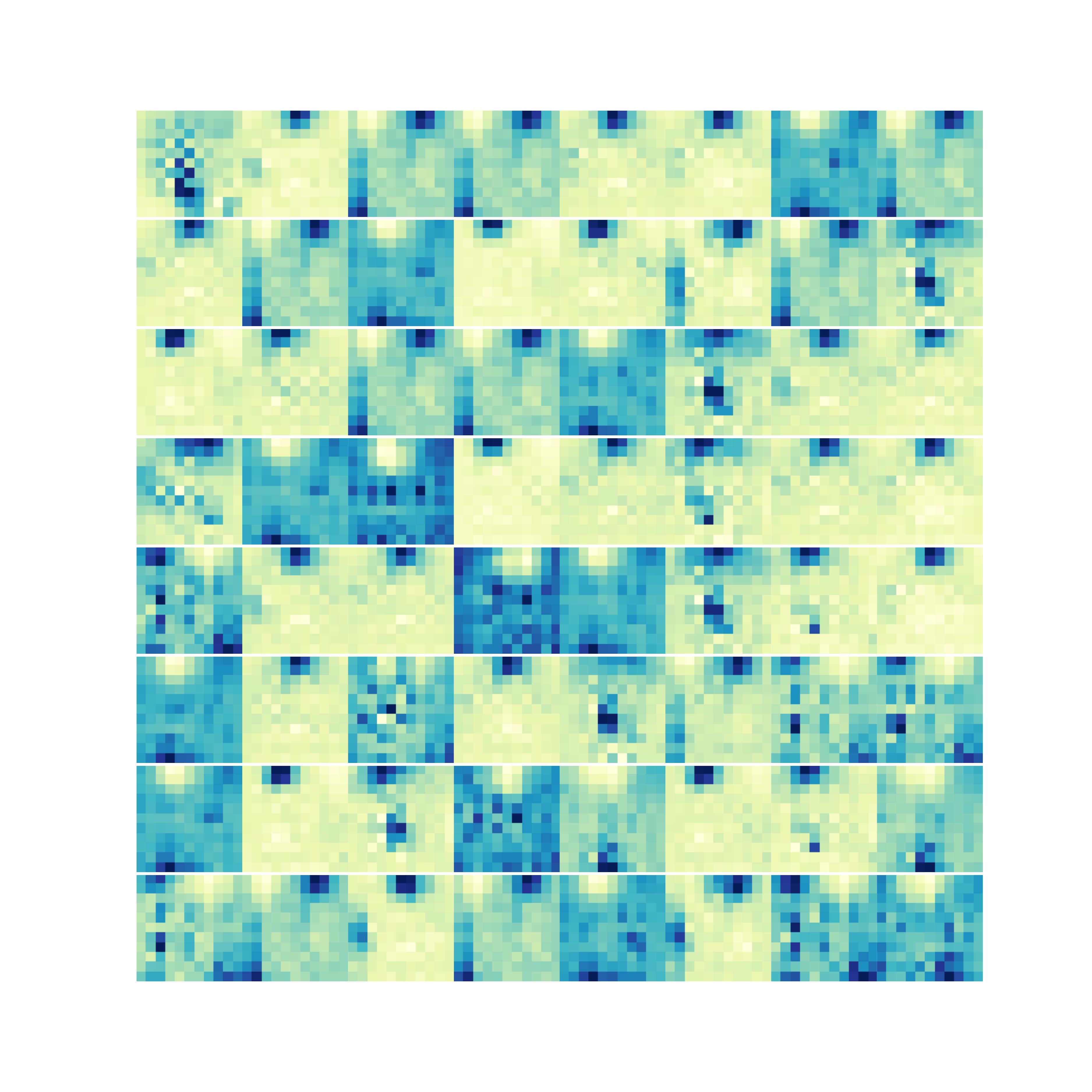}
\includegraphics[width=0.23\linewidth]{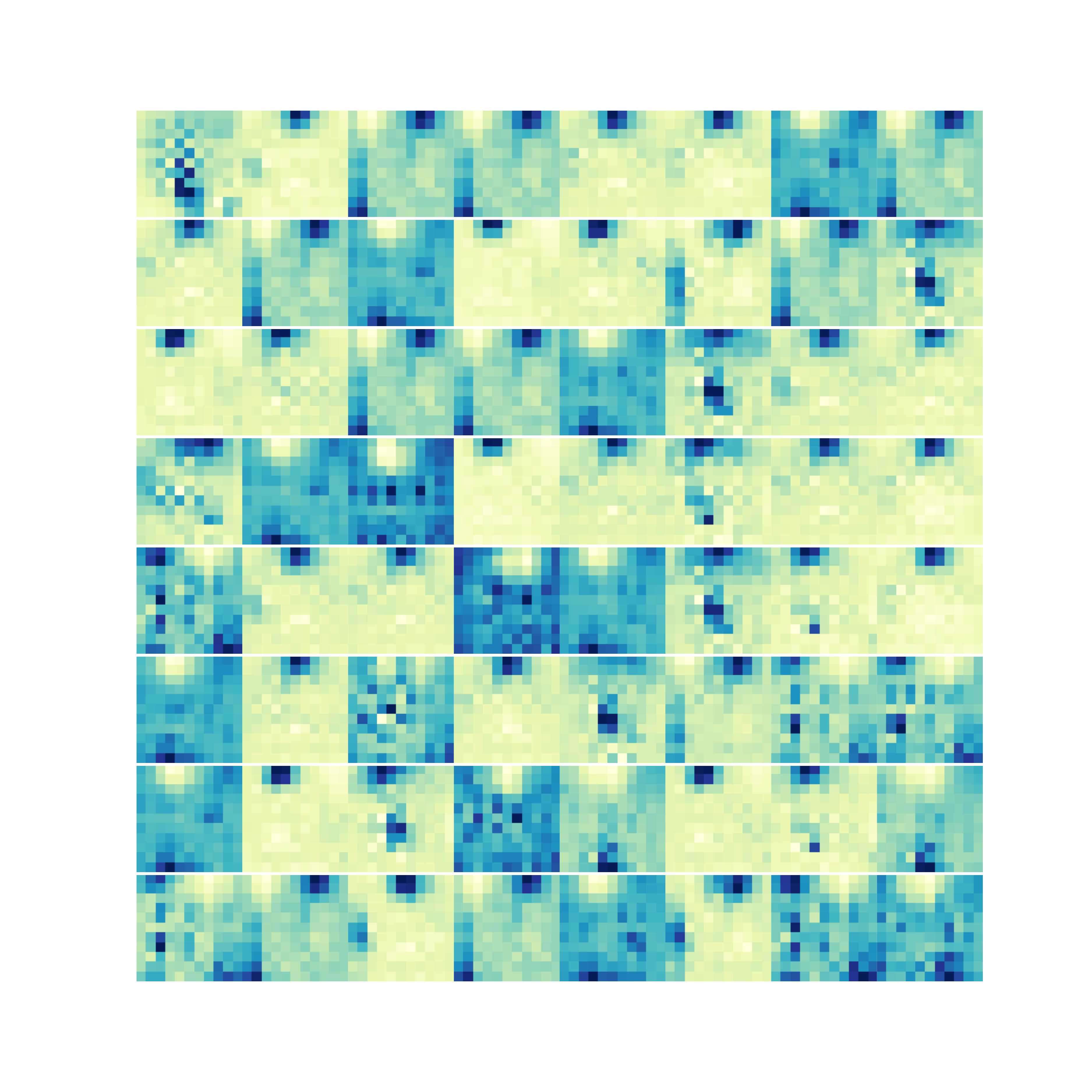}
\includegraphics[width=0.23\linewidth]{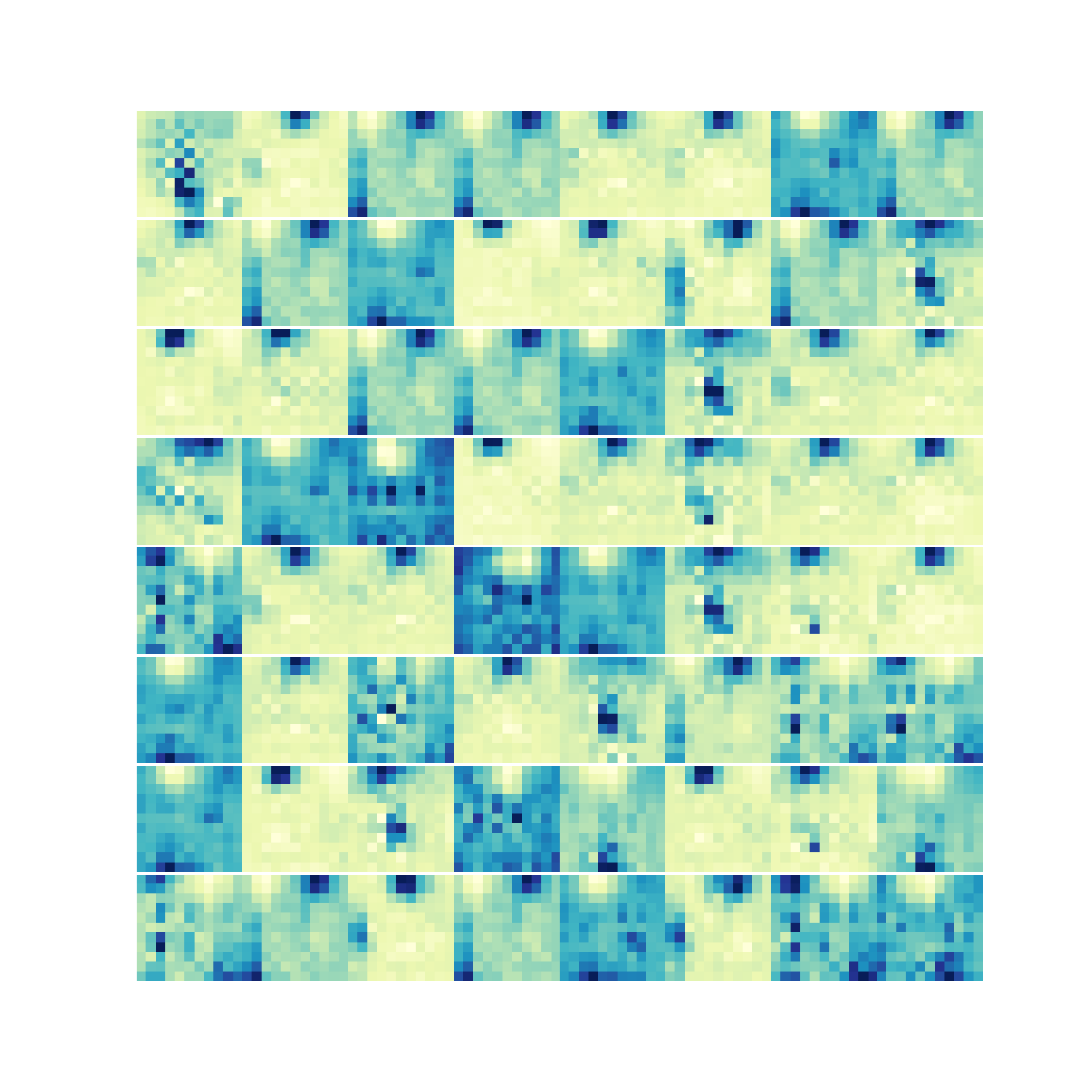}
\\
\includegraphics[width=0.23\linewidth]{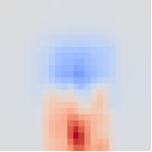}
\includegraphics[width=0.23\linewidth]{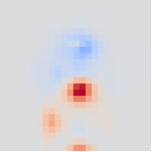}
\includegraphics[width=0.23\linewidth]{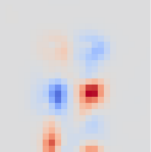}
\includegraphics[width=0.23\linewidth]{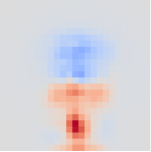}
\caption{}
\label{subfig:wcnn_filt}
\end{center}
\end{subfigure}
\begin{subfigure}[h]{0.49\linewidth}
\begin{center}
\includegraphics[width=0.23\linewidth]{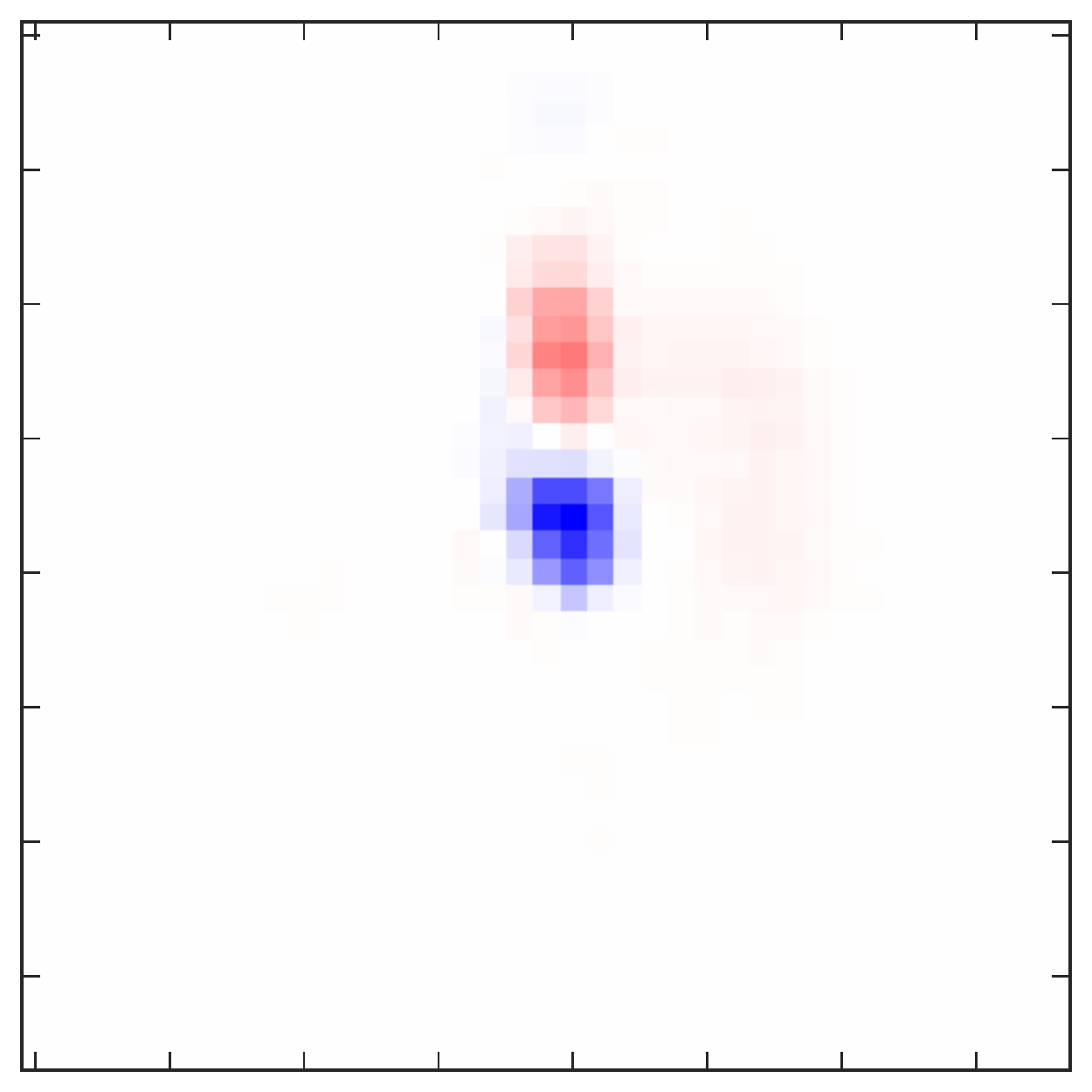}
\includegraphics[width=0.23\linewidth]{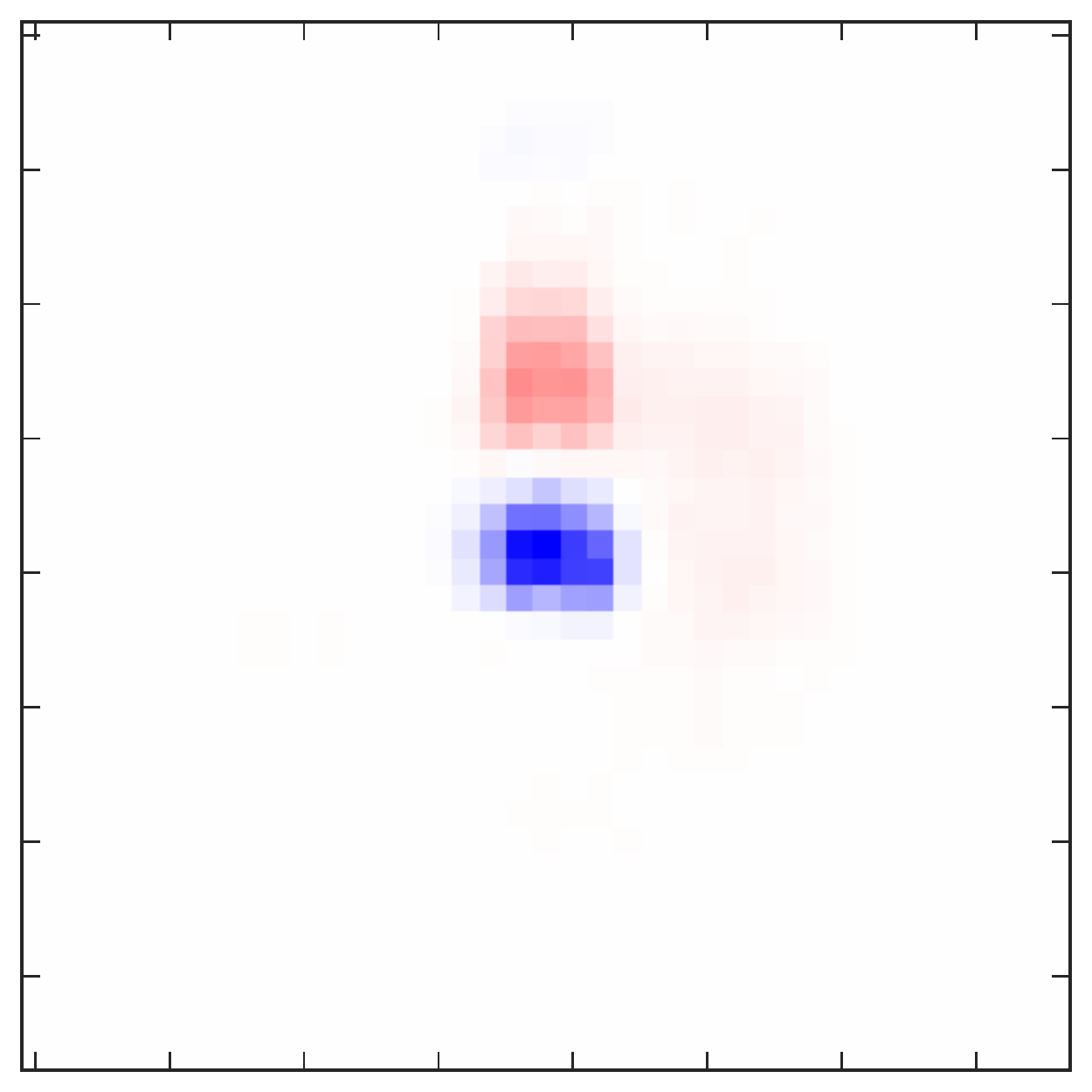}
\includegraphics[width=0.23\linewidth]{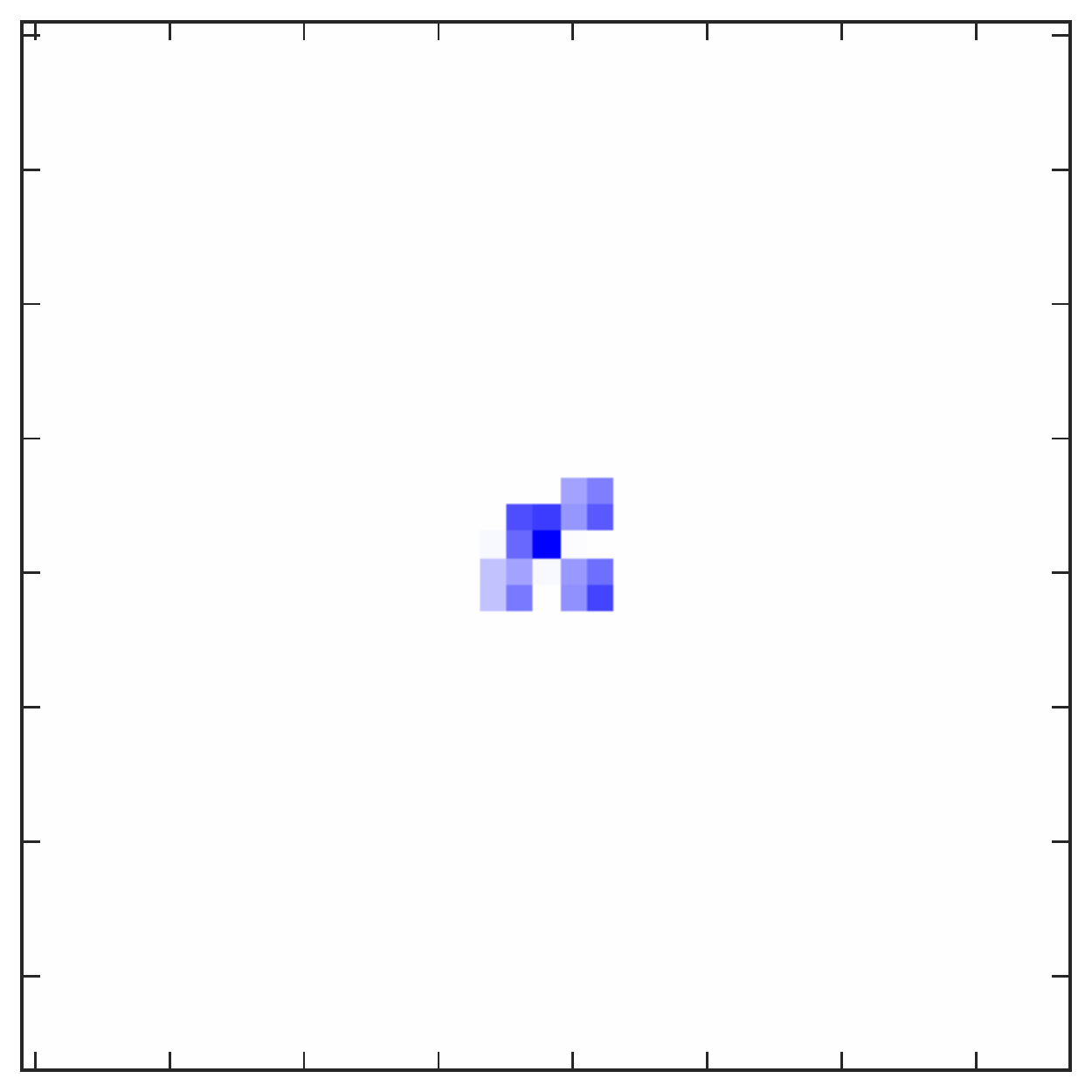}
\includegraphics[width=0.23\linewidth]{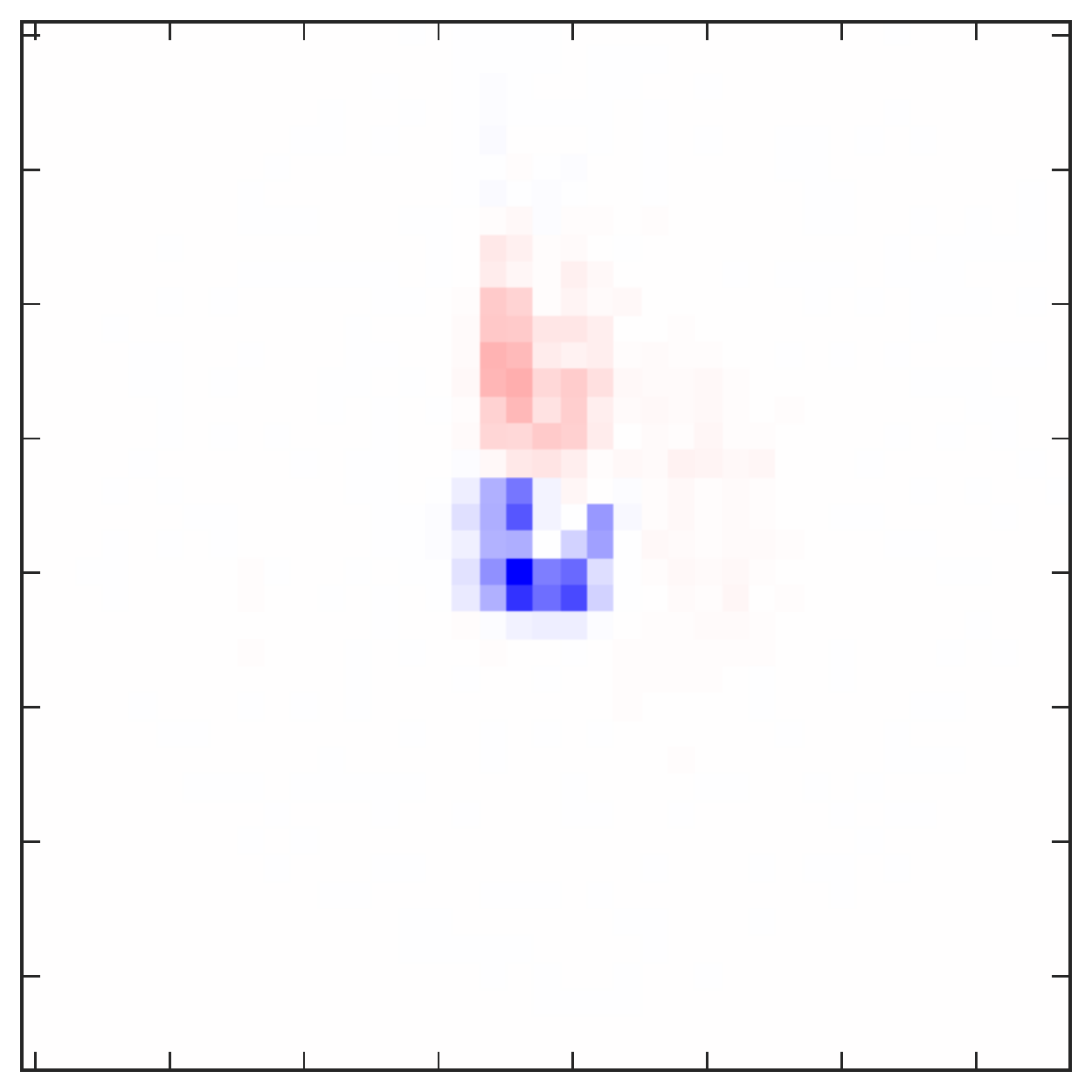}
\\
\includegraphics[width=0.23\linewidth]{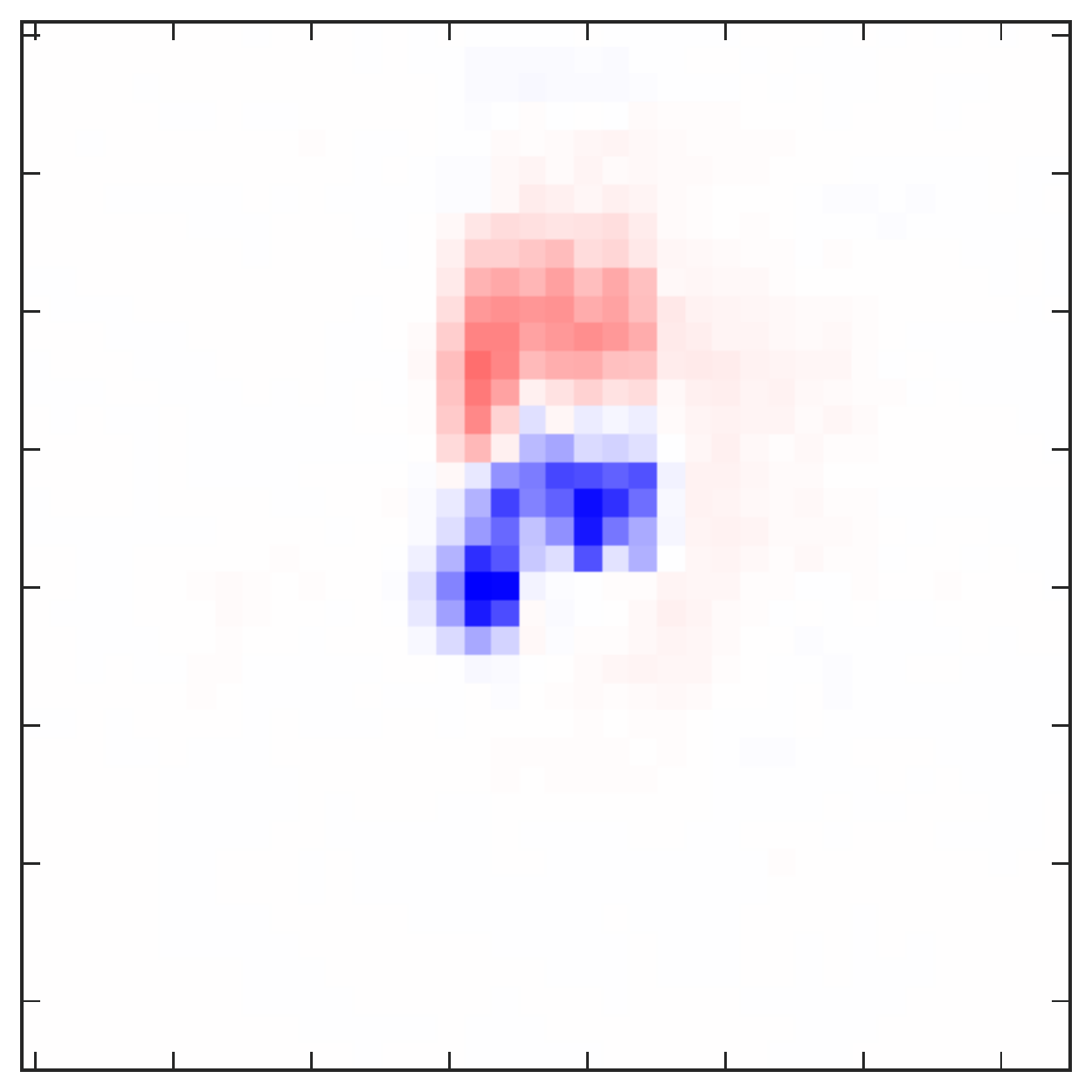}
\includegraphics[width=0.23\linewidth]{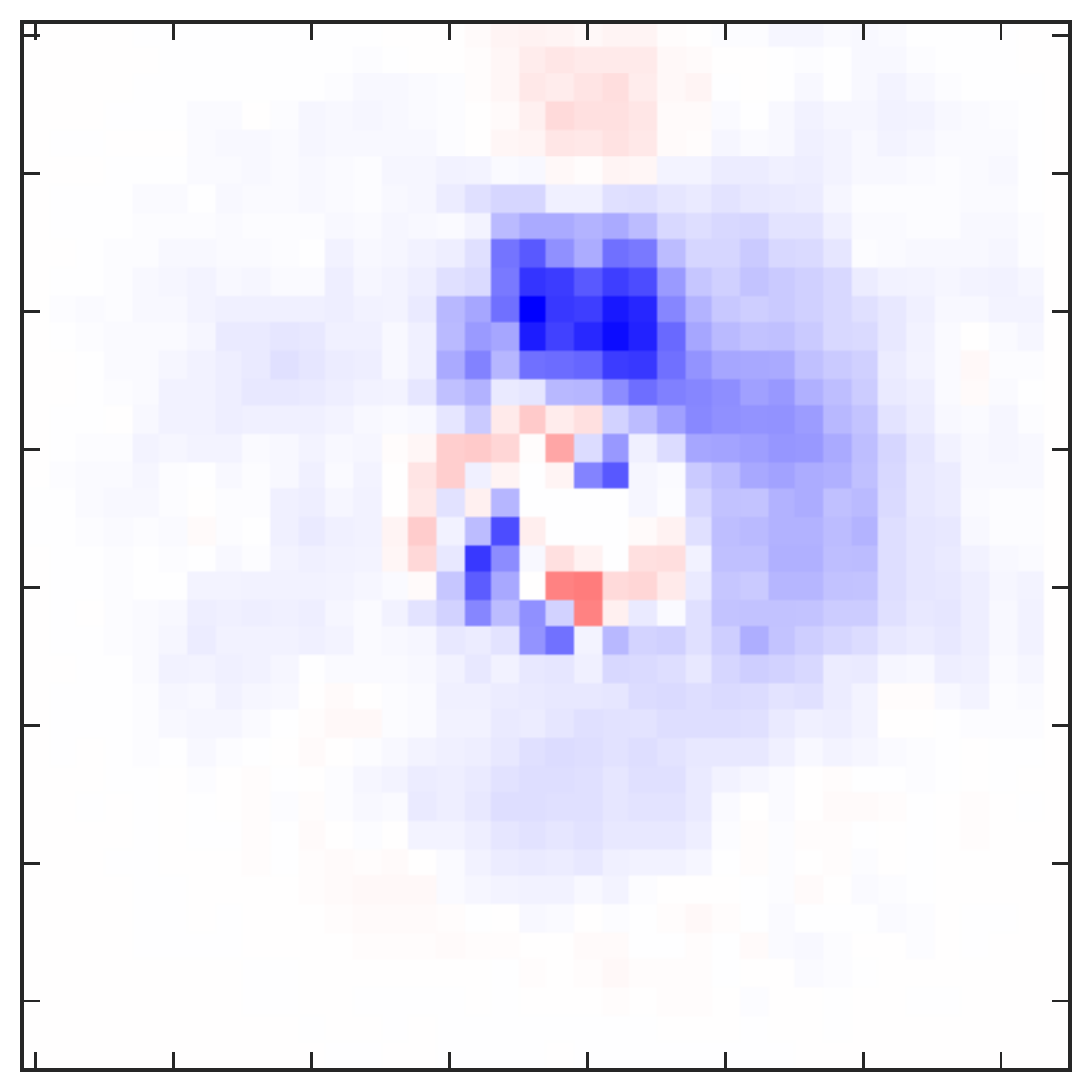}
\includegraphics[width=0.23\linewidth]{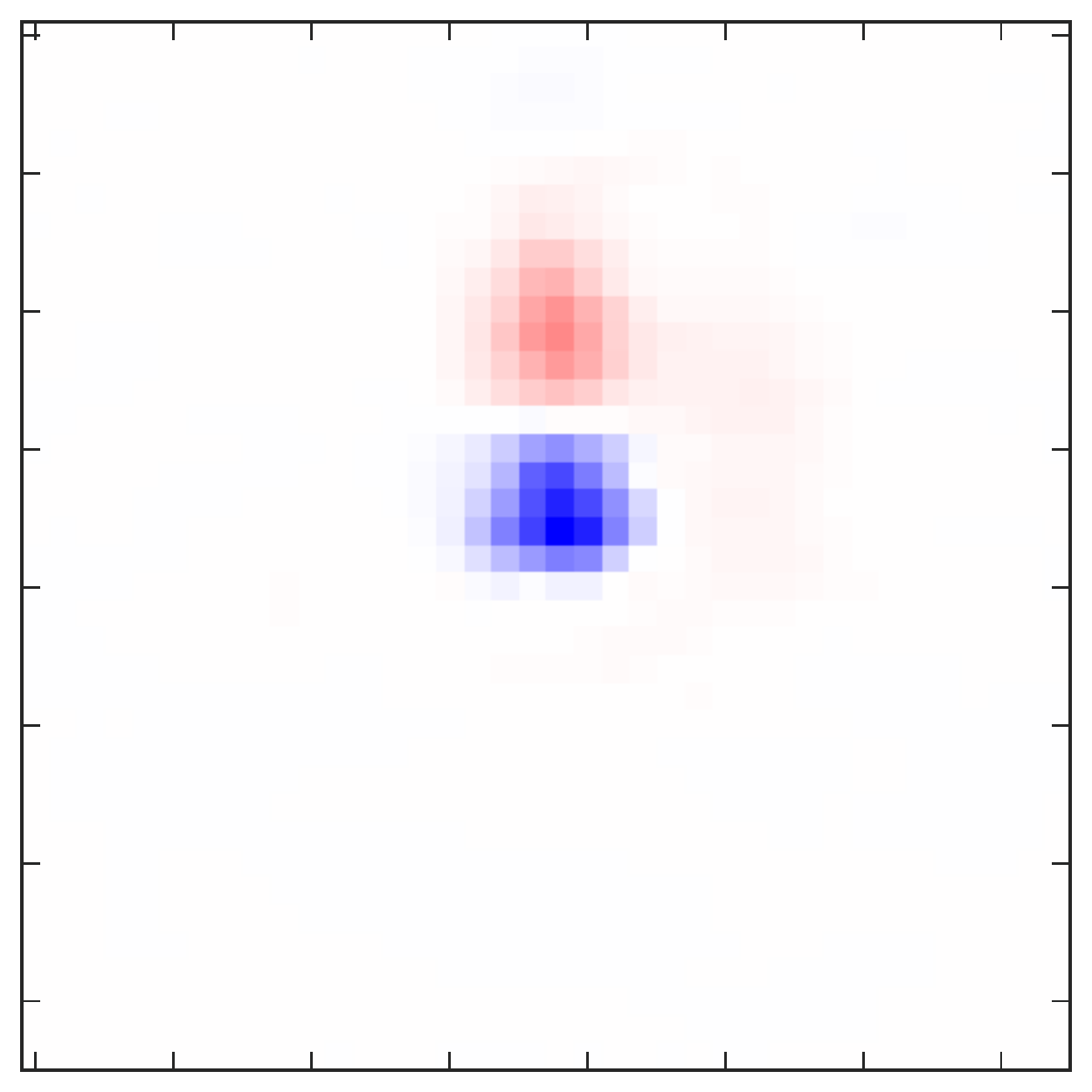}
\includegraphics[width=0.23\linewidth]{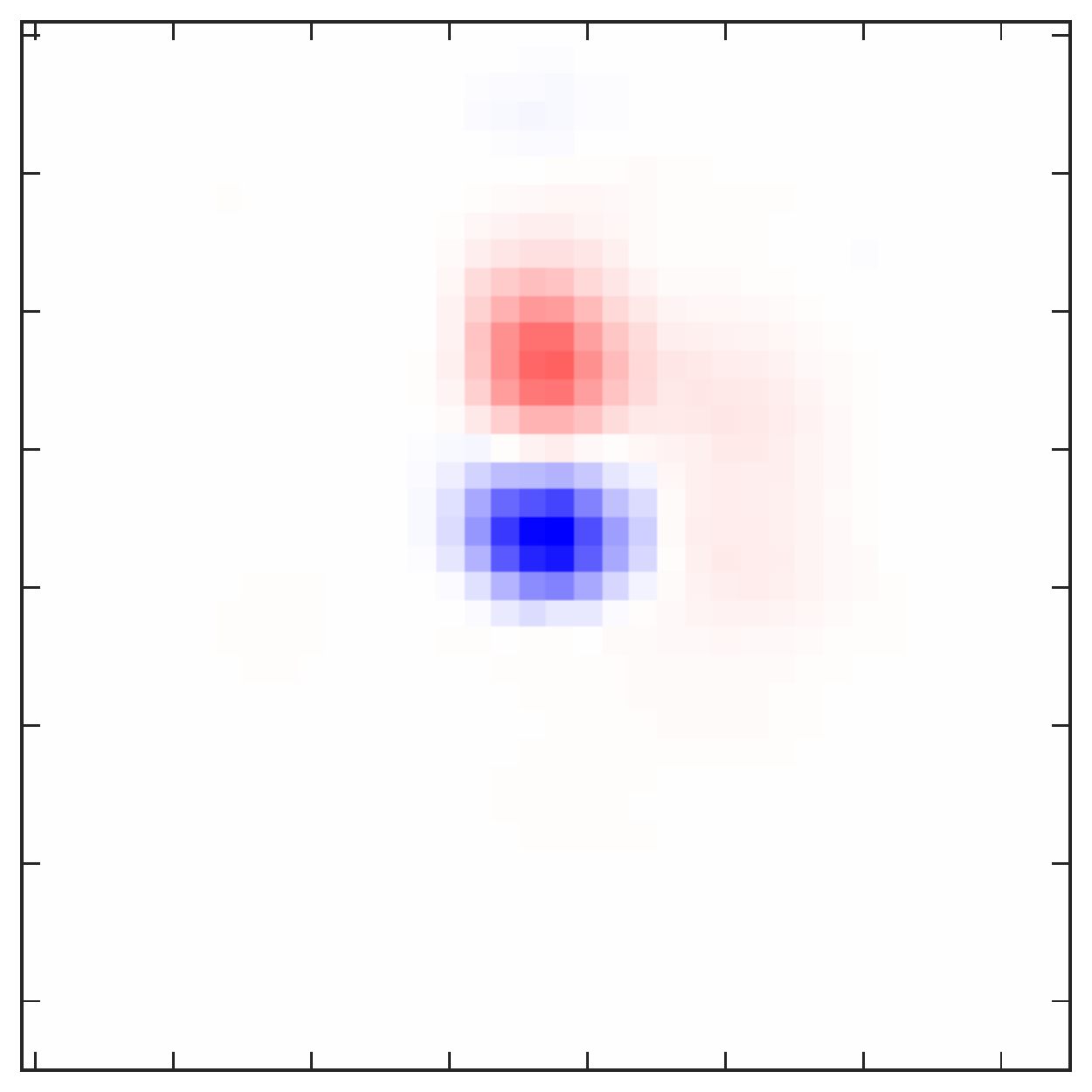}
\\
\includegraphics[width=0.23\linewidth]{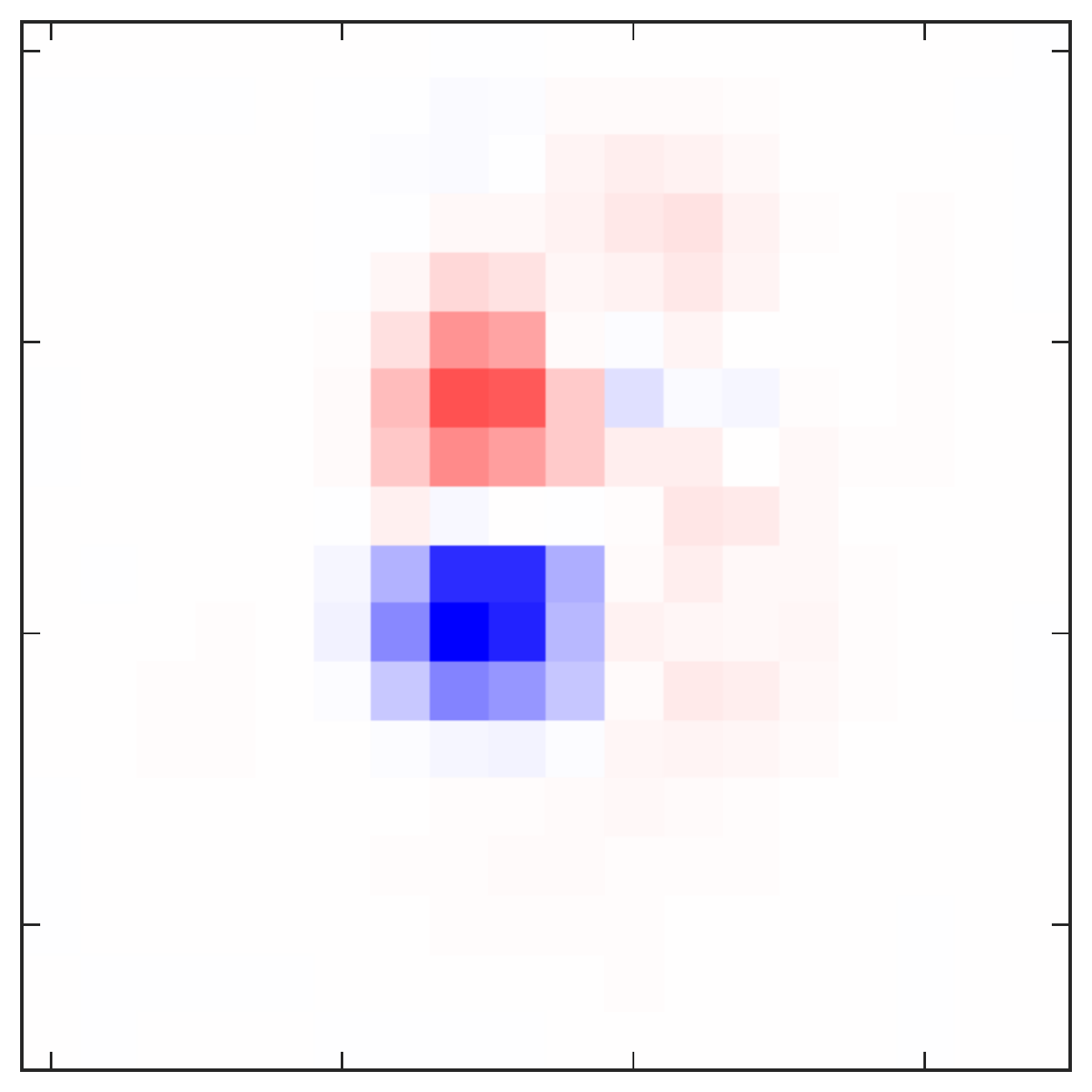}
\includegraphics[width=0.23\linewidth]{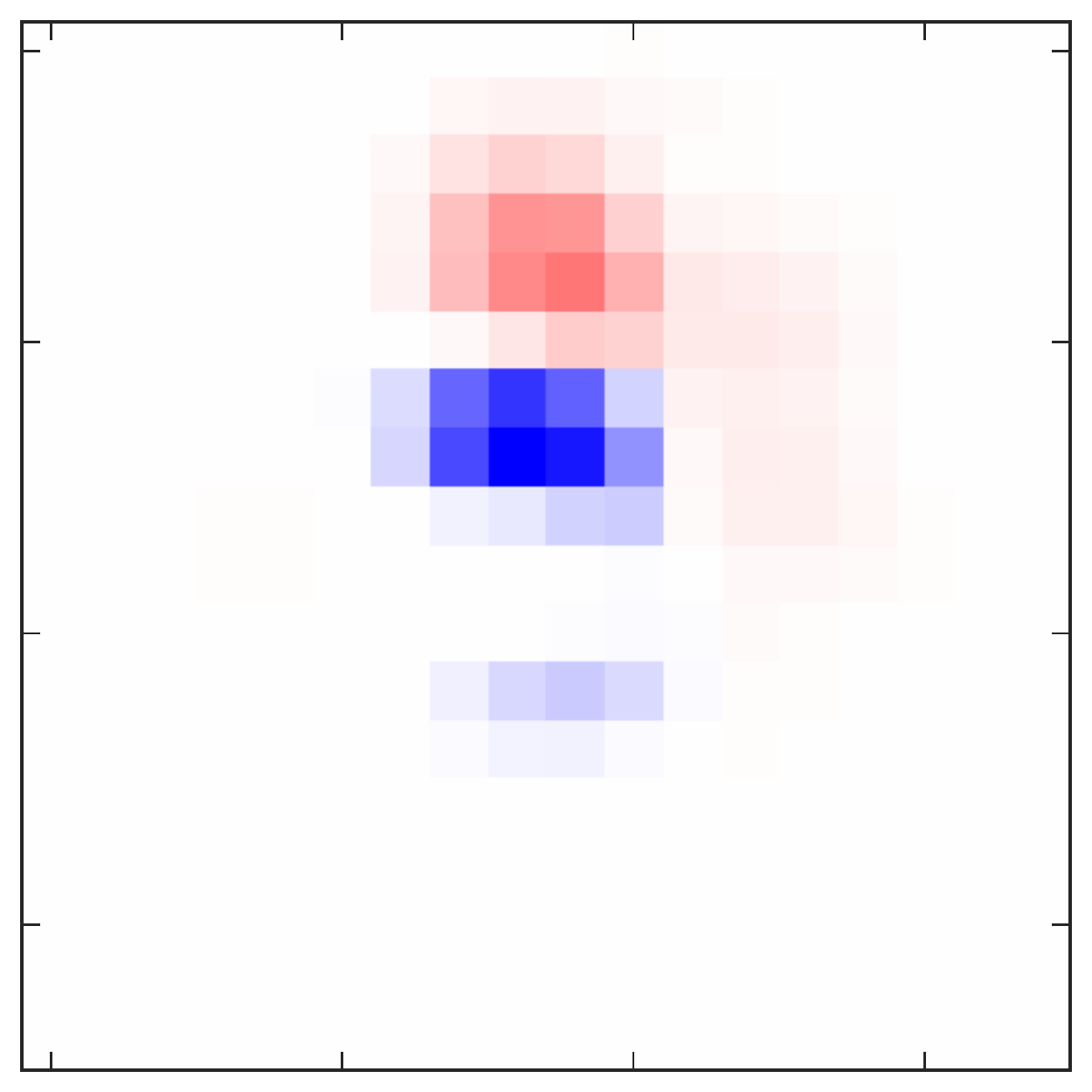}
\includegraphics[width=0.23\linewidth]{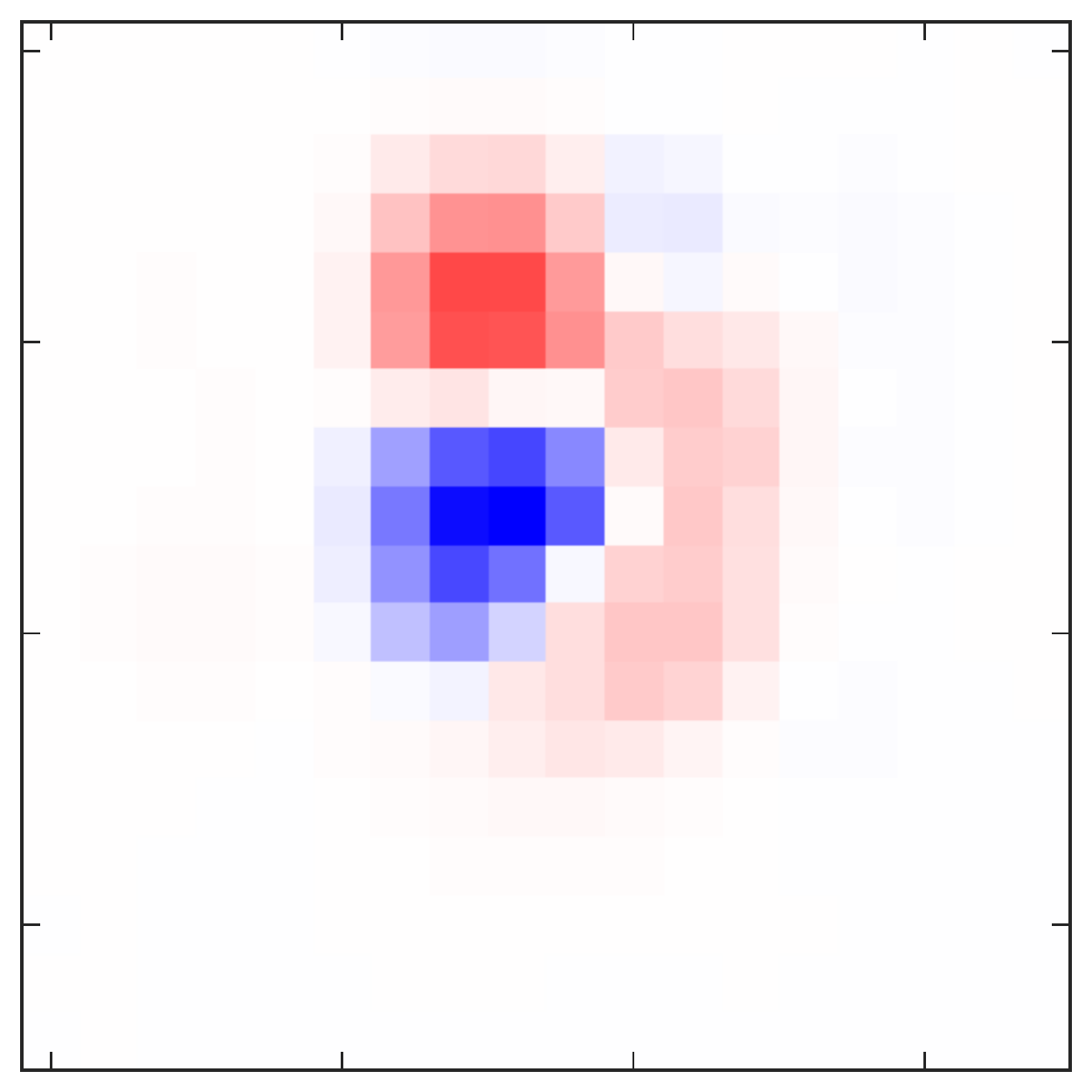}
\includegraphics[width=0.23\linewidth]{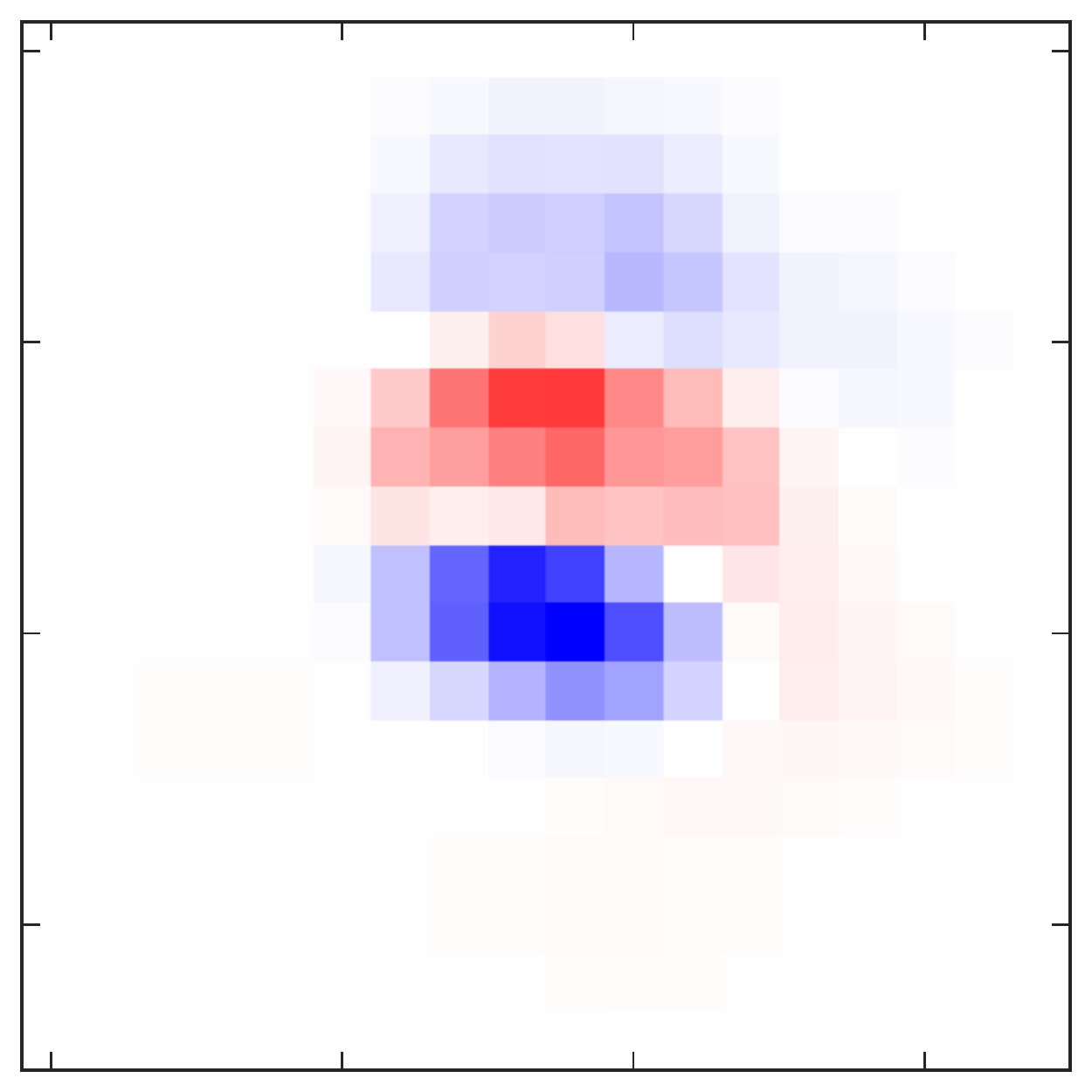}
\\
\includegraphics[width=0.23\linewidth]{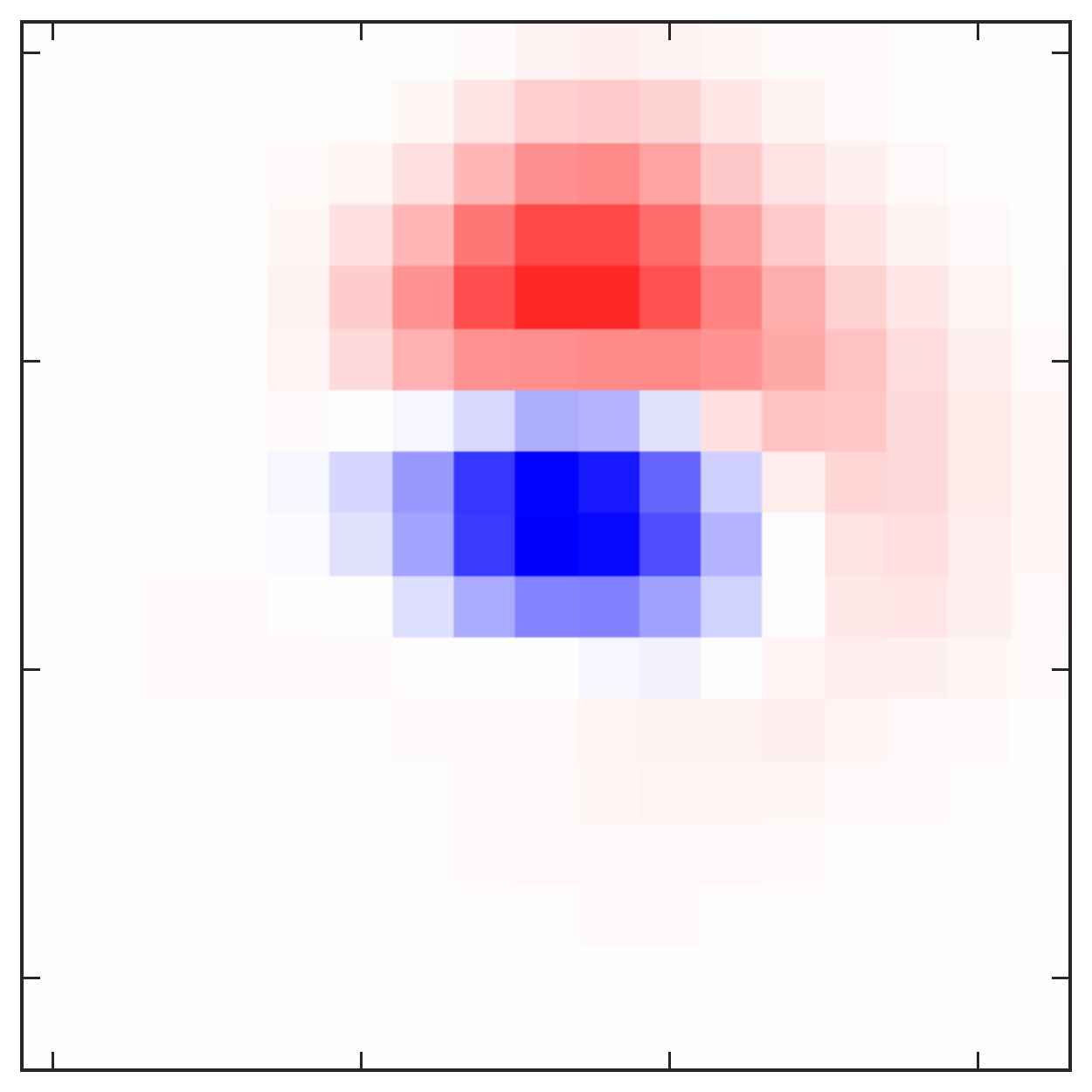}
\includegraphics[width=0.23\linewidth]{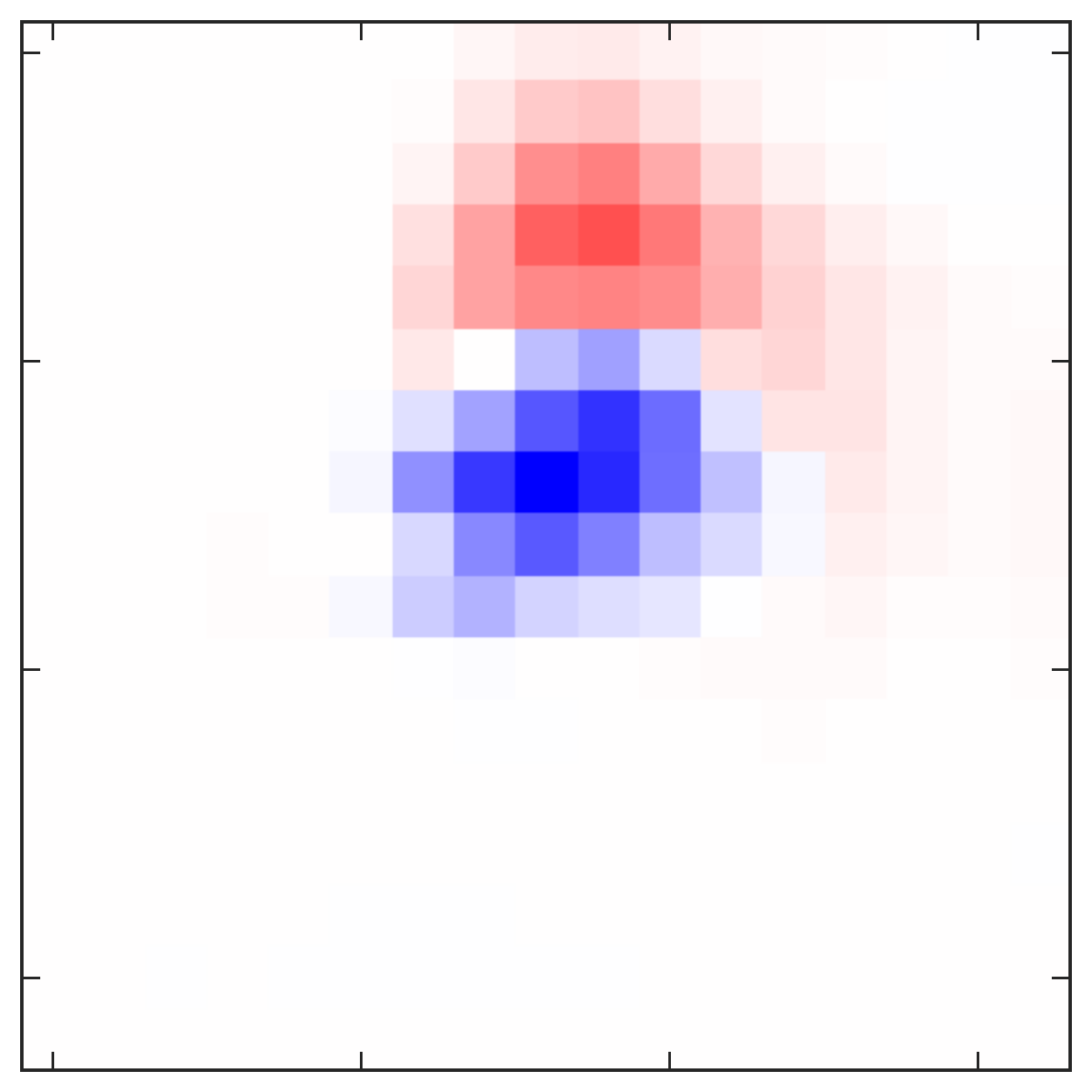}
\includegraphics[width=0.23\linewidth]{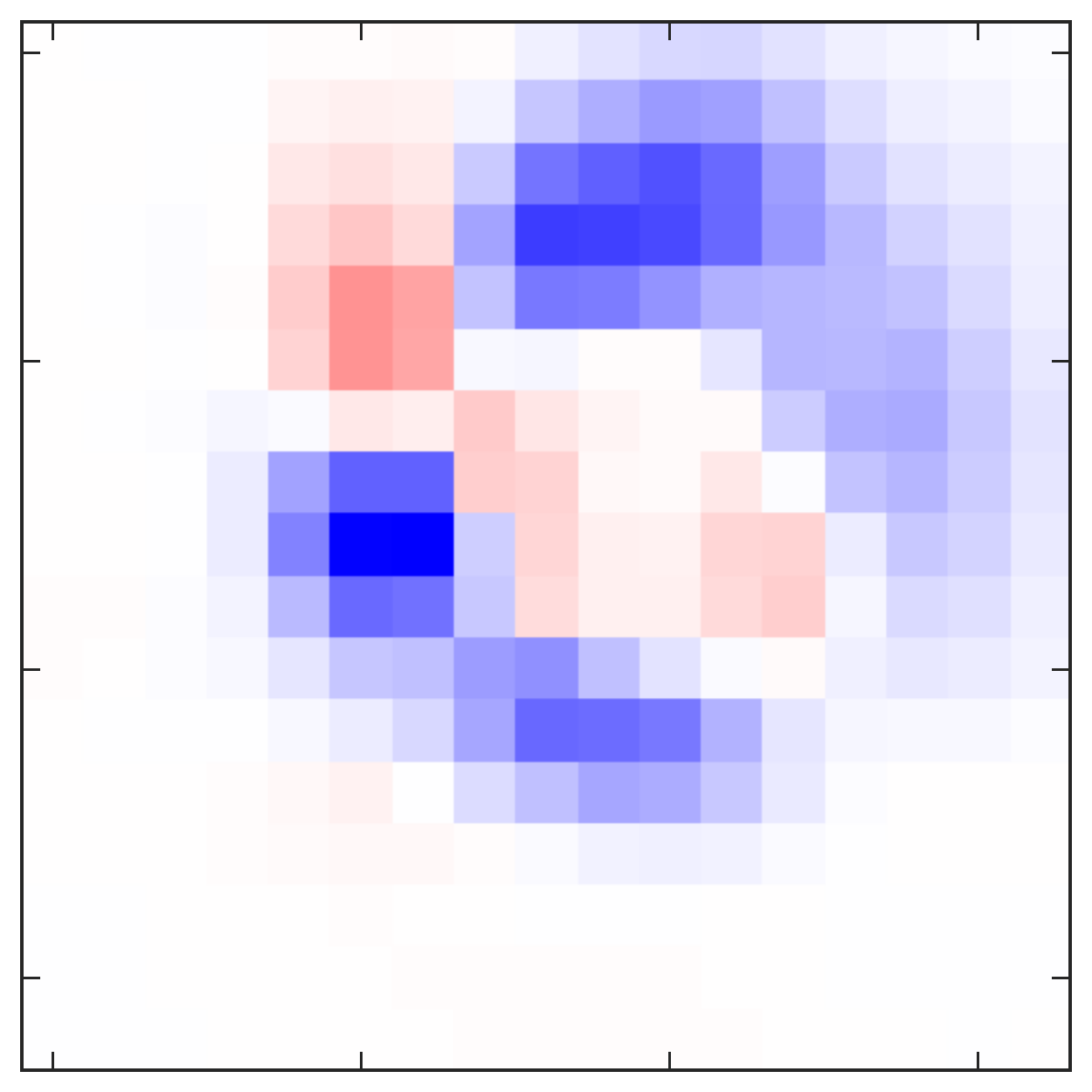}
\includegraphics[width=0.23\linewidth]{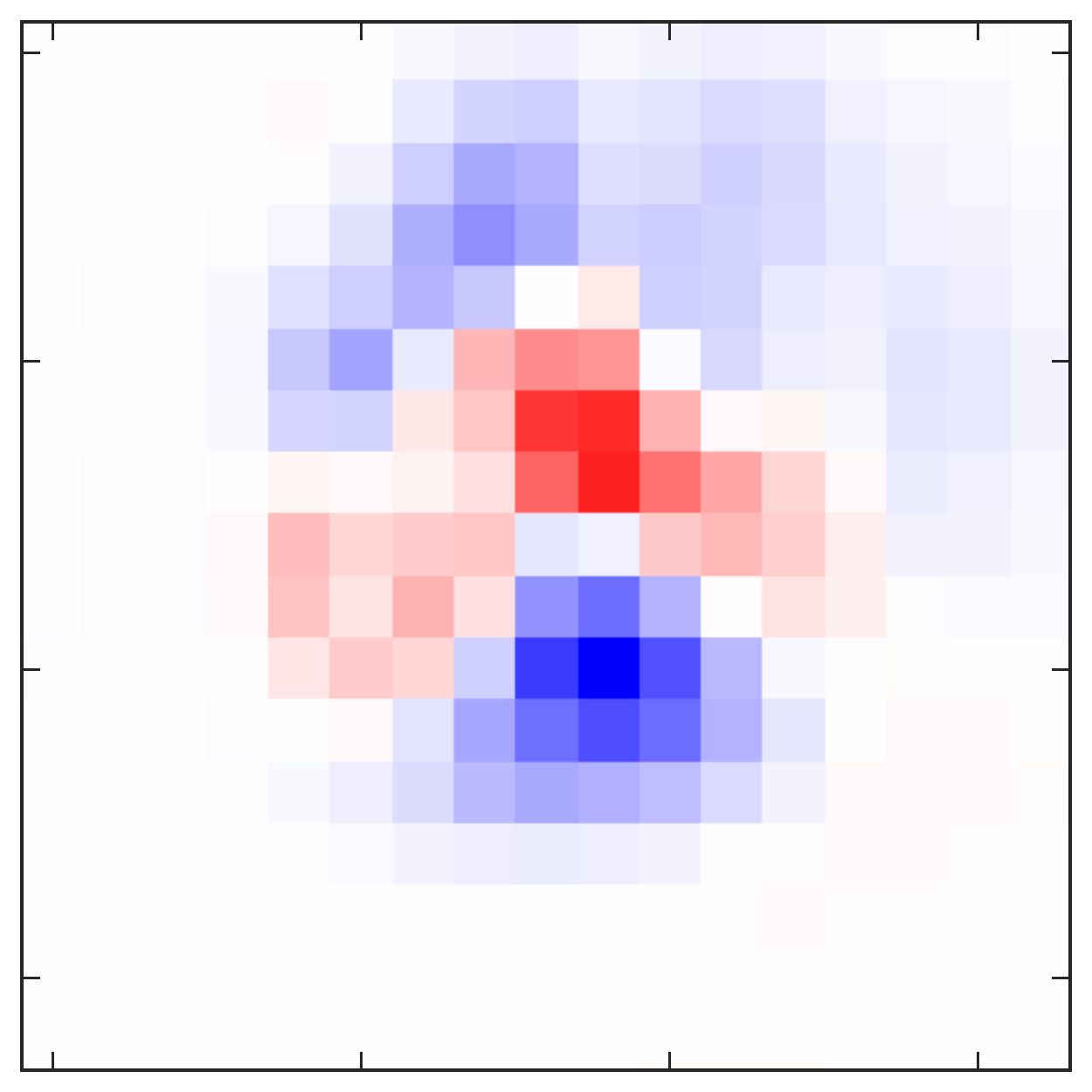}
\caption{}
\label{subfig:topcnn_filt}
\end{center}
\end{subfigure}
\end{center}
\caption{(a) Filters from the first convolutional layer for boosted $W$ tagging with a jet image based CNN tagger~\cite{Oliveira2016JetimagesD} are shown in the top row, while the bottom row shows the average difference between signal and background convolved images from the corresponding filter in the top row. (b) The average difference between signal and background convolved images for several filters of the DeepTop jet image based CNN tagger for boosted top jets~\cite{Kasieczka2017DeeplearningTT}.}
\label{fig:cnn_filt}
\end{figure}

In Figure~\ref{fig:cnn_filt}, the filters in the CNNs for $W$ tagging~\cite{Oliveira2016JetimagesD} and top tagging~\cite{Kasieczka2017DeeplearningTT} with jet images are examined. Several filters from the first convolutional layer of the CNN for $W$ tagging are shown in the top row of Figure~\ref{subfig:wcnn_filt}, and the bottom row shows the corresponding difference between the average convolved image resulting from applying each filter. While the filters are not easy to interpret, one can see dark regions of the filters corresponding to relative locations of large energy depositions in the jet image as well as some intensity gradients that help identify regions where additional radiation may be expected. After applying the filters to sets of signal and background images and taking the difference of the average convolutions to each sample, one can explore how each filter is finding different information in signal and background-like images. The more signal-like regions are shown in red while the more background-like regions are shown in blue.  The blue region at the centers identifies wider energy depositions at the center of the jet image, whilst the signal-like regions at the bottom of such images identify common locations for the subleading energy deposition. There is a strong focus on identifying signal-like radiation between the leading two energy depositions.  Similarly for the DeepTop model, in Figure~\ref{subfig:topcnn_filt} one can see the convolved average image difference for several filters at each layer of the model, where the rows correspond to layer depth from top to bottom.  We again see the tendency for the central region to be background-like, whilst the signal-like regions correspond to different locations of the subleading subjet and radiation between the two leading subjets. One can also see broader radiation patterns which vary depending on the location of the subleading subjet and attempt to identify likely locations of additional subjets in the image.

\begin{figure}[htbp]
\begin{center}
\includegraphics[width=0.24\linewidth]{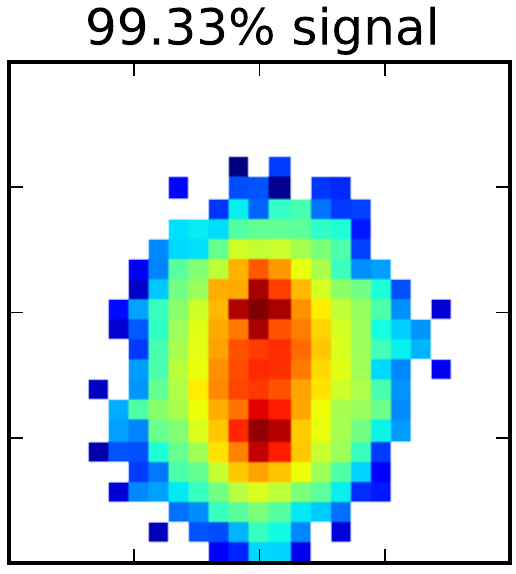}
\includegraphics[width=0.24\linewidth]{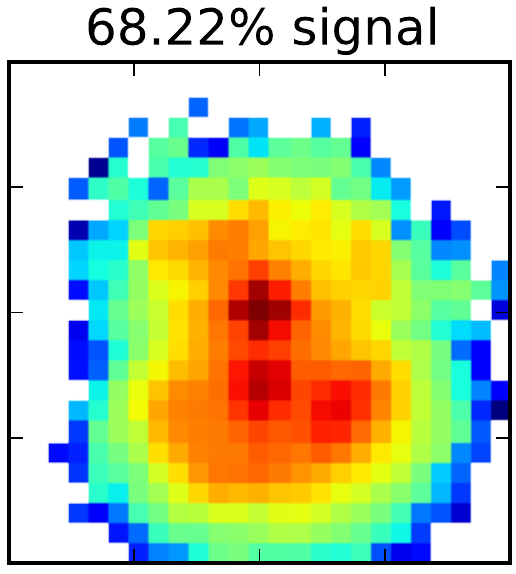}
\includegraphics[width=0.24\linewidth]{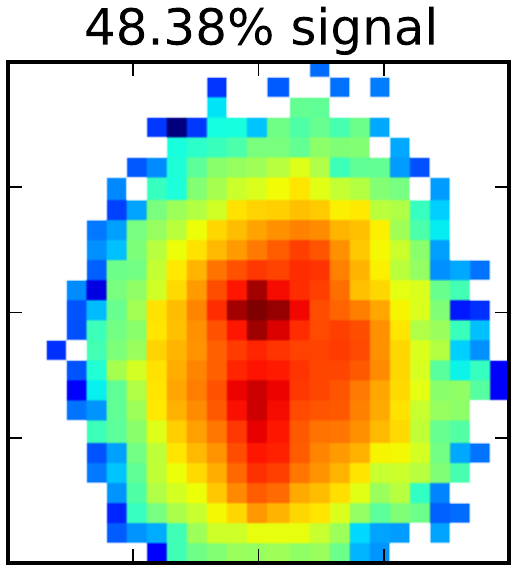}
\includegraphics[width=0.24\linewidth]{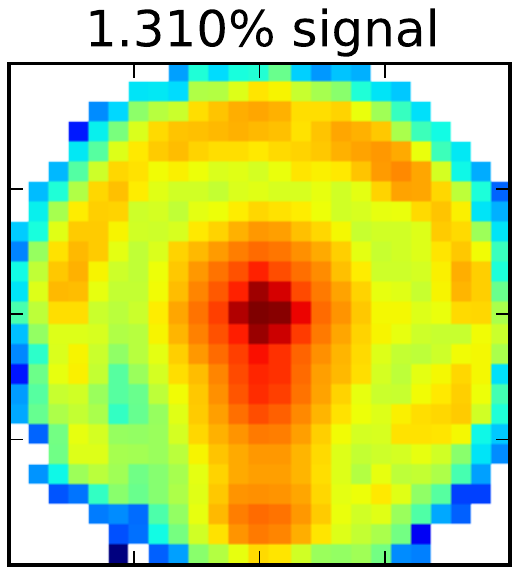}
\end{center}
\caption{The average jet image which most activates a neural in the final layer of a jet image based CNN for boosted $W$ taggeing~\cite{Oliveira2016JetimagesD}. The fraction of signal events for each neuron is noted, thus indicating if the neuron was most activated by signal or background-like images.}
\label{fig:Wcnn_actImg}
\end{figure}

Figure~\ref{fig:Wcnn_actImg} examines the average of the 500 images that lead to the highest activation for each of several neurons in the last (dense) layer of the CNN for $W$ tagging~\cite{Oliveira2016JetimagesD}. The fraction of signal jet images in this sample is also noted, and the images are ordered left to right in terms of this signal fraction. The neuron that activates predominantly on signal jet images has a clear two prong structure and a tight core between these two prongs where radiation is expected. The neuron activating predominantly  on background jet images shows a very different pattern, with a much broader central region where energy may be found and a broad ring around the central region where additional wide angle radiation may be present. These features are in agreement with the known physics of such jets.

\begin{figure}[htbp]
\begin{center}
\begin{subfigure}[h]{0.48\linewidth}
\includegraphics[width=\linewidth]{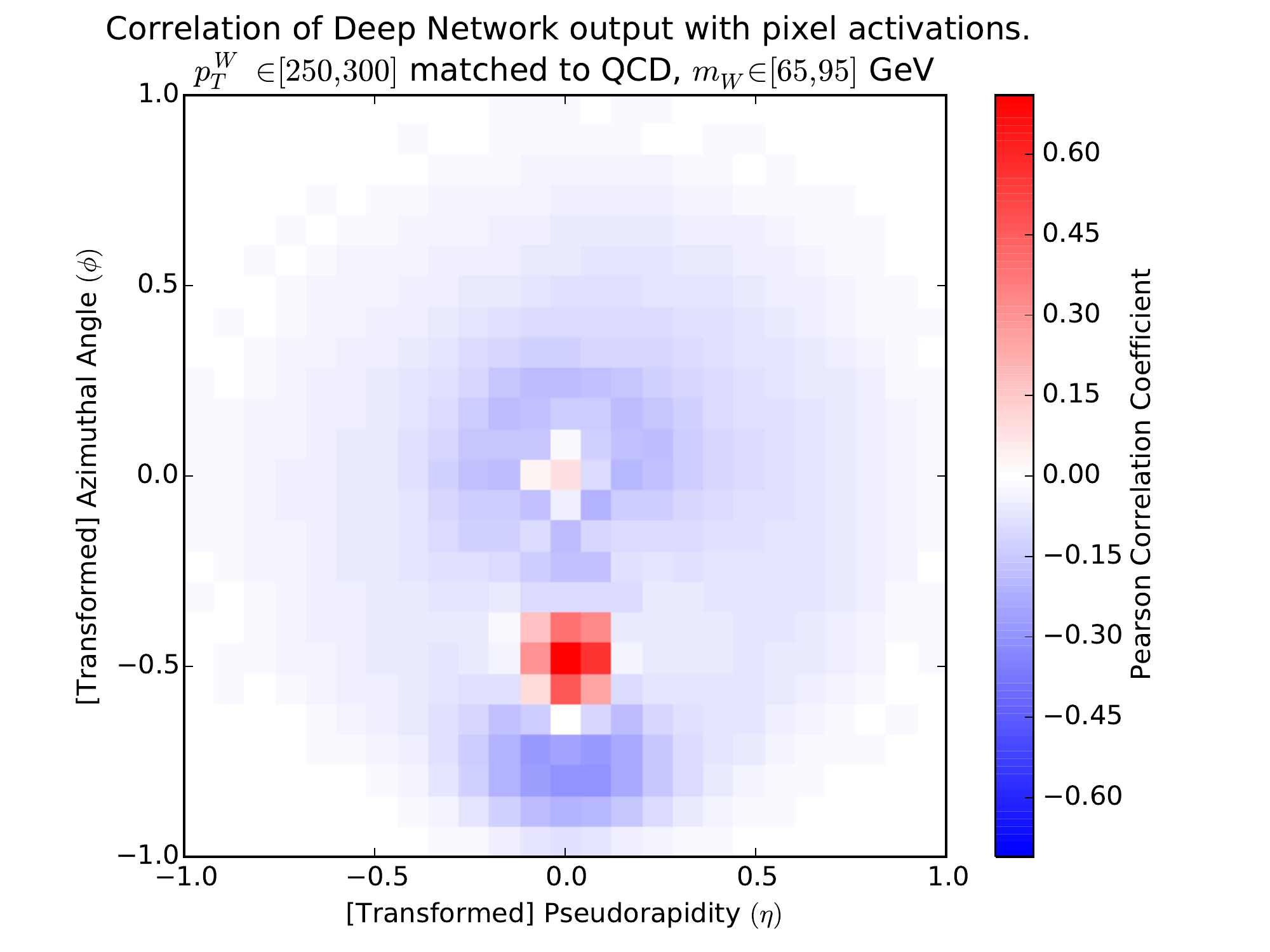}
\caption{}
\label{subfig:W_corr}
\end{subfigure}\quad
\begin{subfigure}[h]{0.48\linewidth}
\includegraphics[width=0.93\linewidth]{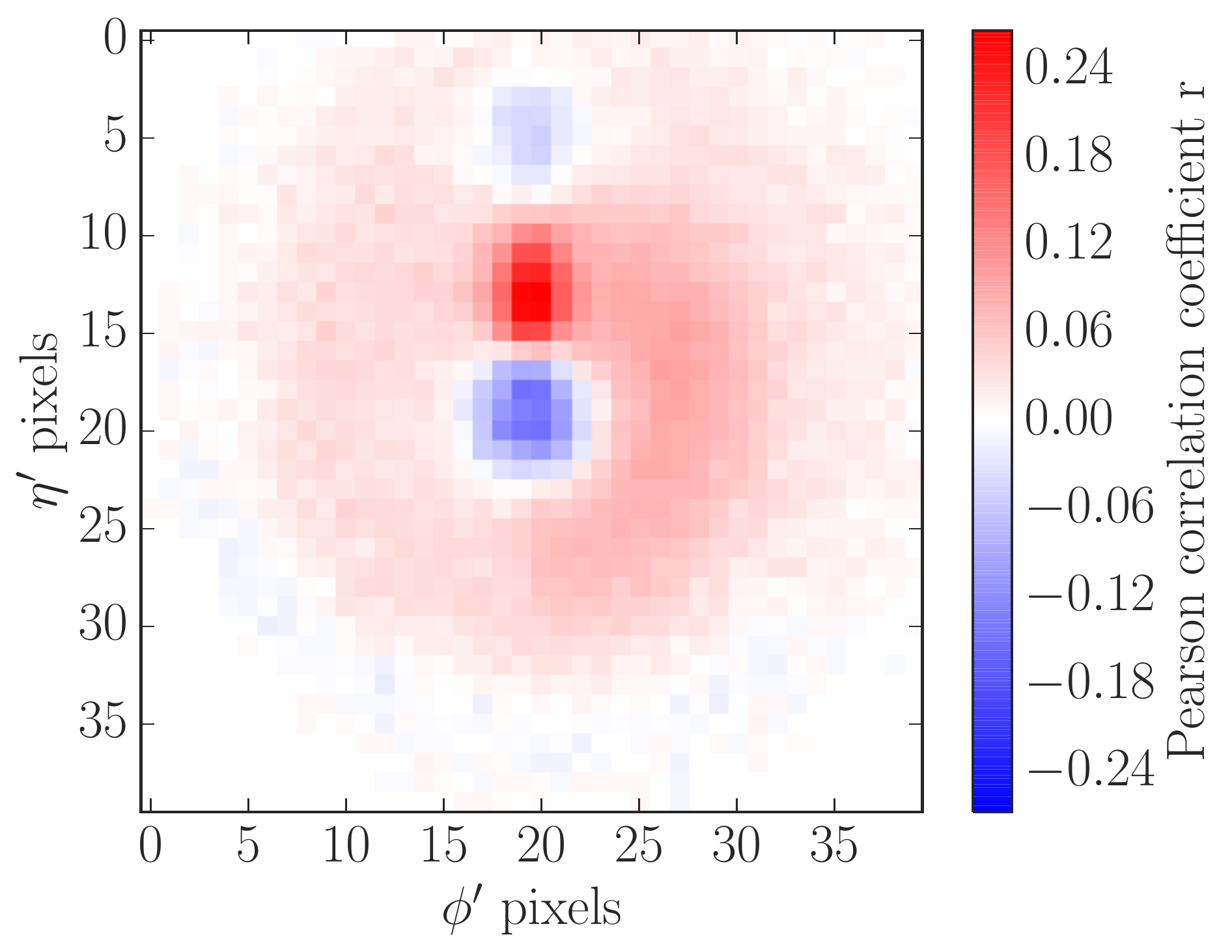}
\caption{}
\label{subfig:top_corr}
\end{subfigure}
\end{center}
\caption{Correlation images showing the Pearson linear correlation coefficient per pixel between jet image pixels and a jet image based CNN tagger output for (a) boosted $W$ tagging~\cite{Oliveira2016JetimagesD}, and (b) boosted top tagging with DeepTop~\cite{Kasieczka2017DeeplearningTT}.}
\label{fig:pcc}
\end{figure}

For a more global view of what the discriminant has learned, one can examine the correlation maps for the CNN $W$ tagging~\cite{Oliveira2016JetimagesD} and the DeepTop model~\cite{Kasieczka2017DeeplearningTT} using full preprocessing in Figure~\ref{fig:pcc}. Structurally they are quite similar\footnote{The relative location of the second subjet was rotated to be below the leading subjet in the case of $W$-tagging and above the leading subjet for DeepTop which leads to the apparent flip in the correlation images over the horizontal axis.}, however the the regions of signal (red) and background (blue) correlation appear inverted. For $W$ tagging, the location of the subleading subjet at the bottom of the image is a strong indicator of signal owing to the fact that $W$ jets have a two particle decay structure which strongly restricts the relative location of the two subjets for a fixed jet $p_T$. This relative location is not as strict in quark/gluon jets and may vary due to additional radiation. The region around the central core of the jet is correlated with background-like images where additional radiation may be found. For top tagging, a strong energy deposition above the central leading energy deposition as well as addition energy depositions, i.e. the third expected subjet in a top quark decay, are correlated with signal-like images. This correlation pattern indicates that the discriminant relies heavily on the identification of the third subjet, as would be expected in a top quark decay.

\section{Other Applications of Jet images}\label{sec:other}

In addition to classification tasks, the approach of using jet images and convolutional layers for processing have also been explored in several other data analysis challenges. We briefly examine some of these applications, showing how this computer vision to jet analysis can be powerful in a variety of settings.

\subsection{Jet Energy Regression and Pileup removal}

Among the major challenges facing analyses utilizing jets at high luminosity hadron colliders is the presence of pileup, or interactions occurring in the same bunch crossing as the primary hard scattering. Pileup interactions lead to additional particles which may fall within the catchment area of a jet and thus are effectively ``noise" in the estimation of jet properties. A variety of techniques have been proposed for pileup mitigation in jets~\cite{Soyez_2019} ranging from subtracting an average pileup energy density from a jet to techniques targeting the classification of each particle in a jet as pileup or from the hard scatter. 

Within the paradigm of jet images, one  approach to pileup mitigation is to predict the per pixel pileup contributions, as is done in the PUMML method~\cite{Komiske2017PileupMW}.  In this technique, a jet can be considered as composed of four components, the charged and neutral hard scatter contributions and the charged and neutral pileup contributions. While the charged components of the hard scatter and pileup are known from charged particle tracking measurements, the neutral hard scatter and pileup components are only observed together in calorimeter measurements. PUMML performs a per pixel regression of the neutral component of the hard scatter contributions to the jet. A multi-channel jet image was used as input, with one channel for each the hard scatter and pileup charged components of the jet, and one channel for the combined neutral component. As the charged contribution measured by tracking detectors has significantly better resolution than the neutral component, a significantly smaller pixel size of $\Delta \eta \times \Delta \phi = 0.025 \times 0.025$ was used for the charged images than the $\Delta \eta \times \Delta \phi = 0.1 \times 0.1$ pixels sizes used for the calorimeter images. Upsampling was then used to create a finer pixel image that matches the resolution of the charged component. These three channel images were then processed by a three layer CNN with a per pixel output prediction of the neutral hard scatter component of the jet.  

\begin{figure}[htbp]
\begin{center}
\begin{subfigure}[h]{0.48\linewidth}
\includegraphics[width=\linewidth]{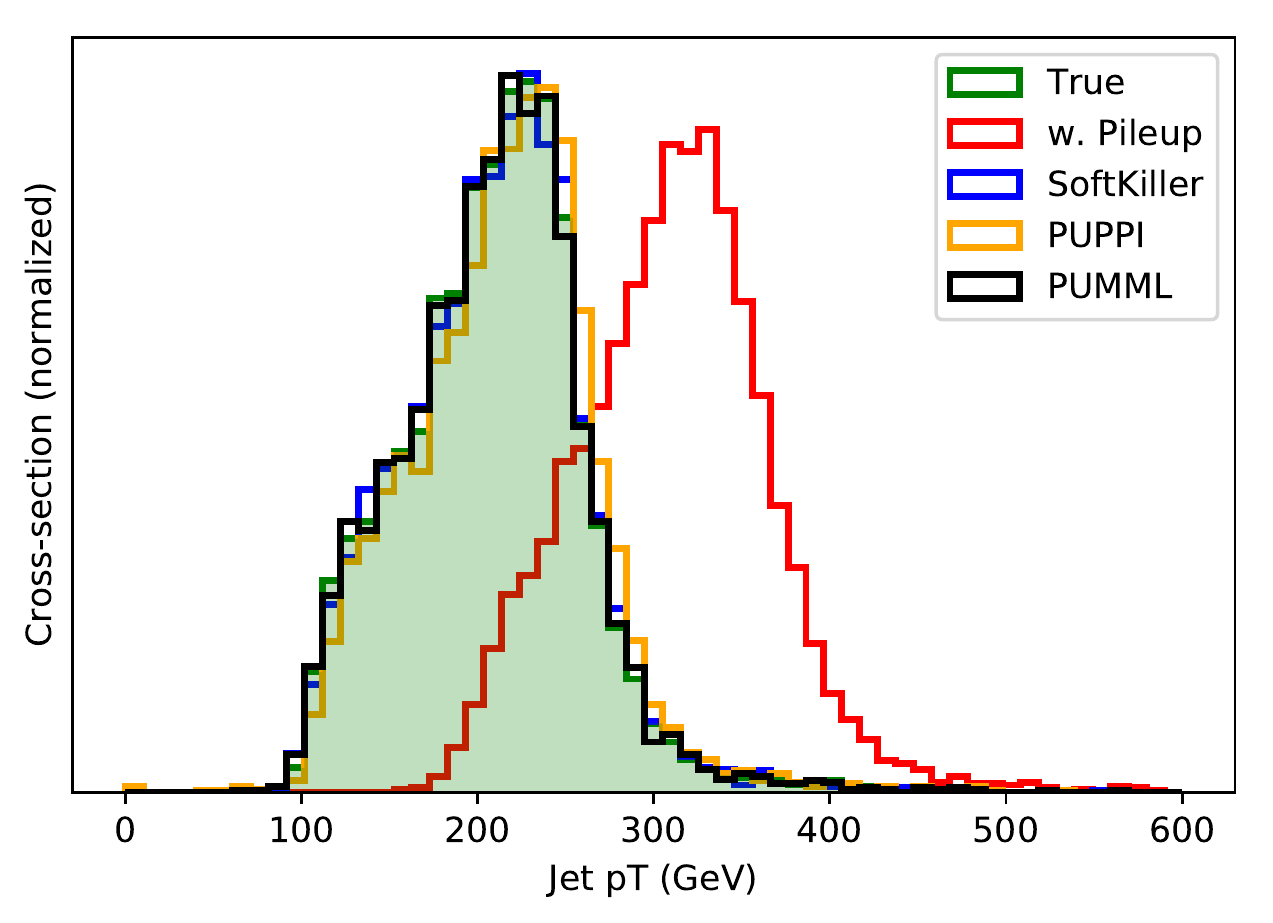}
\caption{}
\label{subfig:pumml_pt}
\end{subfigure}\quad
\begin{subfigure}[h]{0.48\linewidth}
\begin{center}
\includegraphics[width=0.98\linewidth]{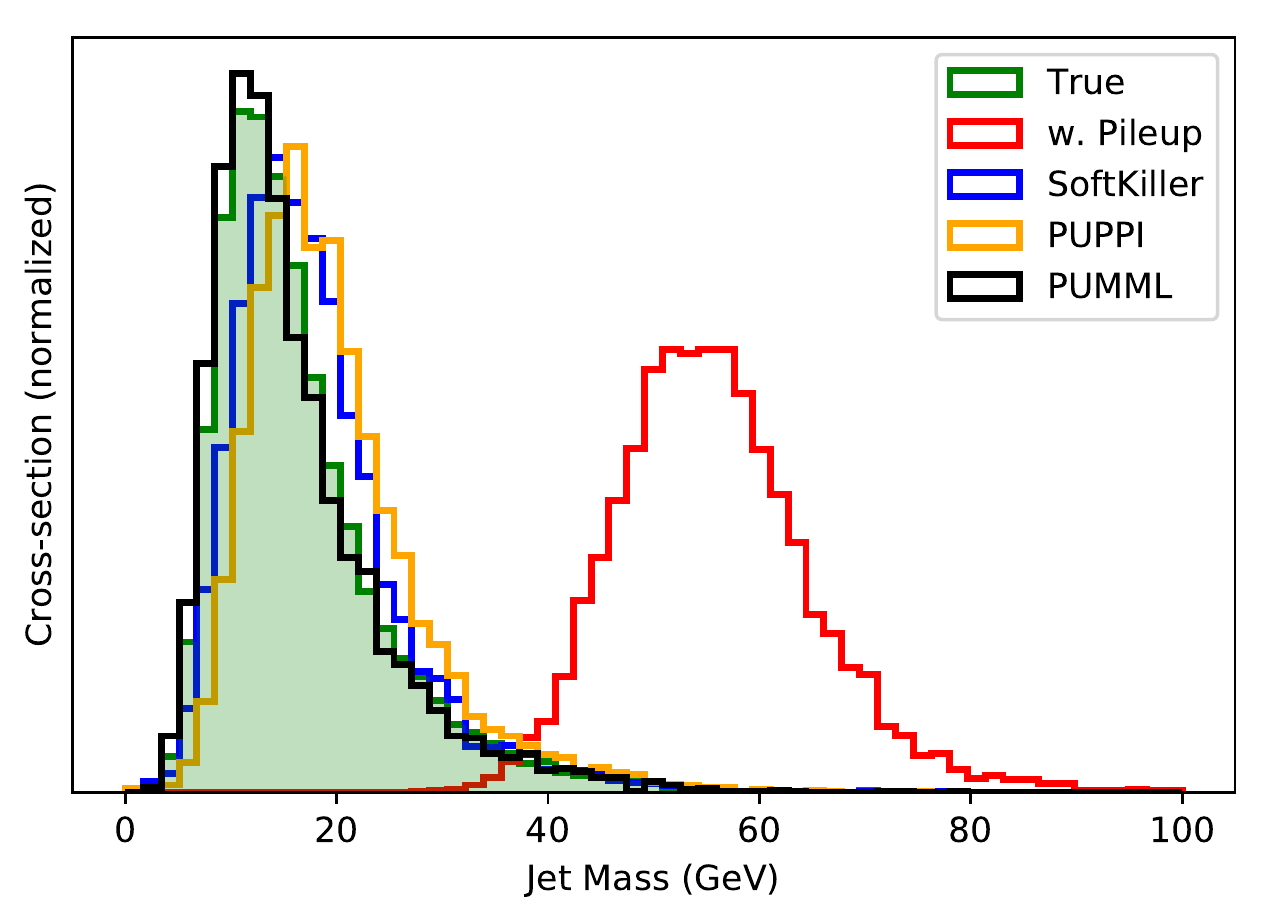}
\caption{}
\label{subfig:pumml_mass}
\end{center}
\end{subfigure}
\end{center}
\caption{The impact of pileup mitigation on (a) jet $p_T$, and (b) jet mass, for various mitigation techniques including the jet image based PUMML algorithm~\cite{Komiske2017PileupMW}.}
\label{fig:pumml}
\end{figure}

The hard scatter neutral component prediction was combined with the known charged component to estimate jet properties and examine the efficacy of the method.  In phenomenological studies using simulated dijet events produced from the decay of a hypothetical new resonance and with an average of 140 additional pileup interactions, the distributions of jet momentum and mass before and after pileup mitigation from PUMML and other methods were compared, as shown in Figure~\ref{fig:pumml}. In terms of momentum prediction, comparing the pileup corrected distributions to the true distribution showed that all methods produced predictions of similar quality, though PUMMPL was seen to have lower per-jet reconstruction error. In terms of jet mass distribution prediction, PUMML was seen to better replicate the underlying true jet mass distribution over other techniques. While not yet applied in an experiment setting, similar ideas applied to pileup reduction for missing energy estimation have been explored on ATLAS in fully simulated events~\cite{ATL-PHYS-PUB-2019-028}  and have shown promising initial results.

\subsection{Generative Models with Jet Images}

\begin{figure}[htbp]
\begin{center}
\begin{subfigure}[h]{0.35\linewidth}
\includegraphics[width=\linewidth]{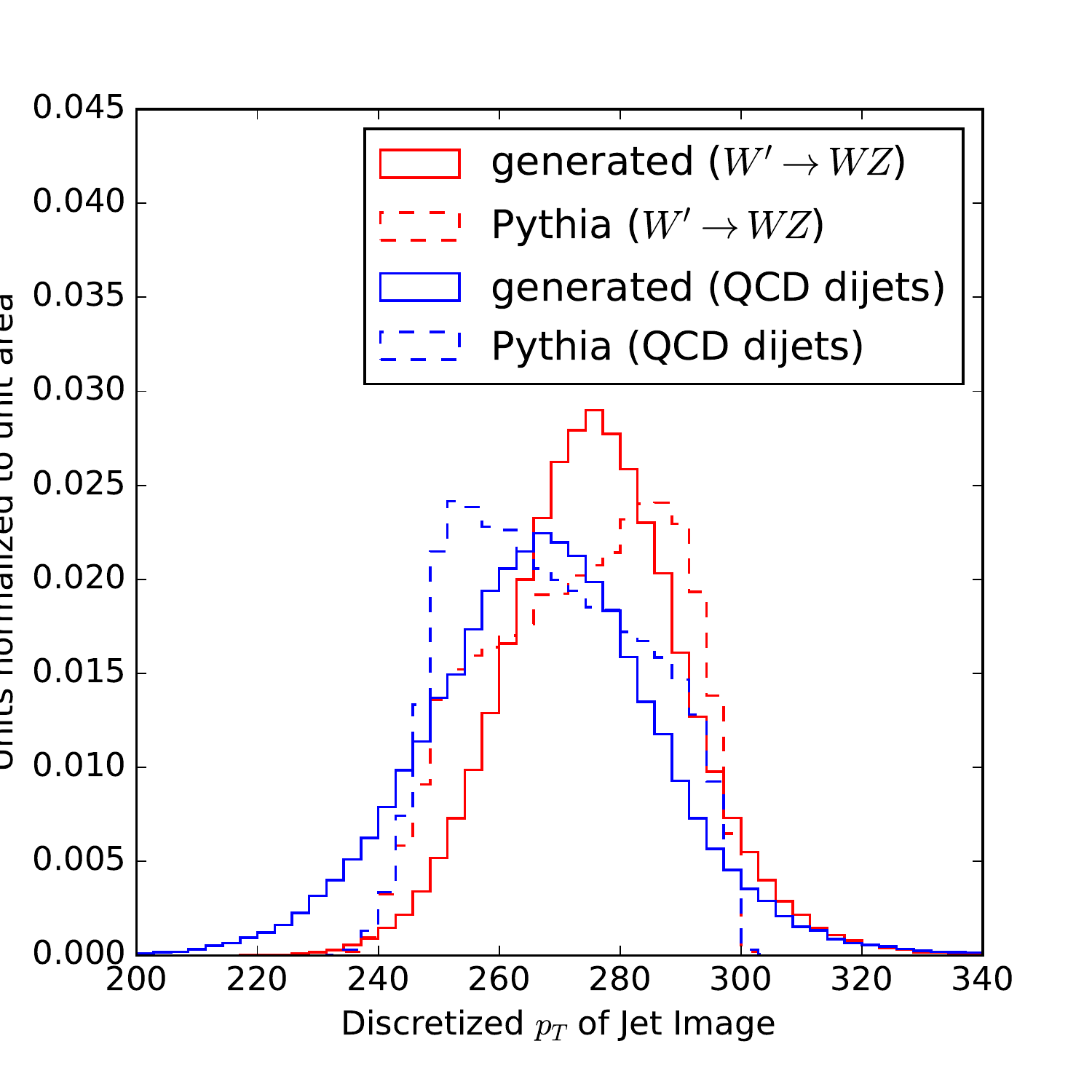}
\caption{}
\label{subfig:gan_pt}
\end{subfigure}\quad
\begin{subfigure}[h]{0.6\linewidth}
\vspace{0.12\linewidth}
\includegraphics[width=\linewidth]{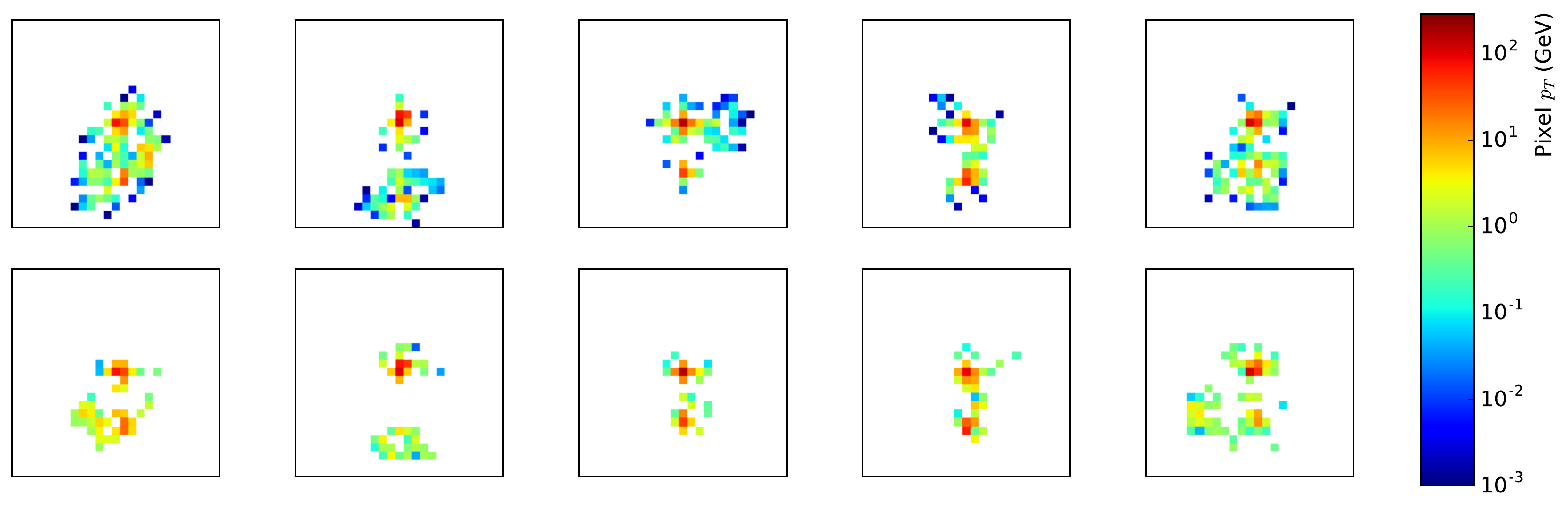}
\vspace{0.07\linewidth}
\caption{}
\label{subfig:gan_neighbor}
\end{subfigure}
\end{center}
\caption{(a) The jet image $p_T$ for $W$ boson jet and quark/gluon jets comparing the GAN generated distribution to the Pythia simulated distribution~\cite{Oliveira2017LearningPP}. (b) A visual comparison of  Pythia simulated (top) and the nearest GAN generated (bottom) jet images.}
\label{fig:gan}
\end{figure}

Among the earliest work applying deep generative models as approximations for HEP high fidelity simulators made use of jet images as the data representation~\cite{Oliveira2017LearningPP}. The aim of this work was to learn the structure of jet images as they may appear in a calorimeter and subsequently draw sample jets from the learned generative model. As a neural generative model can be significantly faster than running a high fidelity simulator, such approaches have the potential to significantly reduce the large simulation times in HEP.  In the phenomenological studies of reference~\cite{Oliveira2017LearningPP}, a generative adversarial network (GAN) setup was used to train a generative model to transform samples from a standard normal distribution into samples of jet images, whilst a second discriminator network was used to penalize the generative model if it could discriminate between real and generated jet images. Locally connected layers as well as convolutional layers were investigated for use in the networks. The distribution of $p_T$ for $W$ boson jets and of quark/gluon jets were compared between the Pythia simulator~\cite{SJOSTRAND2008852,Sj_strand_2006} and the GAN generated images, as seen in Figure~\ref{subfig:gan_pt}.  Figure~\ref{subfig:gan_neighbor} shows a set of Pythia simulated jet images in the top row and their nearest neighbor GAN generated jet images in the bottom row. Both the distribution of jet properties and the general structure of jet images were reasonably well produced by the GAN approach. While not yet reaching the fidelity of HEP simulators, this early work in HEP data generation showed the potential utility in examining fast approximation simulators from deep generative models for HEP.

\subsection{Anomaly Detection}

The use of CNNs to process jet images provides a powerful scheme to learn useful representations of the information contained within a jet. In typical classification tasks, these representations are used for discriminating classes of jets. However, when searching for signs of new physics, one may not know \emph{a priori} the properties of such a new signal but only that such a signal would have properties that deviate from known Standard Model processes. Such anomaly detection tasks are challenging due to the lack of signal knowledge and thus the inability to use standard classifiers for this task. Within the context of a search for jets produced by new particles, recent work has combined the power of CNN representation learning on jet images with autoencoder network architectures~\cite{BALDI198953,10.1145/1390156.1390294} to search for anomalous jets~\cite{Heimel_2019,PhysRevD.101.075021}.  

Autoencoder models are designed to map an input to a compressed latent representation through an ``encoder", and then decompress the latent representation back to original input via a ``decoder". Such models are trained to minimize the ``reconstruction error" computed as the MSE between the original input and the autoencoder output. The reconstruction error can be used to identify inputs that are not well adapted for the compression and reconstruction scheme  learned by the autoencoder. When used for anomaly detection, autoencoders are trained to compress and reconstruct one class of events. Under the assumption that this compression and reconstruction scheme would not be well adapted for inputs from classes different from the training sample, the reconstruction for inputs from new classes is expected to perform poorly and thus lead to a large reconstruction error. 

\begin{figure}[htbp]
\begin{center}
\begin{subfigure}[h]{0.48\linewidth}
\includegraphics[width=\linewidth]{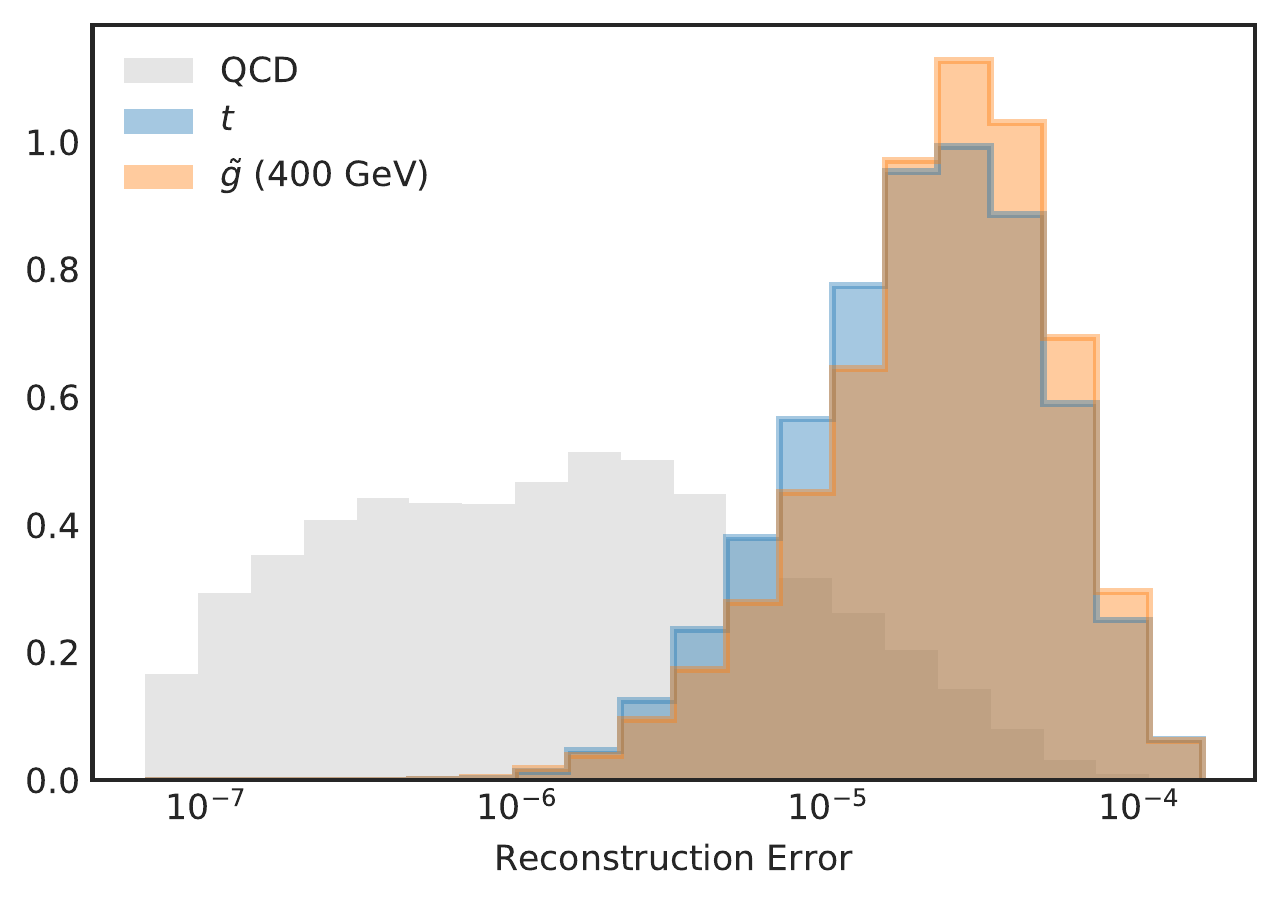}
\caption{}
\label{subfig:anomaly_dist}
\end{subfigure}\quad
\begin{subfigure}[h]{0.48\linewidth}
\begin{center}
\includegraphics[width=0.9\linewidth]{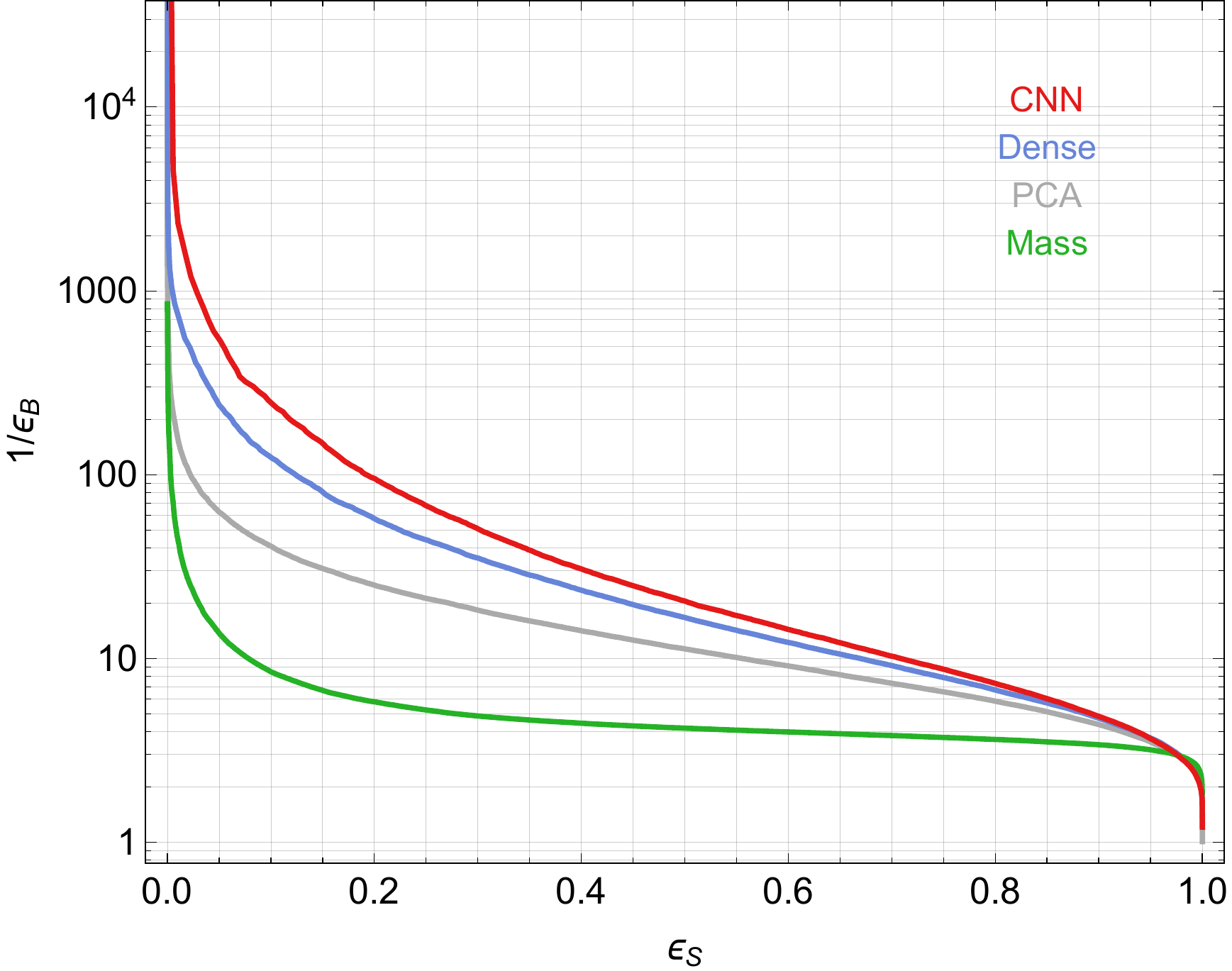}
\caption{}
\label{subfig:anomaly_ROC}
\end{center}
\end{subfigure}
\end{center}
\caption{(a) The distribution of autoencoder reconstruction error trained on quarks / gluon jets, showing potential top or gluino signal distributions. (b) ROC curves for identifying boosted top jets from quarks / gluon jets using autoencoders with various architectures, wherein the jet image based CNN is shown to  outperformance other methods~\cite{PhysRevD.101.075021}.}
\label{fig:anomaly}
\end{figure}

When applied to searches for anomalous jets in phenomenological studies, jet images have been examined as the data representation, and convolutional layers combined with max pooling and with upsampling have been used for the encoder and decoder respectively. In this case, the autoencoder is trained on a background sample of standard quark and gluon jets, and the ability to identify different signal jets was examined. The reconstruction error was used directly to search for excesses of events, as seen in Figure~\ref{fig:anomaly} where the signal was either a sample of top quark jets or jets from a hypothetical new gluino particle.  The distribution of the reconstruction error shows a large separation from the background, denoted QCD, and the potential signal jets. The ROC curve for identifying top jets, produced by scanning a threshold on the reconstruction error, is also shown and compares the CNN based autoencoder with dense architecture based autoencoder applied to a flattened vector of pixel $p_T$'s (denoted Dense), principle components analysis, and applying a threshold only on the jet mass. The jet image+CNN architecture approach dominated the other methods. However, it should be noted that this domination was not seen for gluino jets.

\begin{figure}[htbp!]
\begin{center}
\begin{subfigure}[h]{0.48\linewidth}
\includegraphics[width=\linewidth]{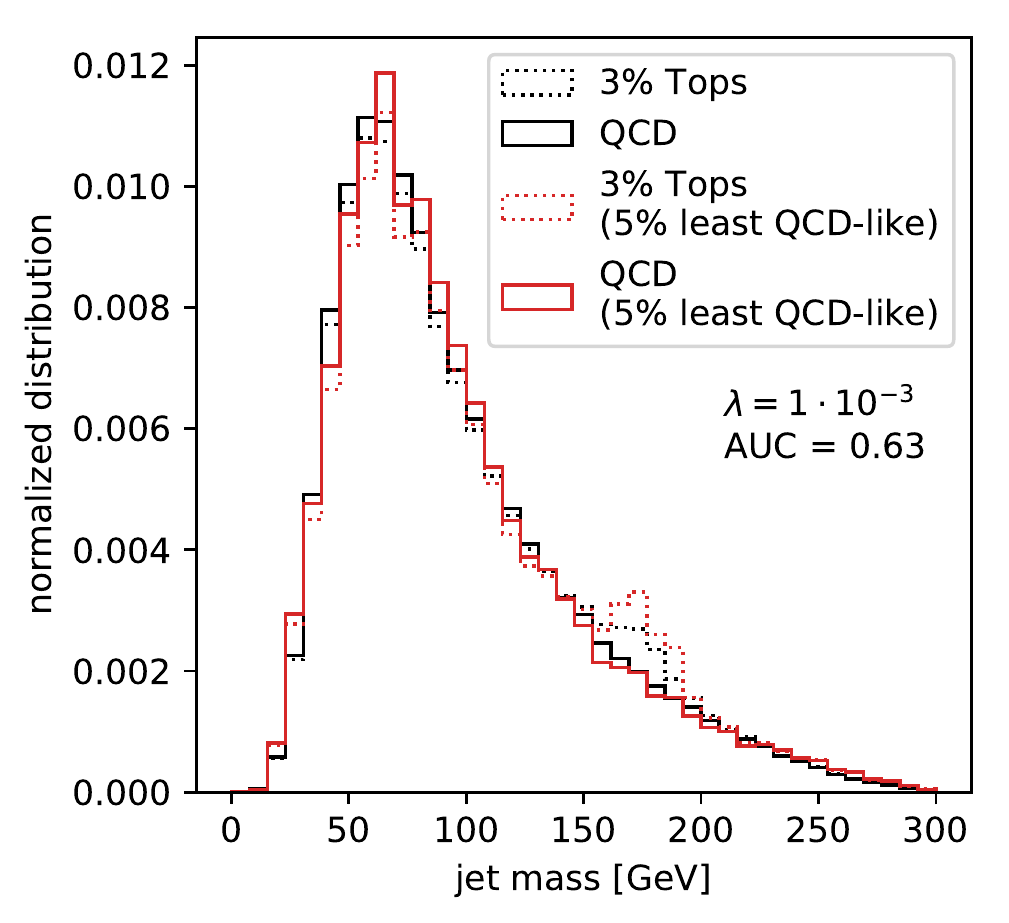}
\caption{}
\label{subfig:anomaly_adv_dist}
\end{subfigure}\quad
\begin{subfigure}[h]{0.48\linewidth}
\begin{center}
\includegraphics[width=0.97\linewidth]{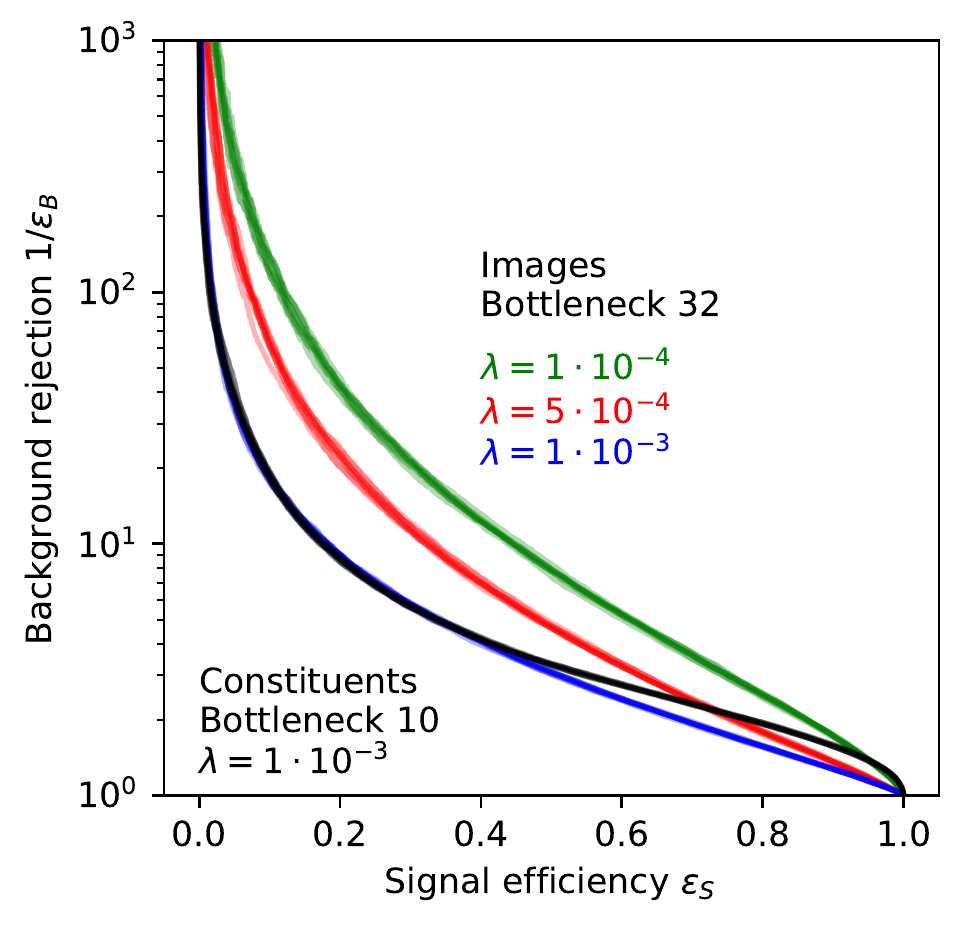}
\caption{}
\label{subfig:anomaly_adv_ROC}
\end{center}
\end{subfigure}
\end{center}
\caption{(a) The jet mass distribution after applying a threshold allowing 3\% or 5\%  quark / gluon jet background efficiency on the jet image based adversarial autoencoder. The background is largely unsculpted and the top jet peaks can be clearly seen~\cite{Heimel_2019}. (b) ROC curves for quark / gluon jet rejection versus top jet efficiency for jet image based adversarial autoencoders with varying strength of adversarial penalty during training~\cite{Heimel_2019}.}
\label{fig:anomaly_adv}
\end{figure}

One challenge with autoencoder approaches for anomalous jet searches is the possibility that the autoencoder reconstruction quality is dependent on the jet mass. In this case the signal identification efficiency could be mass dependent. Moreover, if a bump hunt analysis in the jet mass spectrum is subsequently performed, such a reconstruction error correlation with mass could disturb the jet mass distribution and render the bump hunt strategy infeasible.  To overcome such a challenge, an adversarial approach was investigated in reference~\cite{Heimel_2019}, wherein a second network is simultaneously trained with the autoencoder to predict the jet mass from the autoencoder output whilst the autoencoder is penalized during training if the second network is successful. The resulting adversarial autoencoder performance for identifying a top jet signal can be see in Figure~\ref{fig:anomaly_adv}.  With the adversary in use, the jet mass distribution was kept relatively stable even when applying a threshold on the reconstruction error which only permits a 5\% background jet false positive rate. However, as seen in the ROC curve, increasing the strength $\lambda$ of the adversarial penalty on the autoencoder could significantly decrease the top jet signal sensitivity.

\section{Conclusion}\label{sec:conclusion}

The representation of jets as images has proven highly useful for connecting the fields of high energy physics and machine learning. Through this connection, advanced methods in deep learning and computer vision, primarily with convolutional neural network architectures, have  been applied to the challenges of jet physics and have shown promising performance both in phenomenological studies and in experiments at the LHC. Jet images have seen a broad set of use cases, not only for jet classification but also for energy regression, pileup noise removal, data generation, and anomaly detection.  Image based jet tagging remains an active area of research and broad classes of state of the art deep neural network architectures for computer vision are being explored within the field of high energy physics.

While much of the work presented in this text has been in phenomenological studies using particle level simulations, there remains open question on the applicability of these methods on high-fidelity simulated data and in real experimental data. In more realistic settings, the complexity of the detector and the data-taking conditions, and the challenges of the differences between simulated and real data will be key challenges for understanding and optimizing these models.  Understanding the relationships in realistic data between model accuracy and calibration error, and model complexity / structural assumptions and sensitivity to systematic uncertainties, will be important for the long-term efficacy of these image-based methods. Nonetheless, initial results from both ATLAS and CMS have shown promise, pointing towards the exciting potential for jet imaging in the future.

\section*{Acknowledgments}
This work was supported by the US Department of Energy (DOE) under grant DE-AC02-76SF00515, and by the SLAC Panofsky Fellowship.

\bibliographystyle{tepml}
\bibliography{ChapterJetImage}

%\blankpage
%\printindex[aindx]                 % to print author index
%\printindex                         % to print subject index

\end{document}